\documentclass[]{externalized}
\usepackage[T1]{fontenc}
\usepackage[utf8]{inputenc}
\usepackage{microtype}
\usepackage{inconsolata}
\usepackage{graphicx}
\usepackage{hyperref}
\usepackage{url}
\usepackage{booktabs}
\usepackage{amsfonts}
\usepackage{amsmath}
\usepackage{amssymb}
\usepackage{amsthm}
\usepackage{mathtools}
\usepackage{enumitem}
\usepackage{multirow}
\usepackage{subcaption}
\usepackage{xcolor}
\usepackage{listings}
\usepackage{natbib}
\usepackage{multicol}
\usepackage{pifont}
\usepackage{xspace}
\usepackage{wrapfig}
\AtBeginDocument{%
  \providecommand\BibTeX{{%
    \normalfont B\kern-0.5em{\scshape i\kern-0.25em b}\kern-0.8em\TeX}}}

\makeatletter
\DeclareRobustCommand\onedot{\futurelet\@let@token\@onedot}
\def\@onedot{\ifx\@let@token.\else.\null\fi}

\makeatother

\definecolor{codegreen}{rgb}{0,0.6,0}
\definecolor{codegray}{rgb}{0.5,0.5,0.5}
\definecolor{codepurple}{rgb}{0.58,0,0.82}
\definecolor{backcolour}{rgb}{0.95,0.95,0.92}

\lstdefinestyle{mystyle}{
  backgroundcolor=\color{backcolour},
  commentstyle=\color{codegreen},
  keywordstyle=\color{magenta},
  numberstyle=\tiny\color{codegray},
  stringstyle=\color{codepurple},
  basicstyle=\ttfamily\footnotesize,
  breakatwhitespace=false,
  breaklines=true,
  captionpos=b,
  keepspaces=true,
  numbers=left,
  numbersep=5pt,
  showspaces=false,
  showstringspaces=false,
  showtabs=false,
  tabsize=2
}

\renewcommand\correspondence[1]{\checkdata[\faEnvelopeO\;Email]{#1}}
\usepackage{amsmath,amsfonts}
\usepackage{array}
\usepackage{textcomp}
\usepackage{stfloats}
\usepackage{url}
\usepackage{verbatim}
\usepackage{graphicx}
\usepackage{pifont}

\usepackage{amsthm}

\usepackage{booktabs}
\usepackage{multirow}
\usepackage{lipsum}
\usepackage{hyperref}
\usepackage{breakurl}

\usepackage[edges]{forest}
\definecolor{hidden-draw}{RGB}{20,68,106}
\definecolor{hidden-pink}{RGB}{255,245,247}
\usepackage[framemethod=tikz]{mdframed}
\usepackage{subcaption}
\usepackage{soul}
\usepackage{times}
\usepackage{tabularx}
\usepackage{float}
\usepackage{xcolor} 
\usepackage{latexsym}
\usepackage{bbm}
\usepackage{amsfonts}
\usepackage{dsfont}
\usepackage{graphicx}
\usepackage{caption}
\usepackage{subcaption}
\usepackage{colortbl}
\definecolor{lightgray}{gray}{0.9}
\definecolor{lightgreen}{rgb}{0.9, 1, 0.9}
\usepackage{microtype}
\usepackage{inconsolata}

\usepackage{amsmath,amsfonts,bm}

\def\eqref#1{equation~\ref{#1}}

\def\1{\bm{1}}

\DeclareMathAlphabet{\mathsfit}{\encodingdefault}{\sfdefault}{m}{sl}
\SetMathAlphabet{\mathsfit}{bold}{\encodingdefault}{\sfdefault}{bx}{n}

\usepackage{tikz}
\usetikzlibrary{fadings}
\usetikzlibrary{mindmap,trees}
\usetikzlibrary{arrows,automata,shapes,positioning,shadows,trees}
\usetikzlibrary{shapes,snakes,shadows}
\usetikzlibrary{shapes.arrows}
\usetikzlibrary{calc,shapes, positioning}
\usetikzlibrary{decorations.text}
\usetikzlibrary{matrix,chains,positioning,decorations.pathreplacing,arrows}
\usetikzlibrary{bayesnet}
\usetikzlibrary{shadows.blur}
\usetikzlibrary{shapes.geometric}

\usepackage{pgfplots}
\usepackage{pgfplotstable}
\usepackage{pgfmath,pgffor}
\usepgfplotslibrary{colorbrewer}
\pgfplotsset{compat = 1.14, cycle list/Set1-8}
\usetikzlibrary{pgfplots.statistics, pgfplots.colorbrewer}
\pgfplotsset{compat=1.8}

\tikzstyle{edge}=[-latex',draw=black!90,shorten <=1pt,shorten >=1pt]
\tikzstyle{redge}=[latex'-,draw=black!90,shorten <=1pt,shorten >=1pt]
\tikzstyle{dedge}=[latex'-latex',draw=black!90,shorten <=1pt,shorten >=1pt]

\tikzstyle{block}=[draw, text width=5em,align=center,shape=rectangle, rounded corners, , align=center]
\tikzstyle{nobox}=[align=center]
\definecolor{emb}{RGB}{209,228,252}
\definecolor{hidden-blue}{RGB}{194,232,247}
\definecolor{hidden-orange}{RGB}{224,224,224}
\definecolor{hidden-yellow}{RGB}{242,244,193}
\definecolor{output-purple}{RGB}{219,203,231}
\definecolor{output-green}{RGB}{204,231,207}
\definecolor{hiddendraw}{RGB}{10,128,122}
\definecolor{myred2}{HTML}{F875AA}
\definecolor{mypurple2}{HTML}{D2E0FB}

\definecolor{myred}{HTML}{F8F6F4}
\definecolor{mypurple}{HTML}{FFDFDF}
\definecolor{myyellow}{HTML}{FFF6F6}
\definecolor{mygreen}{HTML}{D2E0FB}
\usepackage{todonotes}

\usepackage{todonotes}

\usepackage{fontawesome5}
\providecommand{\faEnvelopeO}{\faEnvelopeOpen}
\newcommand{\LinkGH}[1]{\href{#1}{\textcolor{black}{\faGithub}}}
\definecolor{hfyellow}{HTML}{FFD21E}
\definecolor{webblue}{HTML}{0000FF}
\newcommand{\LinkHF}[1]{\href{#1}{\textcolor{hfyellow}{\includegraphics[height=1em]{figures/hf-logo.png}}}}
\newcommand{\LinkWEB}[1]{\href{#1}{\includegraphics[height=1em]{figures/web.png}}}

\title{Graph Engineering in the Era of LLM Agents: \\From Individual Intelligence to System Intelligence}

\author[*]{Yuyuan Feng}
\author[*]{Zhishang Xiang}
\author[*]{Chaobin Yang}
\author[*]{Qichao Ma}
\author{Zerui Chen}
\author{Yujing Zhang}
\author{Ke Huang}
\author{Chuanjie Wu}
\author{Zhaoxu Liu}
\author{Yili Wang}
\author{Xin He}
\author{Jiapu Wang}
\author{Zijin Hong}
\author{Hao Chen}
\author{Yuanchen Bei}
\author{Kun Wang}
\author{Shengyuan Chen}
\author{Ningyu Zhang}
\author{Enyan Dai}
\author{Linhao Luo}
\author{Qingyi Pan}
\author{Qi Wang}
\author{Wenqi Fan}
\author{Guangjing Wang}
\author{Na Zou}
\author{Yangqiu Song}
\author{Xin Wang}
\author{Zechao Li}
\author{Xia Hu}
\author{Qing Li}
\author{Xiao Huang}
\author[\dag]{Zhihong Zhang}
\author[\dag]{Jinsong Su}
\author[\dag,\ddag]{Qinggang Zhang}
\author[\dag]{Yi Chang}

\contribution[*]{Equal Contribution}
\contribution[\dag]{Corresponding Authors}
\contribution[\ddag]{Project Leader}

\correspondence{\email{qinggangzhang@jlu.edu.cn}}

\abstract{
Large language models (LLMs) have rapidly evolved from language generation models into autonomous agents capable of solving increasingly complex and long-horizon tasks. This evolution has been accompanied by a series of emerging engineering paradigms, including Prompt Engineering for eliciting model capabilities, Context Engineering for managing information access, Harness Engineering for organizing external tools and resources, and Loop Engineering for enabling continual reflection and self-improvement. However, as real-world tasks grow in complexity, a fundamental limitation of individual intelligence emerges: many tasks inherently require heterogeneous expertise, interdependent subtasks, parallel execution, independent verification, and persistent state, and these requirements exceed the organizational capacity of any single agent. Simply augmenting an individual agent’s capabilities or context cannot resolve this architectural mismatch. Instead, intelligence must be distributed across multiple specialized agents and organized at the system level. We refer to this capability as System Intelligence: the ability of an agent system to organize and coordinate multiple intelligent components into a coherent, adaptive whole that pursues a shared objective. Achieving System Intelligence, however, demands more than merely increasing the number of agents; it requires explicit structures for organizing work, coordinating heterogeneous agents, and maintaining evolving execution states. In this survey, we introduce \textbf{Graph Engineering}, an emerging paradigm for building next-generation agent systems. Unlike previous paradigms that primarily optimize individual interactions or agent-level behaviors, Graph Engineering focuses on constructing explicit, dynamic, and evolving graph structures that represent tasks, agents, and system states. Such graph-based abstractions provide a unified foundation for organizing complex objectives, orchestrating heterogeneous agents, modeling system dynamics, and enabling scalable agent evolution. In this paper, we systematically review the principles, methodologies, and applications of Graph Engineering in the era of LLM agents. All the related resources, including research papers, open-source data, and projects, are collected for the community at \textcolor{blue}{\url{https://github.com/DEEP-JLU/Awesome-Graph-Engineering}}.
}

\begin{document}
\maketitle
\newpage
\tableofcontents
\newpage




\begin{figure*}[tbp]
\vspace{-5mm}
    \centering
    \includegraphics[width=1.\linewidth, trim=0cm 0cm 0cm 0cm,clip]{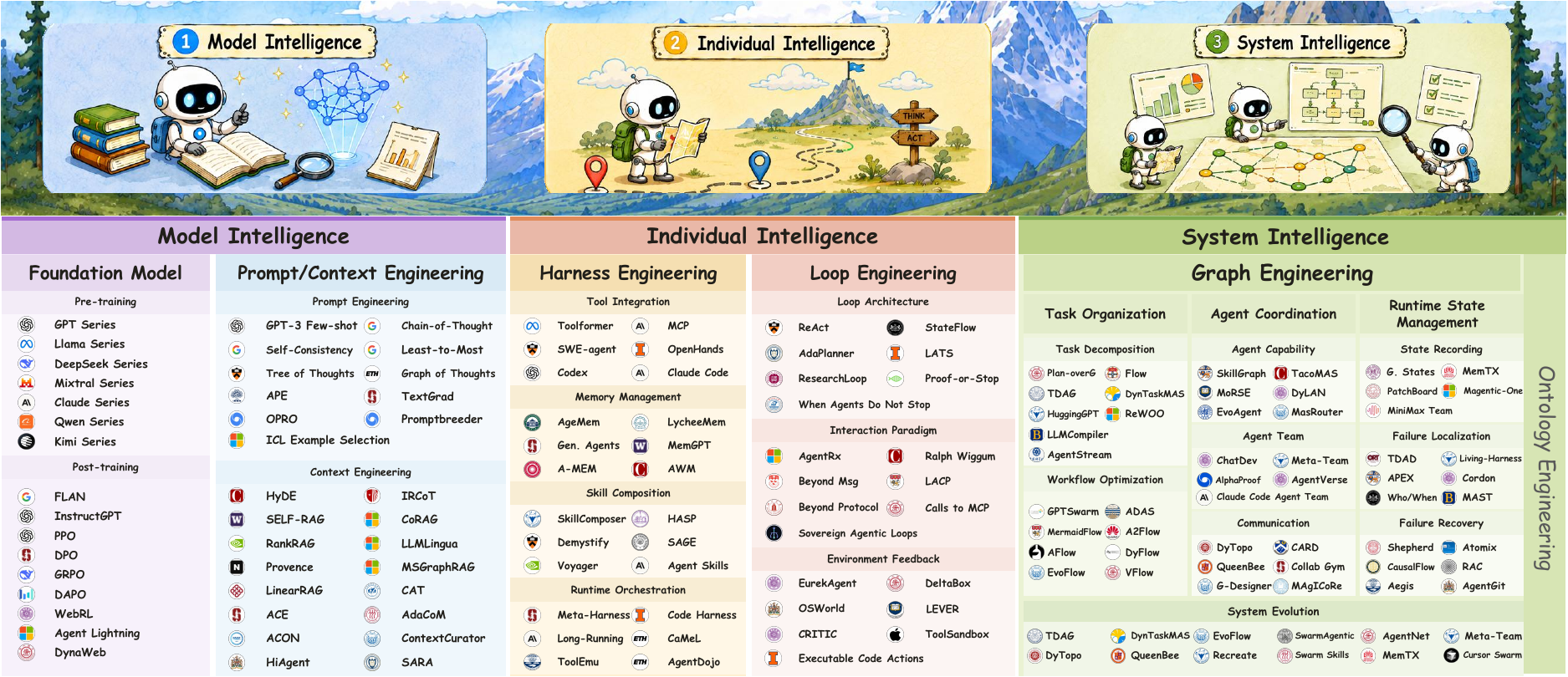}
    \vspace{-5mm}
    \caption{ \textbf{Overview From Model Intelligence to System Intelligence.}
    Prompt and Context Engineering elicit and condition the access of foundation models to realize model intelligence. Harness and Loop Engineering extend and orchestrate agentic capabilities to enable Individual Intelligence.
    Graph Engineering builds on the foundation of individual agents through Task Organization, Agent Coordination, and Runtime State Management, empowering System Intelligence.}
    \label{fig:trend}
\end{figure*}
\vspace{-5mm}

\section{Introduction}
Large language models (LLMs) have rapidly evolved into a foundational component of modern intelligent systems, driven by substantial advances in language understanding, reasoning, generation, and decision making~\cite{touvron2023llama, openai2024gpt4technicalreport, chowdhery2023palm, brown2020language, deepseekai2024deepseekv3,hong2025next}.
This progress has largely followed two complementary directions: strengthening the capabilities encoded in model parameters during training~\cite{dai2024deepseekmoe,liu2025drgrpo,ouyang2022training} and improving how these capabilities are activated and utilized at inference time~\cite{wei2022chain,besta2024graph,anthropic2025harnesses,qin2024toolllm}. Specifically, early research primarily focused on the former, using large-scale pre-training and post-training to expand and refine the knowledge and reasoning capabilities of individual models~\cite{brown2020language, deepseekai2024deepseekv3, wei2022finetuned,liu2025drgrpo}. More recently, increasing attention has shifted toward inference-time engineering, where \texttt{Prompt Engineering} and \texttt{Context Engineering} serve as complementary approaches for shaping model behavior. \texttt{Prompt Engineering}~\cite{wei2022chain,wang2023selfconsistency,besta2024graph} structures task descriptions, instructions, and constraints to guide model reasoning, whereas \texttt{Context Engineering}~\cite{lewis2020retrieval,edge2024graphrag,liu2024lost} determines and organizes the task-relevant information, external knowledge and intermediate results available to the model during inference. Together, these techniques constitute the primary mechanisms for developing and eliciting \textbf{Model Intelligence}, which characterizes the ability of an individual model to leverage its knowledge and reasoning capabilities to solve tasks within a given context.
Despite recent advances in Model Intelligence, its scope remains bounded by what an individual model can access, maintain, and accomplish within a standalone inference process. Many real-world tasks, however, require access to external knowledge and tools, interaction with dynamic environments, and iterative adaptation over extended execution horizons~\cite{lewis2020retrieval,wang2024codeact,schick2023toolformer}. These requirements have motivated a paradigm shift from developing more capable language models toward constructing autonomous systems around these models, leading to the emergence of LLM-based agents. Conceptually, such an agent can be characterized as:
$$\mathrm{Agent} = \mathrm{Loop} (\mathrm{LLM} + \mathrm{Harness}).$$ Here, \texttt{Harness Engineering} extends the capability boundary of the model by connecting it to heterogeneous resources and functional components, including external knowledge~\cite{zhang2025survey, asai2024selfrag}, tools~\cite{qin2024toolllm, qian2025toolrl}, memory~\cite{xu2025amem, yang2026graphmemory}, and skills~\cite{wang2023voyager,zhang2026skillcomposer}. \texttt{Loop Engineering} further organizes these capabilities into a persistent execution process through iterative cycles of planning, action, observation, verification, and adaptation~\cite{xia2026researchloop,hou2026agents}. In this sense, \texttt{Harness Engineering} determines what capabilities and resources are available to the agent and how these capabilities are orchestrated during execution, whereas \texttt{Loop Engineering} defines how the agent continuously interacts with the environment and adapts its behavior in response to evolving task states and external feedback. Together, they transform an LLM from a response-generating model into a goal-directed autonomous entity capable of sustained interaction with environments, giving rise to what we term \textbf{Individual Intelligence}, the ability of an individual agent to extend model-level reasoning into persistent, goal-directed execution through resource orchestration and iterative interaction with the environment.
However, as real-world tasks grow in complexity, a fundamental limitation of individual intelligence emerges: many tasks inherently require heterogeneous expertise, interdependent subtasks, parallel execution, independent verification, and persistent state, and these requirements exceed the organizational capacity of any single agent~\cite{geng2025alas,fu2026benchagent,khan2026concurrency, qian2025scaling, yang2026graph,dong2026towards}. For example, real-world tasks, like scientific discovery and software engineering, often require specialized reasoning to be performed concurrently, intermediate results to be exchanged and validated, and execution states to be maintained across long-running processes. When such tasks are executed within a single agent loop, these heterogeneous processes are forced into a common context and a centralized execution trajectory. This creates several structural bottlenecks: task-relevant information competes for contextual capacity, dependent operations are mediated through a predominantly sequential control process, and the states of different tasks and agents are forced into a single shared context, making it impossible to isolate concurrent work, synchronize on shared results, or recover partial progress independently. 

As tasks become increasingly heterogeneous, interdependent, and long-horizon, simply augmenting an individual agent’s capabilities or context cannot resolve this problem~\cite{guo2025syncmind, liu2024lost,qian2025scaling}. Instead, intelligence must be distributed across multiple specialized agents and organized at the system level~\cite{kim2025towards, qi2026beyondindividual}. We refer to this capability as \textbf{System Intelligence}, as shown in Fig.~\ref{fig:trend}: the ability of an intelligent system to decompose and organize complex objectives, allocate responsibilities across heterogeneous computational agents, coordinate their interdependent execution, and maintain system-level state throughout the task lifecycle. Importantly, \texttt{System Intelligence} is not equivalent to simply increasing the number of agents. A multi-agent system may contain multiple capable agents while still lacking effective work organization, clear responsibility boundaries, coordination mechanisms, or consistent state management~\cite{hao2026evolve,anthropic_claude_code_agent_teams_2026,kimi_k25_2026,minimax_agent_team_2026}. Moving from \texttt{Individual Intelligence} to \texttt{System Intelligence} therefore requires a shift from agent-centric execution to system-level organization, where the central challenge is no longer how to replicate or specialize agents, but how heterogeneous components can be organized and coordinated to operate as a coherent system.

To this end, we introduce \textbf{Graph Engineering}, a novel engineering paradigm in which graph structures are used to organize and control task execution, agent coordination, and runtime state evolution for system-level intelligence.
From a system perspective, as shown in Fig.~\ref{fig:taxonomy}, \texttt{Graph Engineering} addresses three fundamental organizational problems. (i) {Task Organization} determines how a global objective is decomposed into executable units and how dependencies, ordering, concurrency, and verification constraints among them are represented. (ii) {Agent Coordination} determines how these units of work are mapped onto heterogeneous agents and computational components, and how their communication, delegation, synchronization, and result integration are structured. (iii) {Runtime State Management} determines how the evolving state of execution is represented and maintained, enabling the system to track progress, reconcile concurrent updates, preserve provenance, isolate failures, and recover or adapt when execution deviates from plan.
Together, these dimensions transform graph structures from static representations into operational mechanisms for organizing and governing the execution of agent systems, thereby providing a structural foundation for \texttt{System Intelligence}.

Generally, we make the following contributions:

\begin{itemize}
    \item Section~\ref{sec:preliminaries} defines the core concepts of Individual Agents and Agent Systems. It characterizes an Individual Agent through its Foundation Model, Agent Harness, Agent Loop, and local runtime state, and an Agent System through its agent team, shared resources, environment, coordination mechanisms, and system-level state.

    \item Section~\ref{sec:model2individual} traces the evolution from Model Intelligence to Individual Intelligence and establishes the need for System Intelligence. It clarifies the roles of foundation-model development, Prompt and Context Engineering, and Harness and Loop Engineering, and identifies the structural limitations that motivate Graph Engineering.
    
    \item Section~\ref{sec:graph_engineering} formulates Graph Engineering as a system-level engineering paradigm and organizes the field into three interconnected views: \texttt{Task Organization} for structuring tasks, dependencies, and execution processes; \texttt{Agent Coordination} for organizing heterogeneous components and their collaboration; and \texttt{Runtime State Management} for tracking runtime states, supporting recovery, and enabling adaptation.

    \item Sections~\ref{sec:challenges} and~\ref{sec:direction} outline open challenges and future research directions toward ontology engineering, dynamic and self-evolving graph systems, and graph-native agent operating systems.

    \item Section~\ref{sec:benchmarks} synthesizes representative benchmarks, datasets, and executable environments across Model, Individual, and System Intelligence. It further identifies evaluation principles and open challenges concerning structural fidelity, operational correctness, system evolution, and governance.

    \item Section~\ref{sec:open-source-libraries} surveys representative open-source libraries and engineering systems across the three intelligence levels. It examines how existing software stacks support model development, persistent agent runtimes, and multi-component orchestration, while highlighting gaps in interoperability, cross-run structural evolution, and state provenance.

    \item Section~\ref{sec:applications} reviews applications of Graph Engineering across software engineering, scientific discovery, healthcare, enterprise workflows, general-purpose digital agents, and social and economic simulation. It shows that Work Organization, Agent Coordination, and Runtime State Management are increasingly common in practice, whereas persistent System Evolution remains limited.
\end{itemize}

\providecommand{\surveyfull}{\textcolor{green!50!black}{\ding{51}}}
\providecommand{\surveypart}{\textcolor{orange!85!black}{\ensuremath{\circ}}}
\providecommand{\surveynone}{\textcolor{gray}{--}}

\section{Preliminaries}
\label{sec:preliminaries}

We first clarify two fundamental concepts underlying Graph Engineering: the \emph{Individual Agent} and the \emph{Agent System}. The former defines the capabilities and operational mechanism of an autonomous entity, whereas the latter describes how multiple agents and supporting components are organized into a dynamic intelligent system.

\subsection{Individual Agent}

An Individual Agent is an autonomous computational entity that perceives its environment, makes decisions, executes actions, and adapts its behavior according to feedback. It consists primarily of a \emph{Foundation Model} and an \emph{Agent Harness}. The Foundation Model serves as the cognitive core, providing capabilities such as language understanding, reasoning, planning, and content generation. The Agent Harness extends these intrinsic capabilities through interfaces for perception and context construction, memory and knowledge access, tool invocation, reusable skills, and runtime governance.

These components are organized over time by an iterative Agent Loop, which repeatedly performs perception, reasoning, action, feedback processing, and state update. An Individual Agent can therefore be abstracted as
\begin{equation}
    \mathcal{A}_i =
    \operatorname{Loop}
    (\mathcal{F}_i, \mathcal{H}_i; s_i^t),
\end{equation}
where $\mathcal{F}_i$ denotes the Foundation Model, $\mathcal{H}_i$ denotes the Agent Harness, and $s_i^t$ denotes the runtime state of agent $i$ at time $t$. In this formulation, the Foundation Model determines the agent's intrinsic cognitive capabilities, the Harness determines the resources and action spaces it can access, and the Loop determines how these capabilities are continuously employed.

\subsection{Agent System}

An Agent System extends the Individual Agent abstraction to a collection of agents that operate through shared resources, external environments, and coordination mechanisms. At time $t$, an Agent System can be represented as
\begin{equation}
    \mathcal{S}^t =
    \left(
    \mathbb{A}^t,
    \mathcal{R}^t,
    \mathcal{E}^t,
    \boldsymbol{\Pi}^t,
    \mathbf{x}^t
    \right),
\end{equation}
where the system consists of the following elements:

\begin{itemize}
    \item \textbf{Agent Team $\mathbb{A}^t$:} A collection of Individual Agents, each with its own Foundation Model, Harness, Agent Loop, and local runtime state. Agents may assume different roles, possess different capabilities, and undertake different tasks.

    \item \textbf{Shared Resources $\mathcal{R}^t$:} Resources and services accessible to multiple agents, including tools, model services, memory, knowledge bases, verifiers, and human support. Each agent accesses these resources through its Harness.

    \item \textbf{Environment $\mathcal{E}^t$:} The external environment that the Agent System perceives and acts upon. It provides observations and feedback and evolves in response to agent actions.

    \item \textbf{Coordination Mechanisms $\boldsymbol{\Pi}^t$:} Mechanisms that determine how agents assign tasks, exchange information, integrate results, resolve conflicts, and handle failures.

    \item \textbf{System State $\mathbf{x}^t$:} The system-level runtime information, including task progress, shared results, agent availability, resource status, environmental changes, and failure records. Unlike the local state $s_i^t$ of an Individual Agent, $\mathbf{x}^t$ describes the operational condition of the entire system at time $t$.
\end{itemize}

The behavior of an Agent System therefore depends not only on the capabilities of its Individual Agents, but also on how shared resources are used, how agents coordinate with one another, and how the system state evolves throughout execution.

\definecolor{hidden-draw}{RGB}{255,255,255}

\tikzset{
    my-box/.style={
        rectangle,
        draw=hidden-draw,
        rounded corners,
        align=left,
        text opacity=1,
        minimum height=1.5em,
        minimum width=5em,
        inner sep=2pt,
        fill opacity=.8,
        line width=0.8pt,
    },
    leaf-head/.style={
        my-box,
        minimum height=1.5em,
        draw=gray!80,
        fill=gray!15,
        text=black,
        font=\normalsize,
        inner xsep=2pt,
        inner ysep=4pt,
        line width=0.8pt,
    },
    leaf-datasets/.style={
        my-box,
        minimum height=1.5em,
        draw=orange!80,
        fill=orange!15,
        text=black,
        font=\normalsize,
        inner xsep=2pt,
        inner ysep=4pt,
        line width=0.8pt,
    },
    leaf-methods/.style={
        my-box,
        minimum height=1.5em,
        draw=red!70,
        fill=red!15,
        text=black,
        font=\normalsize,
        inner xsep=2pt,
        inner ysep=4pt,
        line width=0.8pt,
    },
    leaf-metrics/.style={
        my-box,
        minimum height=1.5em,
        draw=cyan!70,
        fill=cyan!15,
        text=black,
        font=\normalsize,
        inner xsep=2pt,
        inner ysep=4pt,
        line width=0.8pt,
    },
    modelnode-datasets/.style={
        my-box,
        minimum height=1.5em,
        draw=orange!80,
        fill=white,
        text=black,
        font=\normalsize,
        inner xsep=2pt,
        inner ysep=4pt,
        line width=0.8pt,
    },
    modelnode-methods/.style={
        my-box,
        minimum height=1.5em,
        draw=red!70,
        fill=white,
        text=black,
        font=\normalsize,
        inner xsep=2pt,
        inner ysep=4pt,
        line width=0.8pt,
    },
    modelnode-metrics/.style={
        my-box,
        minimum height=1.5em,
        draw=cyan!70,
        fill=white,
        text=black,
        font=\normalsize,
        inner xsep=2pt,
        inner ysep=4pt,
        line width=0.8pt,
    }
}

\begin{figure*}[t]
    \centering
    \vspace{-5mm}
    \resizebox{1\textwidth}{!}{
        \begin{forest}
            forked edges,
            for tree={
                grow=east,
                reversed=true,
                anchor=base west,
                parent anchor=east,
                child anchor=west,
                base=left,
                font=\normalsize,
                rectangle,
                draw=hidden-draw,
                rounded corners,
                align=left,
                minimum width=1em,
                edge+={darkgray, line width=1pt},
                s sep=3pt,
                inner xsep=0pt,
                inner ysep=3pt,
                line width=0.8pt,
                ver/.style={
                    rotate=90,
                    child anchor=north,
                    parent anchor=south,
                    anchor=center
                },
            },
            [
                 Overall Taxonomy, leaf-head, ver
                [
                    Model Intelligence,
                    leaf-datasets,
                    text width=8em
                    [
                        Parameterized\\ Training,
                        leaf-datasets,
                        text width=7.5em
                        [
                            Pre-Training,
                            leaf-datasets,
                            text width=9em
                            [
                                GPT-3~\cite{brown2020language}
                                Gopher~\cite{rae2021gopher}
                                PaLM~\cite{chowdhery2023palm}
                                LLaMA~\cite{touvron2023llama}
                                Scaling Laws~\cite{kaplan2020scaling}
                                Chinchilla~\cite{hoffmann2022training}
                                \\
                                Switch Transformer~\cite{fedus2022switch}
                                Mixtral~\cite{jiang2024mixtral}
                                DeepSeekMoE~\cite{dai2024deepseekmoe}
                                Llama 3~\cite{grattafiori2024llama3}
                                DeepSeek-V3~\cite{deepseekai2024deepseekv3}
                                \\
                                Deduplication~\cite{lee2022dedup}
                                FineWeb~\cite{penedo2024fineweb}
                                DataComp-LM~\cite{li2024datacomplm}
                                Qwen2.5~\cite{qwen2024qwen25}
                                Qwen3~\cite{yang2025qwen3}
                                Kimi K2~\cite{kimi2025k2},
                                modelnode-datasets,
                                text width=50.7em
                            ]
                        ]
                        [
                            Post-Training,
                            leaf-datasets,
                            text width=9em
                            [
                                FLAN~\cite{wei2022finetuned}
                                T0~\cite{sanh2022multitask}
                                InstructGPT~\cite{ouyang2022training}
                                Flan Collection~\cite{longpre2023flan}
                                Constitutional AI~\cite{bai2022constitutional}
                                RLAIF~\cite{lee2023rlaif}
                                DPO~\cite{rafailov2023direct}
                                \\
                                Tulu 3~\cite{lambert2024tulu3}
                                DeepSeekMath~\cite{shao2024deepseekmath}
                                DeepSeek-R1~\cite{guo2025deepseekr1}
                                DAPO~\cite{yu2025dapo}
                                WebRL~\cite{qi2025webrl}
                                \\
                                Search-R1~\cite{jin2025searchr1}
                                ReTool~\cite{feng2025retool}
                                ToolRL~\cite{qian2025toolrl}
                                RAGEN~\cite{wang2025ragen}
                                Agent-R1~\cite{cheng2025agentr1}
                                Agent Lightning~\cite{luo2025agentlightning}
                                DynaWeb~\cite{ding2026dynaweb},
                                modelnode-datasets,
                                text width=50.7em
                            ]
                        ]
                    ]
                    [
                        Inference-time\\Augmentation,
                        leaf-datasets,
                        text width=7.5em
                        [
                            Prompt Engineering,
                            leaf-datasets,
                            text width=9em
                            [
                                Prompt Programming~\cite{reynolds2021prompt}
                                Demonstrations~\cite{min2022rethinking}
                                In-Context Examples~\cite{liu2022goodexamples}
                                Chain-of-Thought~\cite{wei2022chain}\\
                                Self-Consistency~\cite{wang2023selfconsistency}
                                Least-to-Most~\cite{zhou2023leasttomost}
                                Tree of Thoughts~\cite{yao2023tree}
                                Self-Refine~\cite{madaan2023selfrefine}
                                Graph of Thoughts~\cite{besta2024graph}\\
                                AutoPrompt~\cite{shin2020autoprompt}
                                APE~\cite{zhou2023ape}
                                OPRO~\cite{yang2024opro}
                                Promptbreeder~\cite{fernando2024promptbreeder}
                                TextGrad~\cite{yuksekgonul2024textgrad},
                                modelnode-datasets,
                                text width=50.7em
                            ]
                        ]
                        [
                            Context Engineering,
                            leaf-datasets,
                            text width=9em
                            [
                                DPR~\cite{karpukhin2020dense}
                                RAG~\cite{lewis2020retrieval}
                                FiD~\cite{izacard2021fid}
                                HyDE~\cite{gao2023hyde}
                                IRCoT~\cite{trivedi2023ircot}
                                Self-RAG~\cite{asai2024selfrag}
                                CoRAG~\cite{wang2025corag}
                                \\
                                RankRAG~\cite{yu2024rankrag}
                                LLMLingua~\cite{jiang2023llmlingua}
                                RECOMP~\cite{xu2024recomp}
                                GraphRAG~\cite{edge2024graphrag}
                                Provence~\cite{chirkova2025provence}
                                Lost in the Middle~\cite{liu2024lost}
                                \\
                                FaithfulRAG~\cite{zhang2025faithfulrag}
                                MemGPT~\cite{packer2023memgpt}
                                HiAgent~\cite{hu2024hiagent}
                                ACON~\cite{kang2026acon}
                                ACE~\cite{zhang2026ace}
                                ContextCurator~\cite{li2026contextcurator}
                                AdaCoM~\cite{yi2026adacom}
                                \\
                                LinearRAG~\cite{zhuang2026linearrag}
                                MemGraphRAG~\cite{wu2026memgraphrag}
                                LogicRAG~\cite{chen2026you}
                                LogicPoison~\cite{xiao2026logicpoison}
                                LegalGraphRAG~\cite{chen2026legalgraphrag},
                                modelnode-datasets,
                                text width=50.7em
                            ]
                        ]
                    ]
                ]
                [
                    Individual\\Intelligence,
                    leaf-methods,
                    text width=8em
                    [
                        Harness\\Engineering,
                        leaf-methods,
                        text width=7.5em
                        [
                            Tool Integration,
                            leaf-methods,
                            text width=9em
                            [
                                MRKL~\cite{karpas2022mrkl}
                                TALM~\cite{parisi2022talm}
                                ReAct~\cite{yao2023react}
                                Toolformer~\cite{schick2023toolformer}
                                API-Bank~\cite{li2023apibank}
                                ToolLLM~\cite{qin2024toolllm}
                                Gorilla~\cite{patil2024gorilla}
                                \\
                                MCP~\cite{anthropic2024mcp}
                                CodeAct~\cite{wang2024codeact}
                                SWE-agent~\cite{yang2024sweagent}
                                OpenHands~\cite{wang2024openhands}
                                ToolMaker~\cite{wolflein2025toolmaker}
                                \\
                                Codex~\cite{openai2025codex}
                                Claude Code~\cite{anthropic2025claudecode}
                                Gemini CLI~\cite{google2025geminicli}
                                Copilot Coding Agent~\cite{github2025copilotagent}
                                Symphony~\cite{openai2026symphony},
                                modelnode-methods,
                                text width=50.7em
                            ]
                        ]
                        [
                            Memory Management,
                            leaf-methods,
                            text width=9em
                            [
                                MemSyco-Bench~\cite{xiang2026memsyco}
                                Generative Agents~\cite{park2023generative}
                                MemoryBank~\cite{zhong2024memorybank}
                                MemGPT~\cite{packer2023memgpt}
                                A-MEM~\cite{xu2025amem}
                                Mem0~\cite{chhikara2025mem0}
                                Zep~\cite{rasmussen2025zep}
                                \\
                                MemoryOS~\cite{kang2025memoryos}
                                Memoria~\cite{sarin2025memoria}
                                AgeMem~\cite{yu2026agemem}
                                Memori~\cite{borro2026memori}
                                LycheeMemory V2~\cite{li2026lycheememory}
                                Agent Workflow Memory~\cite{wang2024awm},
                                modelnode-methods,
                                text width=50.7em
                            ]
                        ]
                        [
                            Skill Composition,
                            leaf-methods,
                            text width=9em
                            [
                                Voyager~\cite{wang2023voyager}
                                CRAFT~\cite{yuan2024craft}
                                Agent Skills~\cite{anthropic2025agentskills}
                                SAGE~\cite{wang2025sage}
                                HASP~\cite{liu2026hasp}
                                SSL Skills~\cite{liang2026ssl}
                                \\
                                SkillComposer~\cite{zhang2026skillcomposer}
                                Generative Skill Composition~\cite{zhao2026generativeskill}
                                Skill-Use~\cite{han2026skilluse}
                                HDSO~\cite{shang2026hdso}
                                Demystifying Agent Skills~\cite{jiang2026demystifyingskills},
                                modelnode-methods,
                                text width=50.7em
                            ]
                        ]
                        [
                            Runtime Orchestration,
                            leaf-methods,
                            text width=9em
                            [
                                $A^2E$~\cite{arxiv:Wang_2026}
                                Long-Running Harness~\cite{anthropic2025harnesses}
                                Externalization~\cite{zhou2026externalization}
                                Harness Engineering~\cite{zhong2026runtime}
                                Code as Agent Harness~\cite{ning2026codeharness}
                                \\
                                Harness Configuration~\cite{galster2026harness}
                                Harness-Bench~\cite{yao2026harnessbench}
                                Prompts to Contracts~\cite{ahn2026contracts}
                                ToolSandbox~\cite{lu2024toolsandbox}
                                ToolEmu~\cite{ruan2023toolemu}
                                RHO~\cite{pan2026rho}
                                \\
                                CaMeL~\cite{debenedetti2025camel}
                                MCP Security Bench~\cite{zhang2025mcpsecurity}
                                OpenAI Harness Engineering~\cite{openai2026harnessengineering}
                                Anthropic Harness Design~\cite{anthropic2026harnessdesign}
                                \\
                                Meta-Harness~\cite{lee2026metaharness}
                                Agentic Harness~\cite{lin2026ahe}
                                Self-Harness~\cite{zhang2026selfharness}
                                HarnessFix~\cite{chen2026harnessfix}
                                HARBOR~\cite{sengupta2026harbor}
                                AgentDojo~\cite{debenedetti2024agentdojo}
                                \\
                                Harness Updating~\cite{lin2026harnessupdating}
                                Adaptive Auto-Harness~\cite{liu2026adaptiveharness}
                                LongHorizon-Harness~\cite{ma2026longhorizonharness}
                                OneDayAgent~\cite{zheng2026onedayagent}
                                Evo-Harness~\cite{wei2026evoharness}
                                \\
                                Harness Handbook~\cite{wang2026harnesshandbook}
                                HarnessOpt-Bench~\cite{ursekar2026harnessopt}
                                The Scaffold Effect~\cite{vats2026scaffoldeffect}
                                Harness-IF~\cite{huang2026harnessif}
                                Evo-Bench~\cite{huang2026evobench},
                                modelnode-methods,
                                text width=50.7em
                            ]
                        ]
                    ]
                    [
                        Loop Engineering,
                        leaf-methods,
                        text width=7.5em
                        [
                            Loop Architecture,
                            leaf-methods,
                            text width=9em
                            [
                                StateFlow~\cite{wu2024stateflow}
                                Magentic-One~\cite{fourney2024magentic}
                                AdaPlanner~\cite{sun2023adaplanner}
                                When Agents Do Not Stop~\cite{hou2026when}
                                \\
                                Stop Hand-Holding Your Coding Agent~\cite{macedo2026stop}
                                ResearchLoop~\cite{xia2026researchloop}
                                Proof-or-Stop~\cite{huang2026proof} LoopsBench~\cite{li2026loopsbench},
                                modelnode-methods,
                                text width=50.7em
                            ]
                        ]
                        [
                            Interaction Paradigm,
                            leaf-methods,
                            text width=9em
                            [
                                Beyond Message Passing~\cite{yuan2026beyondmessagepassing}
                                Internet of Agents~\cite{wang2025internetofagents}
                                LACP~\cite{li2025lacp}
                                Beyond the Protocol~\cite{song2026mcp}
                                AgentRx~\cite{barke2026agentrx}\\
                                Supervising Ralph Wiggum~\cite{xu2026supervising}
                                Sovereign Agentic Loops~\cite{he2026sovereign}
                                The Log is the Agent~\cite{nakajima2026log},
                                modelnode-methods,
                                text width=50.7em
                            ]
                        ]
                        [
                            Environment Feedback,
                            leaf-methods,
                            text width=9em
                            [
                                EurekAgent~\cite{xin2026eurekagent}
                                CRITIC~\cite{gou2023critic}
                                Executable Code Actions~\cite{wang2024executable} ToolSandbox~\cite{lu2024toolsandbox} OSWorld~\cite{xie2024osworld}\\ AppWorld~\cite{trivedi2024appworld}
                                MCP-Universe~\cite{luo2025mcp}
                                Terminal-Bench~\cite{merrill2026terminal}   AutoWebWorld~\cite{wu2026autowebworld} DeltaBox~\cite{dong2026deltabox},
                                modelnode-methods,
                                text width=50.7em
                            ]
                        ]
                    ]
                ]
                [
                    System Intelligence,
                    leaf-metrics,
                    text width=8em
                    [
                        Graph Engineering,
                        leaf-metrics,
                        text width=7.5em
                        [
                            Task Organization,
                            leaf-metrics,
                            text width=9em
                            [
                                HuggingGPT~\cite{shen2023hugginggpt}
                                ReWOO~\cite{xu2023rewoo}
                                LLMCompiler~\cite{kim2024llmcompiler}
                                Plan-over-Graph~\cite{planovergraph2025}
                                TDAG~\cite{tdag2025}
                                Flow~\cite{flow2025}
                                VFlow~\cite{wei2025vflow}
                                \\
                                GPTSwarm~\cite{zhuge2024gptswarm}
                                ADAS~\cite{hu2025adas}
                                AutoFlow~\cite{autoflow2024}
                                AFlow~\cite{zhang2025aflow}
                                A2Flow~\cite{zhao2025a2flow}
                                MermaidFlow~\cite{zheng2025mermaidflow}
                                DynTaskMAS~\cite{yu2025dyntaskmas}
                                \\
                                DyFlow~\cite{dyflow2025}
                                EvoFlow~\cite{evoflow2025}
                                QualityFlow~\cite{qualityflow2025}
                                FlowSteer~\cite{flowsteer2026}
                                AgenticLab~\cite{kim2025towards}
                                ScalingAgent~\cite{kim2025towards},
                                modelnode-metrics,
                                text width=50.7em
                            ]
                        ]
                        [
                            Agent Coordination,
                            leaf-metrics,
                            text width=9em
                            [
                                DyLAN~\cite{liu2023dynamic}
                                Agent-Oriented Planning~\cite{li2025agent}
                                MasRouter~\cite{yue2025masrouter}
                                AutoAgents~\cite{chen2023autoagents}
                                EvoAgent~\cite{yuan2025evoagent}
                                Collaborative Gym~\cite{shao2026collaborative}
                                \\
                                AOrchestra~\cite{ruan2026aorchestra}
                                Captain Agent~\cite{li2025adaptive}
                                MaAS~\cite{zhang2025multi}
                                SkillGraph~\cite{nie2026skillgraph}
                                MetaGPT~\cite{hong2024metagpt}
                                ChatDev~\cite{qian2024chatdev}
                                DyTopo~\cite{lu2026dytopo}
                                \\
                                Magentic-One~\cite{fourney2024magentic}
                                AgentVerse~\cite{chen2024agentverse}
                                Puppeteer~\cite{dang2026multi}
                                AgentNet~\cite{yang2026agentnet}
                                MacNet~\cite{qian2025scaling}
                                SwarmAgentic~\cite{zhang2025swarmagentic}
                                AMAS~\cite{leong2025amas}
                                \\
                                Mixture-of-Agents~\cite{wang2025mixture}
                                G-Designer~\cite{zhang2024g}
                                AgentPrune~\cite{zhang2025cut}
                                AgentDropout~\cite{wang2025agentdropout}
                                Alphaproof Nexus~\cite{tsoukalas_alphaproof_nexus_2026}\\
                                Claude Code Agent Team~\cite{anthropic_claude_code_agent_teams_2026}
                                Kimi K2.5 Agent Swarm~\cite{kimi_k25_2026}
                                EvoMap~\cite{evomap_epigenetic_engineering_2026},
                                modelnode-metrics,
                                text width=50.7em
                            ]
                        ]
                        [
                            State Management,
                            leaf-metrics,
                            text width=9em
                            [
                                StateFlow~\cite{wu2024stateflow}
                                AutoGRAMS~\cite{krause2024autograms}
                                Magentic-One~\cite{fourney2024magentic}
                                Graph of States~\cite{luo2026graphstates}
                                LangGraph~\cite{langgraph2026}
                                Burr~\cite{burr2026}
                                Aegis~\cite{song2025aegis}
                                \\
                                LlamaIndex Workflows~\cite{llamaindexworkflows2026}
                                Pydantic AI~\cite{pydanticai2026}
                                AutoGen~\cite{wu2024autogen}
                                Sovereign Agentic Loops~\cite{he2026sovereign}
                                Who \& When~\cite{zhang2025whowhen}
                                \\
                                PatchBoard~\cite{zhang2026patchboard}
                                MemTX~\cite{li2026memtx}
                                Cordon~\cite{chen2026cordon}
                                Atomix~\cite{mohammadi2026atomix}
                                SagaLLM~\cite{chang2025sagallm}
                                ALAS~\cite{geng2025alas}
                                ReflexGrad~\cite{kadu2026within}
                                ProPlay~\cite{ma2026proplay}
                                \\
                                DART~\cite{yang2026dart}
                                AgentGit~\cite{li2025agentgit}
                                Shepherd~\cite{yu2026shepherd}
                                The Log is the Agent~\cite{nakajima2026log}
                                Concurrency Anomaly Prevention~\cite{khan2026concurrency}
                                \\
                                CausalFlow~\cite{bonagiri2026causalflow}
                                RAC~\cite{perera2026robust}
                                TDAD~\cite{alonso2026tdad}
                                LATS~\cite{zhou2023lats}
                                MAST~\cite{cemri2025mast}
                                Living-Harness~\cite{du2026livingharness}
                                APEX~\cite{li2026apex}
                                MiniMax Team~\cite{minimax_agent_team_2026},
                                modelnode-metrics,
                                text width=50.7em
                            ]
                        ]
                        [
                            System Evolution,
                            leaf-metrics,
                            text width=9em
                            [
                                QueenBee Planner~\cite{tian2026queenbee}
                                ReCreate~\cite{hao2026recreate}
                                SkillGraph~\cite{nie2026skillgraph}
                                Swarm Skills~\cite{zhang2026swarm}
                                MemTX~\cite{li2026memtx}
                                The Log is the Agent~\cite{nakajima2026log}
                                \\
                                SwarmAgentic~\cite{zhang2025swarmagentic}
                                DyTopo~\cite{lu2026dytopo}
                                DynTaskMAS~\cite{yu2025dyntaskmas}
                                DyFlow~\cite{dyflow2025}
                                EvoFlow~\cite{evoflow2025}
                                QualityFlow~\cite{qualityflow2025}
                                FlowSteer~\cite{flowsteer2026}
                                \\
                                TDAG~\cite{tdag2025}
                                AgentNet~\cite{yang2025agentnet}
                                Self-Organizing Agents~\cite{dochkina2026drop}
                                Meta-Team~\cite{hao2026evolve}
                                Flow~\cite{flow2025}
                                CARD~\cite{wu2026card}
                                Cursor Swarm~\cite{lin_self_driving_codebases_2026},
                                modelnode-metrics,
                                text width=50.7em
                            ]
                        ]
                    ]
                    [
                        Ontology\\Engineering,
                        leaf-metrics,
                        text width=7.5em
                        [
                            OntoExtend~\cite{lippolis2026ontoextendframeworkrequirementdrivenscalable}
                            OntoSpecification~\cite{gruber1993translation}
                            iCARE~\cite{wiratunga2025icare}
                            OG-MAR~\cite{seo2026toward}
                            CAPAS~\cite{wang2026agentic}
                            LaMAS4PD~\cite{recker2026lamas4pd}
                            Agentology~\cite{ortacc2026agentology}
                            \\
                            OntoCodex~\cite{feng2026ontocodex}
                            AgentO~\cite{ekelhart2026agento}
                            Ontology-to-Tools~\cite{zhou2026ontology}
                            Ontology SLR~\cite{li2026large}
                            Palantir Ontology~\cite{palantir2026ontology},
                            modelnode-metrics,
                            text width=61.4em
                        ]
                    ]
                ]
            ]
        \end{forest}
    }
    \vspace{-5mm}
    \caption{ \textbf{A Taxonomy of Evolving Techniques in the Era of LLM Agents.}}
    \label{fig:taxonomy}
    \vspace{-5mm}
\end{figure*}
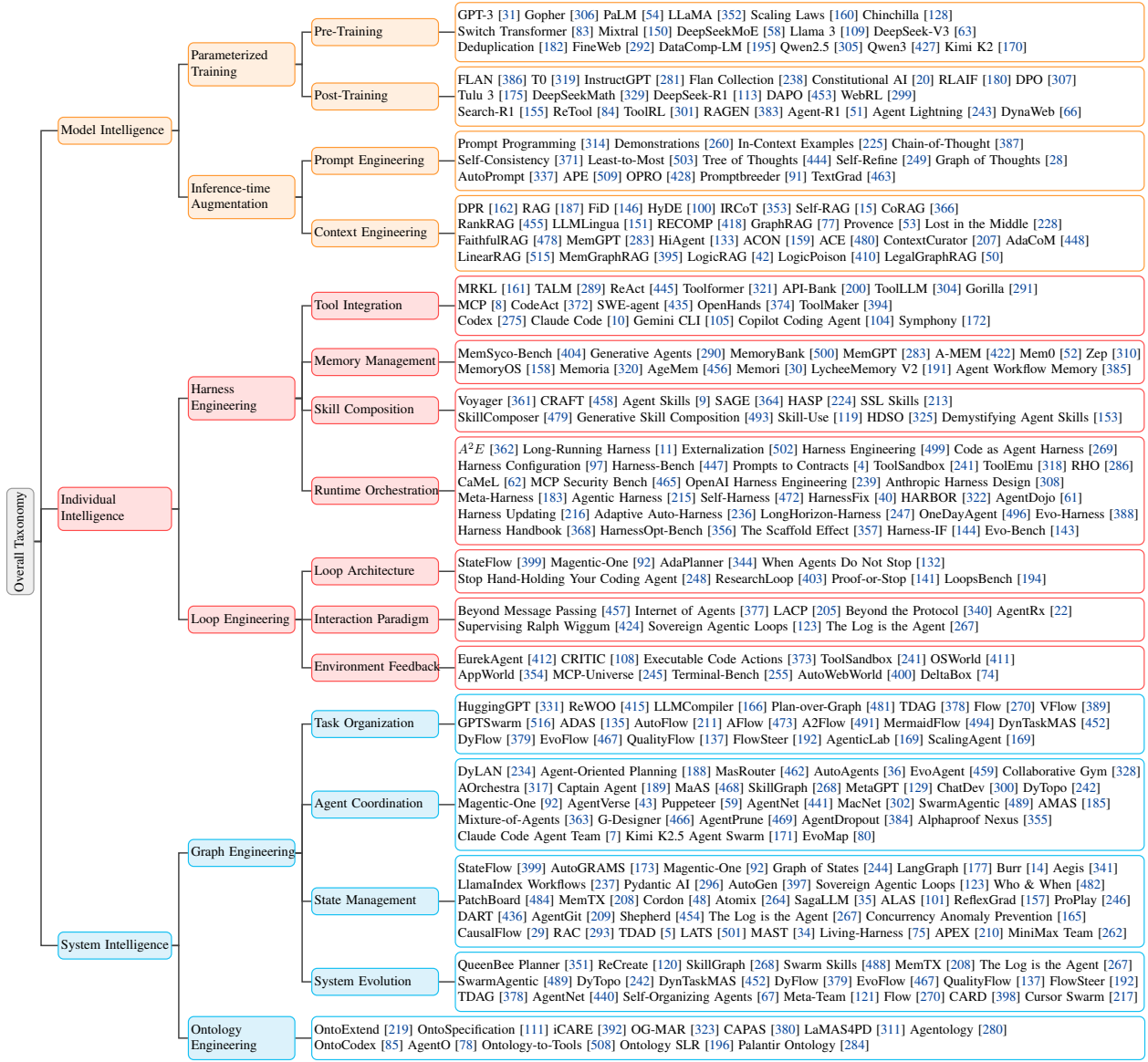

\section{From Model Intelligence to Individual Intelligence}
\label{sec:model2individual}

\begin{figure*}[t]
    \centering
    \includegraphics[width=\textwidth]{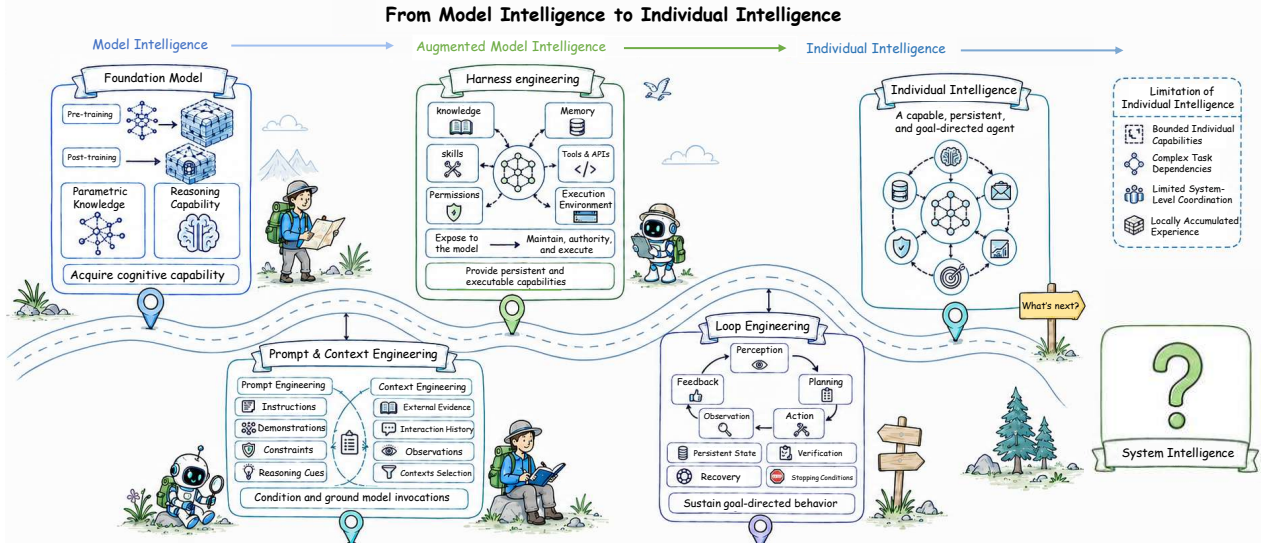}
    \caption{ \textbf{From model intelligence to individual intelligence.} Foundation-model capability is progressively transformed through task conditioning, persistent execution support, and feedback-controlled interaction into a capable, persistent, and goal-directed agent. The limitations of individual intelligence motivate the subsequent transition toward system intelligence.}
    \label{fig:model-to-individual-intelligence}
\end{figure*}


LLM-based intelligent systems have increasingly evolved from improving problem solving within individual inference processes toward constructing autonomous systems capable of sustained goal pursuit, external resource use, and environmental interaction. In the first stage, pre-training and post-training encode knowledge and general reasoning capabilities into model parameters, while Prompt Engineering and Context Engineering guide these capabilities toward effective use in specific tasks and contexts, giving rise to \textbf{Model Intelligence}. However, such intelligence remains bounded by relatively self-contained inference processes, limiting the model's ability to maintain persistent state, perform external actions, and continuously adapt to environmental feedback. To overcome these limitations, \emph{Harness Engineering} connects the model to external knowledge, memory, tools, skills, and execution environments, thereby expanding its accessible and executable capabilities. \emph{Loop Engineering} further organizes model reasoning and external capabilities into persistent cycles of planning, action, observation, verification, and adaptation. Through this transition, intelligence extends beyond problem solving within a given context toward \textbf{Individual Intelligence}, whereby an agent can use resources over time, adapt to its environment, and autonomously pursue goals. Fig.~\ref{fig:model-to-individual-intelligence} summarizes this progression from the parametric capabilities of foundation models, through inference-time capability activation, external capability extension, and closed-loop execution, to individual intelligence.

\subsection{Foundation Models: Establishing Model Intelligence}
\label{sec:foundation-models}

As LLMs have become increasingly capable~\cite{brown2020language,chowdhery2023palm,openai2023gpt4}, they have acquired general knowledge, reasoning, and problem-solving abilities that can be transferred across a wide range of tasks. These capabilities are largely encoded in model parameters through large-scale pre-training and subsequent Post-training~\cite{grattafiori2024llama3,deepseekai2024deepseekv3}, forming the internal capability base of Model Intelligence. We refer to this process as \emph{parameter-level capability development}, which typically involves two major stages: Pre-training and Post-training. Pre-training establishes a broad and reusable capability base, while Post-training further shapes how these capabilities are expressed and extends them toward desired behaviors and more complex task capabilities.

\subsubsection{Pre-training}

The development of Pre-training has been largely guided by scaling laws, which show that model performance improves as model size, training data, and computational resources are increased in a balanced manner rather than through parameter growth alone~\cite{kaplan2020scaling,hoffmann2022training}. Under this scaling paradigm, representative models such as GPT-3~\cite{brown2020language}, Gopher~\cite{rae2021gopher}, LLaMA~\cite{touvron2023llama}, Llama 3~\cite{grattafiori2024llama3}, and DeepSeek-V3~\cite{deepseekai2024deepseekv3} progressively strengthened general knowledge and problem-solving capabilities through larger-scale and more effective training. Its effectiveness depends on how scaling is realized through data, model architecture, and training strategy. High-quality and carefully curated data improve the quality and efficiency of knowledge acquisition~\cite{lee2022dedup,penedo2024fineweb,li2024datacomplm}. Scalable architectures, including dense models and sparse mixture-of-experts models such as Switch Transformer~\cite{fedus2022switch}, Mixtral~\cite{jiang2024mixtral}, and DeepSeekMoE~\cite{dai2024deepseekmoe}, determine how model capacity can be expanded under practical computational constraints. Training strategies further determine how data scale, model capacity, and computational resources are allocated and coordinated during optimization. Advances in these areas have progressively strengthened the general capability base established during Pre-training.

\subsubsection{Post-training}

Post-training further updates model parameters so that the general capabilities acquired through Pre-training can be expressed as more controllable and reliable behaviors, while also developing capabilities for increasingly complex tasks~\cite{qwen2024qwen25,kimi2025k2}. Based on their primary roles in modern training pipelines~\cite{lambert2024tulu3,yang2025qwen3}, their development can be broadly discussed along three directions: Supervised Fine Tuning (SFT), Preference Alignment, and Reinforcement Learning (RL) for capability development. SFT evolved from early instruction tuning methods such as FLAN~\cite{wei2022finetuned}, T0~\cite{sanh2022multitask}, and FLAN-PaLM~\cite{chung2024scaling} toward larger and more diverse instruction collections such as the Flan Collection~\cite{longpre2023flan}, improving instruction following, response formatting, reasoning patterns, and task adaptation~\cite{dong2024abilities}. Preference Alignment introduced explicit human or model feedback: InstructGPT~\cite{ouyang2022training} established the influential RLHF pipeline, Constitutional AI~\cite{bai2022constitutional} and RLAIF~\cite{lee2023rlaif} extended alignment toward AI-generated principles and feedback, and DPO~\cite{rafailov2023direct} simplified preference optimization by directly learning from preference pairs. More recently, RL has increasingly shifted from preference alignment toward direct capability development through verifiable rewards and environmental feedback. DeepSeekMath~\cite{shao2024deepseekmath} introduced GRPO for mathematical reasoning, while DeepSeek-R1~\cite{guo2025deepseekr1} demonstrated that large-scale outcome-based RL can induce extended reasoning behaviors; subsequent methods such as DAPO~\cite{yu2025dapo}, Dr.~GRPO~\cite{liu2025drgrpo}, and GSPO~\cite{zheng2025gspo} further improve the stability and effectiveness of reasoning-oriented RL. This paradigm has also expanded toward \emph{agentic RL}: Search-R1~\cite{jin2025searchr1} trains models to interleave reasoning with search, while ReTool~\cite{feng2025retool}, ToolRL~\cite{qian2025toolrl}, and ToRL~\cite{feng2025torl} extend RL toward tool-integrated reasoning. WebRL~\cite{qi2025webrl}, RAGEN~\cite{wang2025ragen}, and WebAgent-R1~\cite{wei2025webagentr1} further extend RL to interactive and multi-turn agent trajectories, while Agent Lightning~\cite{luo2025agentlightning} and DynaWeb~\cite{ding2026dynaweb} explore more general and scalable training frameworks for agents interacting with external environments. Modern Post-training pipelines therefore combine demonstrations, preference signals, verifiable rewards, and interaction feedback to jointly improve instruction following, behavioral alignment, reasoning, and agentic capabilities.

Together, Pre-training and Post-training determine the knowledge, reasoning abilities, and behavioral capabilities available to a model at the parameter level. However, a general capability base does not automatically translate into effective performance on a specific task. At inference time, the model still requires appropriate task descriptions, behavioral constraints, and task-relevant information to identify and apply the capabilities needed for the current problem.

\subsection{Prompt and Context Engineering: Eliciting and Conditioning Model Intelligence}
\label{sec:prompt-context-engineering}

Pre-training and Post-training establish general capabilities at the parameter level, but these capabilities do not automatically translate into effective performance on specific tasks. Without modifying model parameters, Prompt Engineering and Context Engineering adapt these capabilities by shaping the control signals and information environment available at inference time. Prompt Engineering primarily concerns how tasks and expected behaviors are specified, whereas Context Engineering concerns what task-relevant information is provided and how it is organized and maintained. The former determines what the model should do and how it should approach the task, while the latter supplies the knowledge, evidence, and working state needed to complete it.

\subsubsection{Prompt Engineering}

The development of Prompt Engineering can be broadly characterized by three directions: task specification, reasoning organization, and automatic optimization. Task specification uses instructions, demonstrations, constraints, and output formats to support zero-shot and few-shot adaptation~\cite{brown2020language,reynolds2021prompt,min2022rethinking,liu2022goodexamples}. Reasoning organization further structures how models solve complex problems: Chain-of-Thought~\cite{wei2022chain} introduces intermediate reasoning, Self-Consistency~\cite{wang2023selfconsistency} aggregates multiple reasoning paths, and Least-to-Most~\cite{zhou2023leasttomost} decomposes difficult problems, while Tree of Thoughts~\cite{yao2023tree}, Graph of Thoughts~\cite{besta2024graph}, and Self-Refine~\cite{madaan2023selfrefine} extend reasoning toward search and iterative refinement. Automatic prompt optimization moves prompt design from manual construction toward systematic search and improvement. Representative methods progress from AutoPrompt~\cite{shin2020autoprompt}, APE~\cite{zhou2023ape}, and OPRO~\cite{yang2024opro} to planning-, evolutionary-, and program-level optimization~\cite{wang2023promptagent,guo2024evoprompt,fernando2024promptbreeder,yuksekgonul2024textgrad,opsahlong2024mipro}. More recently, GEPA~\cite{agrawal2025gepa} uses execution trajectories and natural-language reflection to evolve prompts from task feedback. Overall, Prompt Engineering has expanded from specifying tasks, to organizing reasoning, and ultimately to optimizing the control interface itself.

\subsubsection{Context Engineering}

Context Engineering extends this focus to the broader information environment used during task execution, including context acquisition, processing, and management. Context acquisition has progressed from dense retrieval~\cite{karpukhin2020dense} and retrieval-augmented generation~\cite{lewis2020retrieval,izacard2021fid} toward retrieval coupled with reasoning. HyDE~\cite{gao2023hyde}, IRCoT~\cite{trivedi2023ircot}, Self-RAG~\cite{asai2024selfrag}, and CoRAG~\cite{wang2025corag} progressively integrate query transformation, iterative retrieval, and reflection, while recent work further studies when retrieval should occur during reasoning and how retrieval itself can become an agentic process~\cite{guo2026realmretrieve,ming2026llmwiki,xiao2025lag,zhang2025faithfulrag,wu2026memgraphrag,chen2026legalgraphrag}. Context processing improves the relevance, compactness, and structure of acquired information through ranking~\cite{yu2024rankrag}, compression~\cite{jiang2023llmlingua,xu2024recomp}, pruning~\cite{chirkova2025provence}, and restructuring~\cite{edge2024graphrag}. Recent methods such as SARA~\cite{jin2026sara} and BRIEF-Pro~\cite{gu2026briefpro} further improve information-preserving compression under constrained context budgets, while Lost in the Middle~\cite{liu2024lost} shows that longer context alone does not guarantee effective information use. Context management maintains useful working information as execution progresses. MemGPT~\cite{packer2023memgpt} introduced explicit hierarchical context management, followed by hierarchical and adaptive approaches such as HiAgent~\cite{hu2024hiagent}, ACON~\cite{kang2026acon}, ACE~\cite{zhang2026ace}, ContextCurator~\cite{li2026contextcurator}, and AdaCoM~\cite{yi2026adacom}. Context as a Tool~\cite{liu2025cat} further treats context maintenance as an explicit agent action, enabling proactive compression during long-horizon execution.

Prompt Engineering and Context Engineering together constitute an inference-time mechanism for adapting model capabilities. Prompt Engineering establishes the control structure for task execution, while Context Engineering provides and maintains the task-relevant information base. Rather than altering the general capabilities encoded in model parameters, they determine which capabilities are invoked, how reasoning unfolds, and what information conditions model outputs, thereby translating general capabilities into task-specific behavior.

\subsection{Harness Engineering: Orchestrating Agent Capabilities}
\label{sec:harness-engineering}

Model Intelligence takes the model call as its basic unit of operation. Foundation model training establishes general capabilities at the parameter level, while Prompt Engineering and Context Engineering adapt them to specific tasks. However, a model call alone cannot maintain persistent resources, execute external operations, or sustain interaction with an environment over time. Extending Model Intelligence toward \emph{Individual Intelligence} therefore requires persistent and executable capabilities that remain available across calls. Harness Engineering provides and manages these capabilities, while Loop Engineering organizes how they are repeatedly invoked and adapted during task execution. Recent work increasingly treats the harness as the runtime layer surrounding the model, connecting memory, tools, skills, execution environments, state, verification, and other supporting mechanisms into an operational agent system~\cite{anthropic2025harnesses,zhou2026externalization,zhong2026runtime,macedo2026whatharness,ning2026codeharness}.

\subsubsection{Tool Integration}

Many mechanisms now associated with Harness Engineering appeared before the term itself became widely established. Early systems such as MRKL~\cite{karpas2022mrkl}, TALM~\cite{parisi2022talm}, and Toolformer~\cite{schick2023toolformer} connected models to external tools. Tool use subsequently expanded toward large API ecosystems through API-Bank~\cite{li2023apibank}, ToolLLM~\cite{qin2024toolllm}, and Gorilla~\cite{patil2024gorilla}, while MCP~\cite{anthropic2024mcp} provides a standardized interface to external tools and data sources. CodeAct~\cite{wang2024codeact}, SWE-agent~\cite{yang2024sweagent}, and OpenHands~\cite{wang2024openhands} further extend execution to code, files, shells, browsers, and computing environments, moving external capability access from function invocation toward richer agent-computer interaction~\cite{li2026deepagent,dong2025agentic}.

\subsubsection{Memory Management}

Persistent memory allows agents to retain information and experience beyond individual model calls. Generative Agents~\cite{park2023generative}, MemoryBank~\cite{zhong2024memorybank}, and MemGPT~\cite{packer2023memgpt} established early mechanisms for persistent and long-term memory, while later systems such as A-MEM~\cite{xu2025amem}, Mem0~\cite{chhikara2025mem0}, Zep~\cite{rasmussen2025zep}, and MemoryOS~\cite{kang2025memoryos} improve memory organization, consolidation, and reuse. Recent work increasingly moves memory from a passive storage component toward an actively managed agent capability. AgeMem~\cite{yu2026agemem} integrates short- and long-term memory operations into the agent policy, allowing the model to decide when to store, retrieve, update, summarize, or discard information, while graph-based systems such as GAM~\cite{wu2026gam} and HeLa-Mem~\cite{zhu2026helamem} organize evolving experiences through explicit relational structures. MAGE~\cite{chen2026mage} further treats memory as execution-state management for long-horizon tasks, supporting state reconstruction and recovery, while Text2Mem~\cite{wang2026text2mem} introduces typed and executable memory operations for more controllable memory management. Other recent work explores efficient consolidation, filesystem-based persistent memory, and the reliability of memory addition and deletion~\cite{zhang2026lightmem,zhou2026filesystemmemory,xiong2026memorymanagement}. Together, these developments shift memory engineering from storing past information toward actively organizing, governing, and maintaining reusable experience across extended execution.

\subsubsection{Skill Composition}

Beyond individual tools and memories, skill-based methods externalize successful procedures as reusable capabilities. Voyager~\cite{wang2023voyager} introduced an executable skill library accumulated from experience, while Agent Workflow Memory~\cite{wang2024awm} reuses recurring action workflows and Agent Skills~\cite{anthropic2025agentskills} packages instructions, scripts, and supporting resources into reusable procedural capabilities. Subsequent methods such as SAGE~\cite{wang2025sage}, HASP~\cite{liu2026hasp}, SkillComposer~\cite{zhang2026skillcomposer}, and Skill-Use~\cite{han2026skilluse} improve skill construction, composition, evolution, and invocation. More recent work increasingly treats the skill library itself as an adaptive engineering object. SkillX~\cite{wang2026skillx} automatically constructs hierarchical skill knowledge bases from trajectories, while SkillOpt~\cite{yang2026skillopt} uses execution feedback to systematically optimize reusable skill artifacts. Anything2Skill~\cite{pan2026anything2skill} compiles heterogeneous external knowledge into reusable procedural skills, extending capability acquisition beyond direct trajectory reuse. As skill libraries grow, SkillOps~\cite{pu2026skillops} and SkillWiki~\cite{huang2026skillwiki} address library-level maintenance, provenance, governance, and lifecycle evolution, while SkillZip~\cite{bai2026skillzip} and related approaches~\cite{zhang2026skilltolora} reduce the runtime and maintenance cost of repeatedly using large skill artifacts. These developments expand skill engineering from acquiring individual reusable procedures toward constructing, optimizing, maintaining, and evolving persistent skill ecosystems.

\subsubsection{Runtime Orchestration}

As external capabilities become richer, Harness Engineering increasingly concerns how these resources are organized, governed, verified, and improved as a runtime system. Anthropic's long-running agent harness~\cite{anthropic2025harnesses,anthropic2026harnessdesign}, AI Harness Engineering~\cite{zhong2026runtime}, What Makes a Harness a Harness~\cite{macedo2026whatharness}, Code as Agent Harness~\cite{ning2026codeharness}, and Harness-Bench~\cite{yao2026harnessbench} make this surrounding runtime an explicit research object and clarify its responsibilities and boundaries. In this view, Context Engineering determines what information is presented to a model call, whereas Harness Engineering maintains the persistent resources, interfaces, and execution environments through which information and external capabilities remain available across calls.

Runtime orchestration also introduces configuration, governance, verification, and optimization concerns. ToolEmu~\cite{ruan2023toolemu}, ToolSandbox~\cite{lu2024toolsandbox}, AgentDojo~\cite{debenedetti2024agentdojo}, CaMeL~\cite{debenedetti2025camel}, and MCP Security Bench~\cite{zhang2025mcpsecurity} study failures, security risks, and control mechanisms, while harness configuration~\cite{galster2026harness} and contract-based validation~\cite{ahn2026contracts} address configuration and runtime guarantees. These engineering principles are increasingly reflected in widely used coding agents such as Codex~\cite{openai2025codex,openai2026harnessengineering}, Claude Code~\cite{anthropic2025claudecode,anthropic2026harnessdesign}, Gemini CLI~\cite{google2025geminicli}, and GitHub Copilot coding agent~\cite{github2025copilotagent}. More recent work treats the harness itself as an optimization target: Meta-Harness~\cite{lee2026metaharness}, Agentic Harness Engineering~\cite{lin2026ahe}, Self-Harness~\cite{zhang2026selfharness}, HarnessFix~\cite{chen2026harnessfix}, HARBOR~\cite{sengupta2026harbor}, and Retrospective Harness Optimization~\cite{pan2026rho} explore search, adaptation, diagnosis, repair, and feedback-driven improvement. Related work further evaluates harness effects and optimization~\cite{vats2026scaffoldeffect,lin2026harnessupdating,ursekar2026harnessopt,huang2026harnessif,wang2026optimizers}, while Adaptive Auto-Harness~\cite{liu2026adaptiveharness}, LongHorizon-Harness~\cite{ma2026longhorizonharness}, OneDayAgent~\cite{zheng2026onedayagent}, and Evo-Harness~\cite{wei2026evoharness} extend adaptation toward open-ended and long-horizon execution. Harness Handbook~\cite{wang2026harnesshandbook} further addresses the understandability and maintainability of increasingly complex harnesses. Harness Engineering therefore concerns not only what external capabilities an agent can access, but also how the runtime surrounding the model is structured, governed, maintained, and improved over time.

\subsection{Loop Engineering: Enabling Iterative Agent Execution}

Harness Engineering establishes the capability space available to an agent, whereas Loop Engineering organizes how the agent moves through that space over time. A harness provides persistent resources, executable tools, validation mechanisms, and controlled environments, together with the interfaces and permissions governing their use. However, it does not by itself determine how an active task should proceed after each execution. For example, when a test fails, the harness can return the failure log, but the loop can decide whether to revise the implementation, inspect a dependency, invoke another capability, recover an earlier state, request assistance, or terminate the task. We therefore define \emph{Loop Engineering} as the engineering of a bounded, stateful, and feedback-driven process that coordinates agent operation until the goal achievement is supported by sufficient evidence or continued execution is no longer justified~\cite{li2026loopsbench，macedo2026loop,xia2026researchloop,huang2026proof}. Its defining property is not the repetition of model calls, but the continuous use of execution outcomes to control the subsequent trajectory of the task.

We analyze Loop Engineering through three coupled dimensions:
\emph{Loop Architecture}, \emph{Interaction Paradigm}, and
\emph{Environment Feedback}. Loop Architecture specifies the control structure that maintains task state, evaluates progress, and governs continuation or termination. Interaction Paradigm determines how goals, action requests, observations, and verification results are represented and exchanged across iterations. Environment Feedback grounds these decisions in the external consequences of execution, where actions induce state transitions and return new observations that become inputs to subsequent decisions~\cite{li2026agentic}. Together, these dimensions form a closed control cycle: architectural decisions are expressed through interaction, realized by the harness in an environment, and revised in light of the evidence returned.

\subsubsection{Loop Architecture}
\emph{Loop Architecture} describes the main components of a loop architecture and how they jointly keep goal-directed execution coherent and bounded. A loop is initialized with a goal, acceptance criteria, and terminal conditions, after which a controller maintains the operational state of the task and tracks which requirements remain unresolved. Planning and decomposition mechanisms organize these requirements into executable operations, while progress assessment uses verification evidence to determine whether an
operation has produced a meaningful state change and whether the loop should advance. The architecture must also specify how the loop responds to failure, including revising the plan, selecting another capability, recovering an earlier state, escalating the task, or terminating execution. ResearchLoop~\cite{xia2026researchloop} represents task contracts, evidence objects, claim ledgers, and closeout conditions as durable control state, allowing research activities to advance only when their evidence requirements are satisfied. Proof-or-Stop~\cite{huang2026proof} similarly permits lifecycle transitions only when fresh and mechanically verifiable evidence satisfies the relevant gate. Complementary work examines when additional search has become redundant~\cite{tang2026saas}, while analyses of infinite agentic loops~\cite{hou2026agents} show that progress checks, resource limits, and explicit stopping conditions are necessary to prevent unbounded feedback paths.  Therefore, Loop Architecture determines not only how execution continues, but also when continuation remains justified.

\subsubsection{Interaction Paradigm}
Building on this control structure, \emph{Interaction Paradigm} explains how task state, action requests, observations, and supervisory signals are exchanged across iterations. At each step, the loop communicates the current goal, task state, unresolved requirements, and available operations to the model; the model returns proposed decisions or action intents; and the harness returns execution observations, validation results, and error conditions. These exchanges must preserve sufficient continuity for later decisions to be interpreted relative to earlier actions and outcomes. Sovereign Agentic Loops~\cite{he2026sovereign} formalizes the action boundary by representing model outputs as structured intents that can be checked against system state and policy before execution. Interaction may also introduce diagnostic and supervisory feedback. AgentRx~\cite{barke2026agentrx} transforms execution trajectories into validation records that localize critical failure steps, whereas Supervising Ralph Wiggum ~\cite{xu2026supervising} introduces metacognitive supervision when repeated refinement becomes stagnant. The harness provides communication interfaces, memory mechanisms, and execution records; the loop determines which information is relevant to the current task and how that information changes subsequent control decisions.

\subsubsection{Environment Feedback}
After an operation has been issued through the interaction layer, \emph{Environment Feedback} examines how its external consequences become evidence for the next loop decision. Agentic environments can be modeled as dynamical systems in which an action applied to a current state induces a state transition and returns an observation or reward. Execution becomes closed-loop when each subsequent decision is conditioned on this updated interaction history~\cite{li2026agentic}. Environment feedback therefore includes not only terminal success signals, but also intermediate state changes, execution traces, errors, rewards, and verifier judgments. Interactive environments instantiate this relationship through executable actions and state-based or execution-based evaluation~\cite{xie2024osworld,yao2024taubench,barres2025tau,luo2025mcp,merrill2026terminal}. Because observations may be partial, delayed, uncertain, outdated, or generated by an environment that diverges from the target system, reliable loop progression requires both admissible evidence and confidence in the correctness and fidelity of the environment that produced it~\cite{li2026agentic}. Proof-or-Stop~\cite{huang2026proof} binds accepted evidence to the current source state before allowing lifecycle transitions, while Sovereign Agentic Loops~\cite{he2026sovereign} checks proposed actions against true system state and policy before real-world execution. The harness exposes the executable interface and returns observations from external execution. Within Loop Engineering, Environment Feedback denotes how observed state changes and verification results are incorporated into task state and used by the loop controller to select the next control transition or terminate execution.

Harness Engineering and Loop Engineering consequently address the complementary requirements of \emph{Individual Intelligence}. Harness Engineering determines what persistent and executable capabilities an individual agent can access and under what conditions they can be used. Loop Engineering organizes those capabilities into a bounded, goal-directed process in which actions, observations, feedback, and termination decisions remain connected across time. Together, they extend call-level Model Intelligence into the sustained behavior of an individual agent and thereby establish \emph{Individual Intelligence}.

\begin{figure*}[tbp]
\vspace{-5mm}
    \centering
    \includegraphics[width=1.\linewidth, trim=0cm 0cm 0cm 0cm,clip]{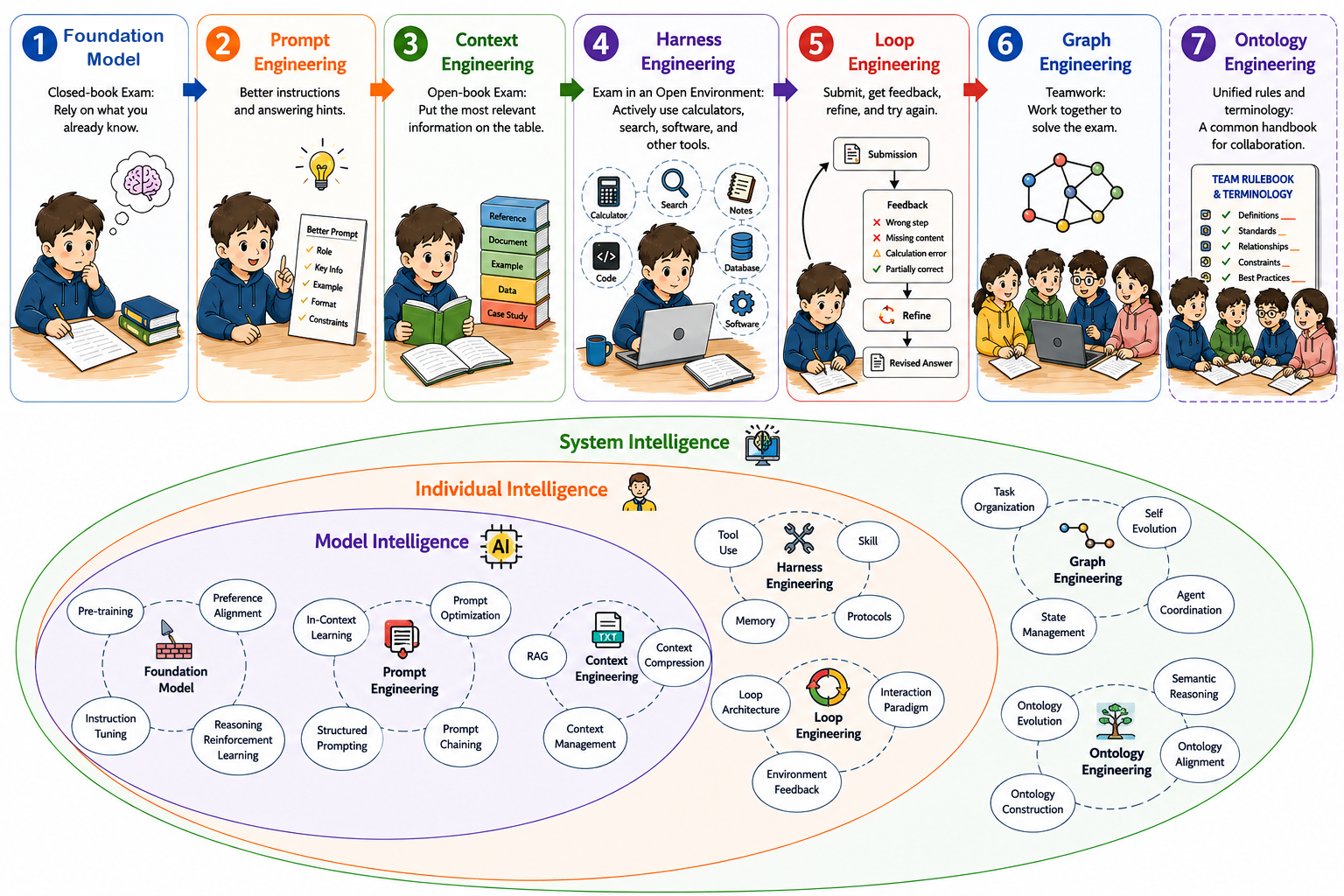}
    \vspace{-5mm}
    \caption{\textbf{An Illustrative Conceptualization of System Intelligence and Its Related Technologies.} The exam analogy depicts the progression from Model, Prompt, and Context Engineering to tool-enabled Harness Engineering, feedback-driven Loop Engineering, team-oriented Graph Engineering, and ontology-based collaboration. The lower panel summarizes representative technologies associated with these layers.}
    \label{fig:trend}
\end{figure*}

\subsection{Limitations of Individual Intelligence}

Despite the advances in Harness and Advances in Harness and Loop Engineering have enabled agents to exhibit individual intelligence, allowing them to pursue goals autonomously through sustained reasoning and interaction with their environment. However, since individual intelligence is typically organized around \emph{a single agent} and its execution loop, it still faces several fundamental limitations when applied to complex real-world tasks:

\ding{182} \textbf{Scheduling parallel and interdependent tasks:}
Real-world tasks often contain subtasks that depend on one another or can be carried out in parallel. A single-agent loop, however, tends to compress them into a serial execution trace~\cite{hu2026agentloops,geng2025alas}. This makes scheduling implicit, wastes the efficiency of parallelism, and makes failure location difficult.
For example, in a software fault diagnosis task, log analysis, failure reproduction, and code inspection can often proceed in parallel as relatively independent branches, whereas repair and testing depend on their results. A single agent, however, tends to serialize these branches within a single execution loop, losing the efficiency of parallelism. Moreover, wrong intermediate results may be propagated to subsequent steps, making the faulty stage difficult to localize.

\ding{183} \textbf{Integrating specialized expertise and verification:}
Many complex tasks require specialized expertise or independent verifiers. Although a single agent can use tools or call specialist models, these capabilities remain coordinated within the same control loop rather than organized into stable and independent roles~\cite{fu2026benchagent,qian2025scaling,cemri2025mast}. This can lead to role confusion and confirmation bias. For example, when the same agent writes and evaluates code, it may mistake its own judgment that the code is correct for evidence that it is actually correct, even when prompts assign it different roles.

\ding{184} \textbf{Maintaining persistent state and handling failure recovery:}
An individual's context is not an organized or persistent state. Once an error enters the execution loop, it can be carried through later steps, making it difficult to repair only the affected parts or to recover in a way that can be traced and checked~\cite{khan2026concurrency,guo2025syncmind,cemri2025mast}. For example, in long-running web or coding tasks, a small mistake made early may remain hidden until the task fails near the end. By then, it is often difficult to determine and localize where the error first appeared.

\begin{figure*}[tbp]
\vspace{-5mm}
    \centering
    \includegraphics[width=1.\linewidth, trim=0cm 0cm 0cm 0cm,clip]{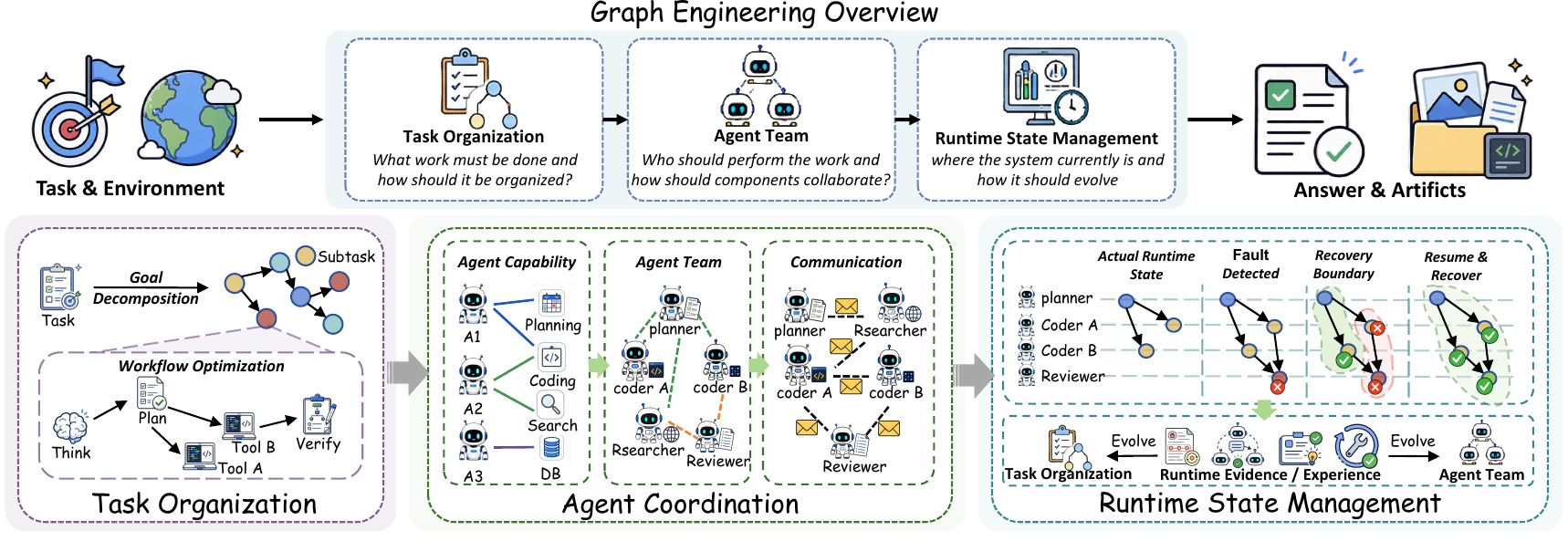}
    \vspace{-5mm}
    \caption{ \textbf{Overview of Graph Engineering.} Task Organization structures the objective into explicit subtasks and executable workflows; Agent Coordination matches capabilities to work, defines team topology, and routes communication among agents; Runtime State Management records execution states, detects and localizes anomalies, and supports recovery and structural updates. Together, these coupled graph views organize work, coordinate agents, and use runtime evidence to evolve the system toward reliable task completion. }
    \label{fig:graph_engineering}
\end{figure*}

\section{Graph Engineering: From Individual Intelligence to System Intelligence}
\label{sec:graph_engineering}

\subsection{Overview of Graph Engineering}
Despite advances in individual intelligence, the unit of \textbf{Individual Intelligence} still faces inherent limitations in scheduling parallel and interdependent tasks, integrating heterogeneous capabilities, and maintaining runtime state.
This motivates the next stage of intelligence toward \textbf{System Intelligence}, in which multiple components with complementary capabilities form an adaptive whole in pursuit of a shared goal~\cite{kim2025towards, fu2026agentic,ke2026mas}.


However, system intelligence does not arise from the mere aggregation of agents and other intelligent components; rather, it depends on how the relationships among tasks, components, and runtime states are explicitly represented, constrained, and optimized.
Specifically, a system must organize objectives into decomposable and schedulable task structures, coordinate heterogeneous components according to their capabilities and roles, and maintain persistent and recoverable runtime state throughout execution. At its core, \emph{system intelligence requires the systematic governance of relationships among tasks, components, and runtime states}.

To address this, graphs provide a natural structure for modeling the system-level relationships, as shown in Fig.~\ref{fig:graph_engineering}.
\emph{First}, graphs organize tasks through objective decomposition, dependency modeling, and workflow refinement, transforming complex objectives into schedulable and executable operations~\cite{kim2025bel, yang2025docagent, liu2025select, zhang2026shapecraft, polat2025xchemagents, gao2025graph}. \emph{Second}, graphs can coordinate intelligent components by representing operational topologies and communication patterns, enabling heterogeneous components to collaborate effectively~\cite{feng2026heterogeneous, duan2026bayesian, xiao2025srefiner, sun2025cortexdebate, yu2024researchtown, yang2025nader, men2025troublemaker, shahroz2025agents, liu2025principle}. 
\emph{Third}, graphs can support runtime state management by recording events, dependencies, and state transitions, converting operational information scattered across contexts and logs into auditable and recoverable system states~\cite{zhang2026patchboard,chen2026mage,chen2026traceelephant}.
To this end, we introduce \textbf{Graph Engineering} as a structure-centered engineering foundation for system intelligence: it uses graph structures as the core substrate for externalizing relationships among tasks, components, and runtime states, thereby supporting system-level organization, coordination, monitoring, recovery, and optimization.

In the following section, we review existing approaches that leverage graphs to organize tasks, coordinate intelligent components, and manage execution states. We further discuss how system evolution leverages execution feedback and state evidence to iteratively improve the structure of Graph Engineering and enable the continual evolution of system intelligence.

\subsection{Task Organization: Structuring What to Do}

\begin{figure*}[tbp]
    \centering
    \includegraphics[width=1.\linewidth, trim=0cm 0cm 0cm 0cm,clip]{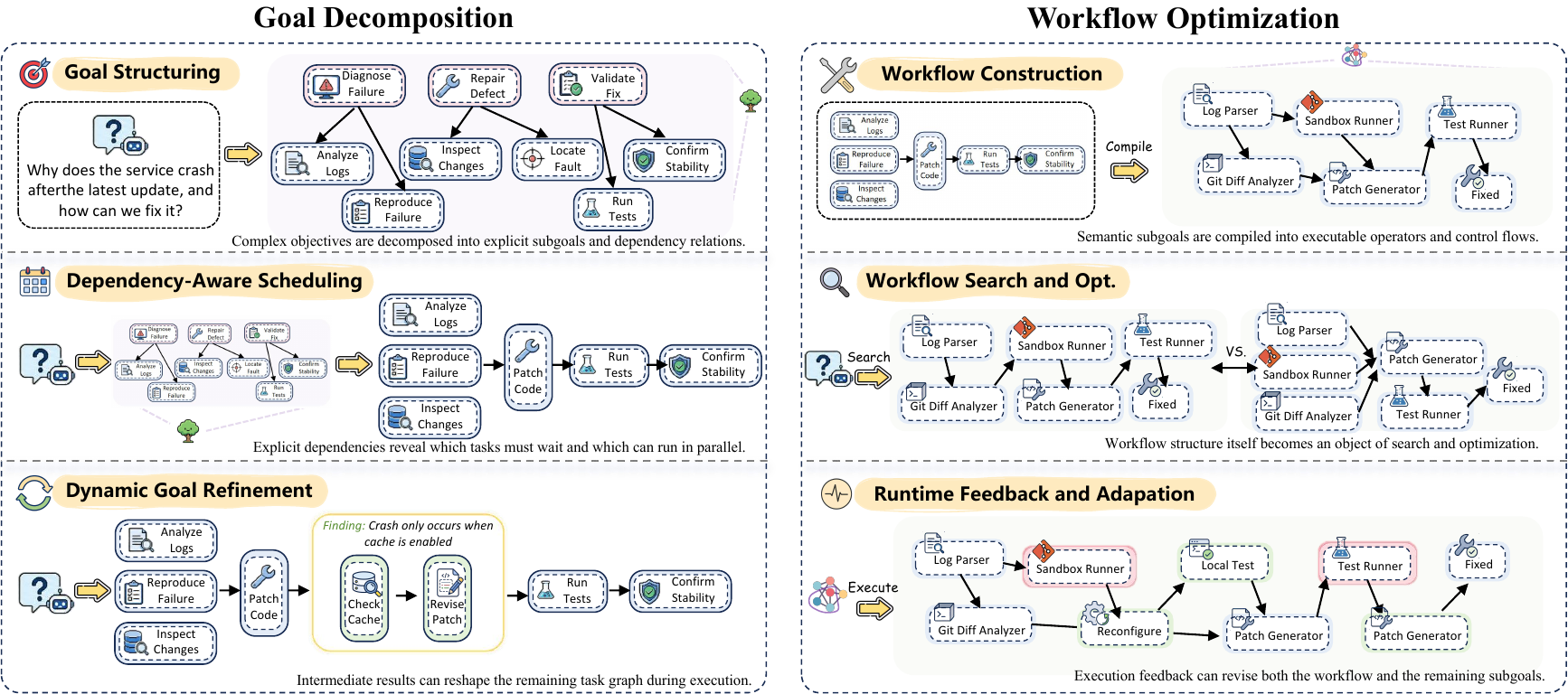}
    \caption{ \textbf{Overview of Task Organization.} Goal Decomposition translates a high-level objective into explicit subtasks, exposes their dependencies for scheduling, and refines the remaining task graph using intermediate execution results. Workflow Optimization compiles semantic subgoals into executable workflows, searches and optimizes alternative control flows, and adapts execution in response to runtime feedback. These mechanisms specify what work must be accomplished, how it should be operationalized, and how the work structure evolves during execution.}
    \label{fig:work_organization_graph}
\end{figure*}

The first challenge in building system intelligence is to transform a high-level objective or task stream into an organized set of subtasks and operations, enabling intelligent components to perform interdependent actions rather than isolated local tasks~\cite{feng2026graphplanner,chen2026gtool}. 
However, subgoals may depend on one another, and their execution may involve parallel branches, verification steps, and dynamic replanning. Relying solely on context makes it difficult to maintain a clear global task structure and determine which operations should be performed.
To address this challenge, existing work externalizes task decomposition and executable operations as graph structures, transforming task organization from implicit reasoning into a schedulable, optimizable, and revisable system structure, as shown in Fig.~\ref{fig:work_organization_graph}. We next discuss how graph structures provide the foundation for goal decomposition and workflow optimization.


\subsubsection{Goal Decomposition}
\label{sec:graph-task-decomposition}

System intelligence requires explicit goal decomposition so that intelligent components can perform coordinated, interdependent actions. However, complex user objectives or task streams~\cite{yan2026agentstream,fu2026agent} are difficult to organize within context alone: their subtasks may depend on one another, some steps may run in parallel, and the execution plan may need revision as intermediate results arrive. Graph-based task decomposition addresses this challenge by representing an objective as a graph of subgoals and dependencies, where nodes denote subtasks or intermediate goals and edges encode precedence, data, or logical relations. This explicit structure supports the scheduling of parallel and dependent branches and provides a basis for workflow refinement and component coordination.


Early work began by making task decomposition and subtask dependencies explicit, rather than leaving them implicit within an execution loop.
HuggingGPT~\cite{shen2023hugginggpt} decomposes multimodal user requests into subtasks and routes them to specialized models, using dependency relations to determine their execution order. ReWOO~\cite{xu2023rewoo} decouples reasoning from tool execution and observations through variable references, making dependencies among planned tool calls explicit.

To enable better task scheduling, subsequent studies represent these task dependencies as explicit and schedulable graphs.
LLMCompiler~\cite{kim2024llmcompiler} compiles function-calling plans into a dataflow DAG, so that ready nodes can be dispatched in parallel once their upstream dependencies are satisfied. Plan-over-Graph~\cite{planovergraph2025} directly studies planning over task graphs and focuses on generating parallelizable agent schedules under dependency constraints. These works not only make the task dependencies explicitly interpretable but also operational for scheduling and coordination. TDAG~\cite{tdag2025} and Flow~\cite{flow2025} further relax the assumption that the task graph is fixed before execution. They show that task decomposition can be dynamically refined according to intermediate results, and in multi-agent settings, such evolving task graphs can also drive agent generation, task assignment, and parallel collaboration~\cite{yu2025dyntaskmas}.

In short, Goal Decomposition Graph divides the objective goal into explicit, schedulable sub-goal graphs. It defines the structured objective space over which later workflow construction and execution adaptation operate.


\subsubsection{Workflow Optimization}

After the subgoals and their dependencies are known, system intelligence still needs to transform them into concrete computational operations, such as LLM calls, specialized agents, retrieval modules, tools, memory operations, aggregators, and verifiers.
However, the operations space is large and must be explicitly structured to construct an effective workflow. To address this challenge, existing methods use graph structures to compile decomposed tasks into executable workflows, where nodes represent concrete operators and edges encode the dependencies needed for scheduling, coordination, and verification. This process transforms task organization from a descriptive decomposition into an executable structure that can be optimized.

A group of studies treats agentic workflows as optimizable graph structures. GPTSwarm~\cite{zhuge2024gptswarm} represents language-agent systems as computational graphs and optimizes both node behavior and edge connections. ADAS~\cite{hu2025adas} automates the design of agentic systems by searching over code-defined workflows, where graph semantics are expressed through executable program structures. AutoFlow~\cite{autoflow2024} and AFlow~\cite{zhang2025aflow} formulate workflow generation as an automatic search problem, reducing reliance on manually designed agent pipelines. In particular, AFlow uses LLM-guided search over executable workflow code, making the workflow structure itself the object of optimization.
Later works refine different parts of this workflow search space. A2Flow~\cite{zhao2025a2flow} learns abstraction operators from demonstrations instead of assuming a fixed operator library, allowing both node semantics and graph topology to evolve. MermaidFlow~\cite{zheng2025mermaidflow} introduces a structured Mermaid-based intermediate representation and safety-constrained evolutionary programming, improving the readability, validity, and controllability of generated workflows. VFlow~\cite{wei2025vflow} incorporates domain-specific verifiers into the workflow search loop, showing how external feedback such as syntax checks, functional correctness, synthesizability, and hardware constraints can guide workflow discovery.



Despite the advances in static workflow optimization, these approaches remain insufficient for open-ended environments. Even well-designed task and workflow graphs may fail or propagate errors during execution due to incorrect intermediate results, tool failures, or ambiguous feedback. To address this limitation, recent approaches have developed dynamic mechanisms that adapt workflow graphs in response to real-time execution feedback.
DyFlow~\cite{dyflow2025} exemplifies this execution-adaptive paradigm. Instead of committing to a fixed workflow before execution, it uses intermediate feedback to dynamically generate and adjust subsequent operator subgraphs. In this sense, runtime adaptation can revise both the local workflow and the remaining subgoal structure. EvoFlow~\cite{evoflow2025} maintains diverse workflow candidates during inference and evolves them on the fly, treating different workflow graphs as competing executable hypotheses. QualityFlow~\cite{qualityflow2025} introduces quality checking as a control mechanism for program synthesis, where the system dynamically selects whether to accept, debug, clarify, roll back, or continue based on intermediate quality signals. FlowSteer~\cite{flowsteer2026} further highlights that workflow structure can be modified inside the execution loop, rather than only optimized before deployment.

In summary, Task Organization provides a unified view of graph-based task and execution management in agentic systems. It shifts the design of LLM agents from implicit reasoning and acting to explicit work structures. 
\begin{figure*}[tbp]
\vspace{-5mm}
    \centering
    \includegraphics[width=1.\linewidth, trim=0cm 0cm 0cm 0cm,clip]{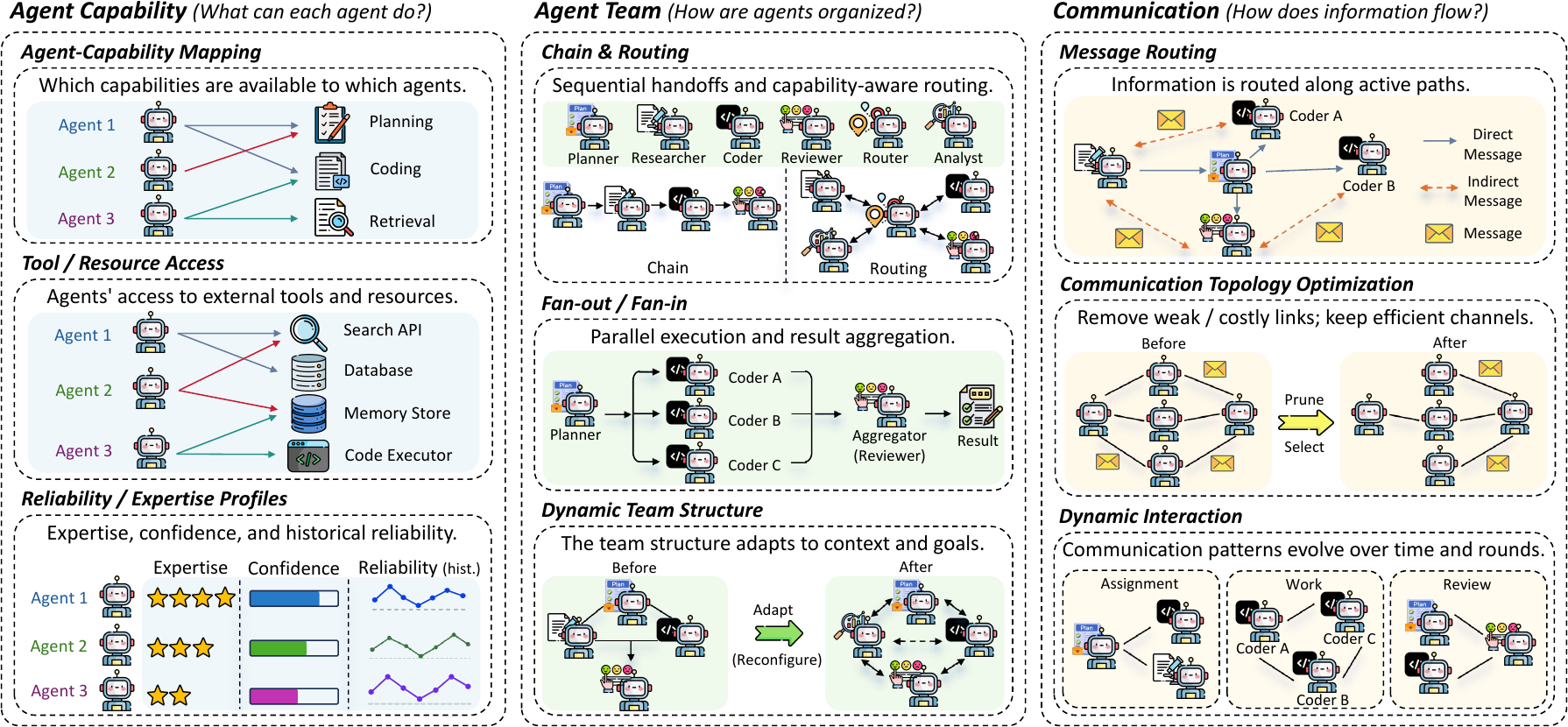}
    \vspace{-5mm}
    \caption{ \textbf{Overview of Agent Coordination.} The Agent Capability Graph maps agents to their capabilities and accessible resources; the Agent Team Graph organizes agents into task-dependent collaboration structures; and the Communication Graph specifies and adapts information flow among agents. These graphs determine who should perform the work and how agents collaborate during execution.}
    \label{fig:coordination_graph}
\end{figure*}

\subsection{Agent Coordination: Structuring Who Works}~\label{sec:agent_graph}

The second challenge in building system intelligence is coordinating heterogeneous agents as a coherent system rather than invoking them within a single control loop. This requires identifying what different agents can do, assigning them appropriate roles and responsibilities, and adapting their interactions as execution unfolds~\cite{mu2026adaptive,li2026assemble,zhang2026learning,dong2026agent}. As shown in Fig.~\ref{fig:coordination_graph}, \emph{Agent Coordination} addresses these requirements through three connected functions that can be represented using graph structures~\cite{yun2026graph,riedl2026emergent,liu2026agentpo,zhou2026multi}. \emph{Agent Capability Modeling} represents agents' skills, resources, permissions, and suitability for different tasks. \emph{Agent Team Organization} arranges selected agents into task-dependent collaboration structures, specifying role assignments, delegation paths, and review responsibilities. \emph{Multi-agent Communication} captures the runtime information exchange and feedback through which agents coordinate actions, evaluate intermediate results, and adapt subsequent execution. Together, these functions determine who should perform the work, how responsibilities should be organized, and how agents should interact as task conditions change.

\subsubsection{Agent Capability Modeling}

System intelligence requires heterogeneous work to be assigned to agents with suitable expertise and resources as the demands of complex tasks evolve. Because task stages are interdependent, a capability mismatch at one stage may delay parallel execution and compromise downstream results. The system must therefore maintain up-to-date information about each agent's skills, available resources, access permissions, and reliability. To address this challenge, graph structures make this information explicit: nodes represent agents, skills, tools, models, and other resources, while typed edges encode capability ownership, resource access, permissions, and reliability~\cite{guo2026agent,belov2026llm}. This representation enables capability-aware task assignment and agent reconfiguration as execution conditions change~\cite{xu2026tacomas,huang2026evolverouter}. For example, in scientific discovery, literature analysis, experiment design, implementation, and independent verification can be assigned to suitable agents. If an agent loses access to a computing resource, the system can query the graph to identify a compatible replacement and reassign the affected task.

Existing methods often infer capability from task-specific behavior. DyLAN~\cite{liu2023dynamic} estimates the contribution of candidate agents and retains those that are more useful for the current task, while Agent-Oriented Planning~\cite{li2025agent} assigns solvable and non-redundant subtasks to suitable agents. MasRouter~\cite{yue2025masrouter} further learns to select collaboration modes, roles, and underlying models according to task difficulty and cost. These methods capture capability differences effectively, but capability is mainly encoded in scores or routing policies rather than explicit and reusable relations~\cite{li2026morse}.

Other methods represent capability through agent configuration. AutoAgents~\cite{chen2023autoagents} creates specialized roles and collaboration plans for a given task, EvoAgent~\cite{yuan2025evoagent} generates diverse specialists through evolutionary operations, and AOrchestra~\cite{ruan2026aorchestra} composes instructions, context, tools, and models to instantiate task-specific agents. Captain Agent~\cite{li2025adaptive} similarly recruits and reorganizes experts as new requirements emerge during interaction.

More recent work connects capability modeling with graph-based organization. SkillGraph~\cite{nie2026skillgraph} explicitly represents agent skills and uses them to guide the construction of communication topologies. MaAS~\cite{zhang2025multi} takes a broader approach by representing agents and operators within an agentic supernet and searching this space for suitable multi-agent structures. However, these representations are typically constructed for a particular task or orchestration process. A persistent and updateable graph representation would instead allow knowledge about agents' expertise, reliability, available resources, and access permissions to be queried, revised, and reused across tasks.

\subsubsection{Agent Team Organization}

System intelligence requires heterogeneous agents to be organized into a team that can execute interdependent work coherently. Capability modeling identifies which agents are suitable for particular tasks, but it does not determine task ownership, output handoffs, delegation paths, or review responsibilities. To address this challenge, these organizational relations can be represented as a graph, in which nodes denote agents, roles, or tasks, and typed edges encode assignment, delegation, supervision, verification, and reporting relations~\cite{chen2025internet,chen2026toward,lee2026agentic,zhou2026multi}. By specifying each agent's position and responsibilities, this representation connects individual capabilities to an executable division of labor~\cite{pappu2026multi,liu2025workteam,zhang2026dynamic,hao2026evolve,song2026webswarm}.

For tasks with clear stage dependencies, agents can be organized into a chain in which the output of one role becomes the input to the next. MetaGPT~\cite{hong2024metagpt} structures software-development agents as an assembly line governed by standard operating procedures, while ChatDev~\cite{qian2024chatdev} connects design, coding, and testing roles through a sequential chat chain. Such structures make execution order, role transitions, and responsibility boundaries explicit, although their paths are largely fixed before execution~\cite{hu2025owl,shang2025agentsquare}.

When subtasks require different expertise, routing structures direct each unit of work to an appropriate agent. Magentic-One~\cite{fourney2024magentic} uses an orchestrator to plan and delegate tasks, monitor progress, and replan after failures. WorkTeam~\cite{liu2025workteam} employs a supervisor that invokes specialized orchestrator and filler agents according to user intent, while AgentVerse~\cite{chen2024agentverse} composes teams of experts according to task requirements. Routing supports specialized division of labor, but centralized designs may impose substantial planning and coordination burdens on the routing agent.

Tasks that benefit from parallel execution or diverse candidate solutions can instead adopt fan-out/fan-in structures. Work is distributed to multiple agents and their outputs are subsequently compared, aggregated, or synthesized. Mixture-of-Agents~\cite{wang2025mixture} uses a layered structure in which several agents generate candidate responses in parallel and agents in the next layer integrate them. MacNet~\cite{qian2025scaling} generalizes this branching and aggregation process through a directed acyclic graph, allowing multiple execution paths to converge at downstream nodes. These structures increase parallelism and reasoning diversity, but also incur additional communication, computation, and aggregation costs.

Static team structures become less effective when task requirements or agent performance change during execution~\cite{liu2026mas}. Puppeteer~\cite{dang2026multi} dynamically selects and sequences agents according to the current task state. AgentNet~\cite{yang2026agentnet} removes the central controller and allows agents to adjust their connections and route tasks based on local expertise and context. Team organization can also be optimized during system construction. SwarmAgentic~\cite{zhang2025swarmagentic} jointly optimizes agent functions and collaboration patterns while generating candidate systems. Studies of self-organizing agents~\cite{dochkina2026drop} further suggest that role specialization and shallow hierarchies can emerge without fully predefined assignments. These methods adapt team organization at different timescales, from design-time optimization to runtime reconfiguration.

Graph-based team organization can combine these structures within a single system. A coordinator may route subtasks to specialists, distribute selected tasks for parallel execution, aggregate their outputs, and pass the combined result through a chain of reviewers. The graph must therefore represent both stable responsibility relations and task-dependent structural changes, specifying who participates, what each participant is responsible for, and how work moves among them.

\subsubsection{Multi-agent Communication}

As task execution unfolds, system intelligence must coordinate information exchange among agents and prevent unreliable intermediate results from propagating downstream. Errors, conflicts, and missing information may require clarification, review, feedback, or human intervention~\cite{zhang2026swarm,evomap_epigenetic_engineering_2026,papadakis2025atlas}. To address this challenge, these runtime interactions can be modeled as a dynamic graph, where nodes represent agents or human participants and activated edges specify who communicates, what information is exchanged, and how it affects subsequent actions. Whereas team organization defines relatively stable roles and responsibilities, communication modeling captures the information flows and feedback relations that emerge during execution.

Communication serves not only to transfer results but also to detect and correct errors. Different agents can generate, evaluate, and revise an output, returning identified problems to the relevant execution stage. MAgICoRe~\cite{chen2025magicore} combines model-generated feedback with external stepwise reward signals to locate reasoning errors and iteratively refine candidate solutions through multi-agent interaction. This process forms a feedback loop among generation, evaluation, and revision rather than a one-way flow of information. Communication structure also determines how correct and incorrect information propagates, so adding more connections does not necessarily improve collaboration~\cite{shen2025understanding}.

Communication structures can be constructed according to task requirements and optimized under multiple objectives. G-Designer~\cite{zhang2024g} generates task-dependent communication graphs by considering candidate agents, performance, communication cost, and structural robustness. AMAS~\cite{leong2025amas} selects interaction structures according to the current input, allowing different tasks to employ different communication patterns. Other methods reduce collaboration overhead by removing low-value relations. AgentPrune~\cite{zhang2025cut} eliminates redundant connections from a spatio-temporal message graph, while AgentDropout dynamically removes low-contribution agents and their communication edges across interaction rounds~\cite{wang2025agentdropout}. These methods indicate that communication modeling should determine not only whether information can be transmitted, but also which information paths are worth maintaining.

Runtime feedback can further be used to adapt subsequent communication. DyTopo~\cite{lu2026dytopo} reconstructs sparse communication edges in each round by matching the information required by one agent with that available from others. CARD~\cite{wu2026card} incorporates environmental signals, including changes in model capabilities, tool availability, and computational resources, enabling communication structures to adapt during both training and execution. QueenBee Planner~\cite{tian2026queenbee} extracts communication design knowledge from execution traces and evaluation results, converting it into structural rules that can be reused and revised in later tasks. These approaches extend communication optimization from one-time topology selection to a feedback-driven process informed by current conditions and previous outcomes.

Not all feedback can be generated reliably by agents. Tasks involving implicit preferences, specialized expertise, or high-risk actions may require humans to clarify requirements, correct errors, review outputs, approve actions, or assume control. Collaborative Gym~\cite{shao2026collaborative} supports asynchronous and bidirectional interaction among humans, agents, and task environments, allowing human participation throughout execution. Graph-based communication modeling can represent humans as explicit participants, with edges denoting assistance requests, feedback, approval, and escalation. Humans are thus incorporated as active collaborators in the feedback loop rather than being limited to evaluating the final result.

Graph-based communication modeling should therefore be distinguished from the relatively stable representation of team organization. Team organization determines who participates and what responsibilities they assume. Communication modeling captures who needs to exchange information at a particular point in execution, how feedback is transmitted, and how that feedback changes subsequent actions.



\begin{figure*}[tbp]
\vspace{-5mm}
    \centering
    \includegraphics[width=1.\linewidth, trim=0cm 0cm 0cm 0cm,clip]{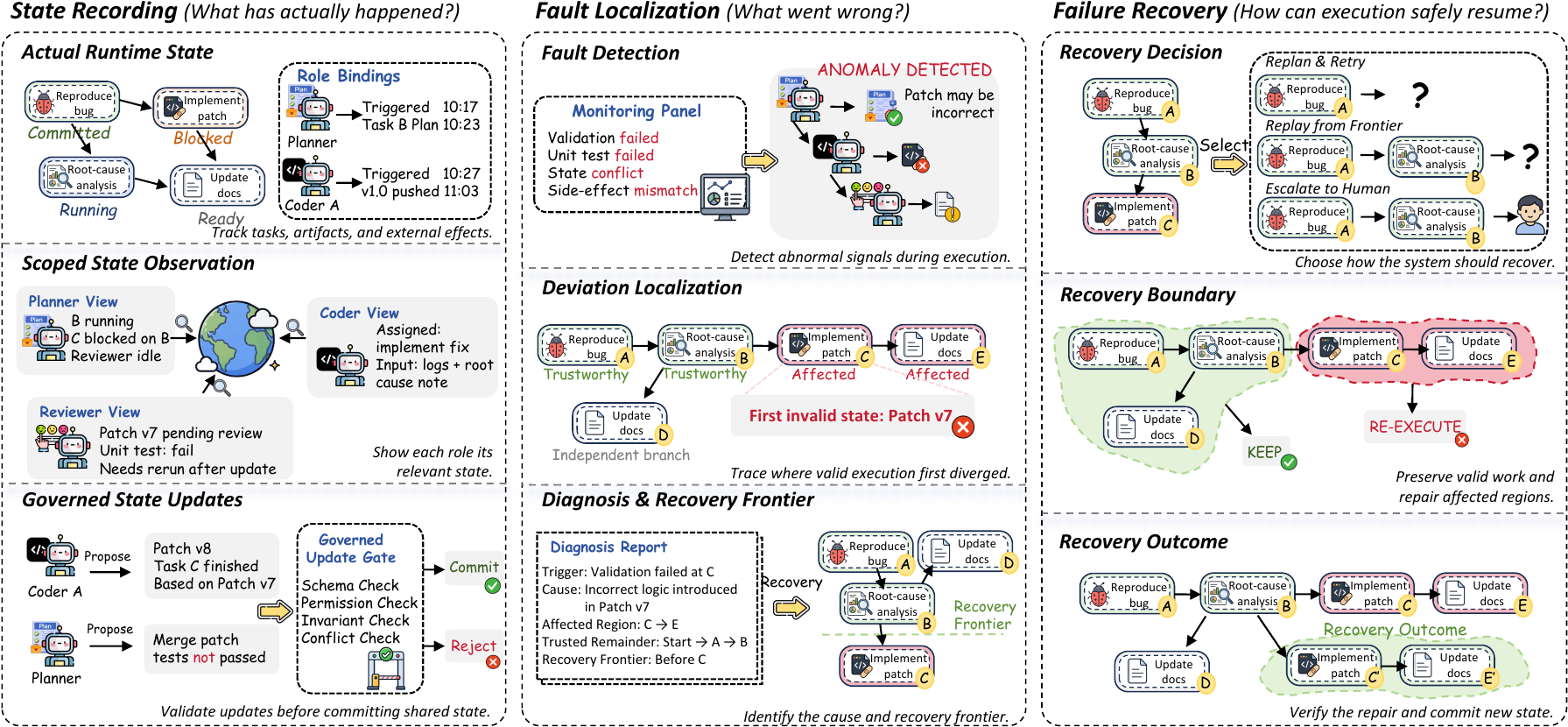}
    \vspace{-5mm}
    \caption{ \textbf{Overview of Runtime State Management.} Runtime State Management structures how to manage state by recording consistent and traceable runtime views, localizing failures from execution evidence, and recovering from validated states. These capabilities turn distributed execution histories into a reliable substrate for monitoring, diagnosis, recovery, and continual system evolution.}
    \label{fig:runtime_state_management}
\end{figure*}

\subsection{Runtime State Management: Structuring How the System Operates}
In an individual agent, runtime state can often remain local to its context, memory, and action history, as observation, decision, and execution are largely unified within a single locus of control. 
However, when intelligence is distributed across a system, execution is split across interdependent tasks and specialized agents, together with partial observations and external effects. Although \emph{Task Organization} and \emph{Agent Coordination} specify what should be executed and by whom, they do not by themselves maintain a reliable account of what has happened, which commitments remain valid, or how one state change affects later decisions. Without such an account, agents may act on inconsistent views, failures are difficult to localize, and valid progress is difficult to recover.
To address this gap, we introduce \emph{Runtime State Management}, which provides three complementary capabilities, as illustrated in Fig.~\ref{fig:runtime_state_management}: \emph{State Recording} maintains consistent and traceable runtime views, \emph{Fault Localization} identifies deviations from intended execution, and \emph{Failure Recovery} restores or redirects execution from validated states.

\subsubsection{State Recording}

Task Organization and Agent Coordination describe what should happen and who should act, but they do not record what has actually happened. During distributed execution, different agents and tools produce partial updates about progress, role bindings, commitments, shared facts, resources, and external effects. If these updates are not kept in a consistent and traceable form, system intelligence cannot maintain a reliable view of the current run or provide evidence for later diagnosis and recovery. 
To address these challenges, \emph{State Recording} preserves the evidence, provenance, and version of each state transition. It turns the planned task and team structures into an explicit, queryable record of the run.

The conditions that a reliable state satisfies usually include \emph{structured representation, governed updation, scoped visibility, and consistency management}. For structured representation, Magentic-One~\cite{fourney2024magentic} externalizes shared execution progress through orchestrator-maintained Task and Progress Ledgers, while Graph of States~\cite{luo2026graphstates} organizes structured belief states and constrains their transitions through causal graphs and state machines. Together, they illustrate a shift from private conversational context toward explicit and traceable runtime state.
For governed updation, PatchBoard~\cite{zhang2026patchboard} validates agent-generated patches against schemas, role permissions, and runtime invariants before commitment, whereas MemTX~\cite{li2026memtx} distinguishes tentative writes from transactional belief commits with explicit provenance and repair semantics. These mechanisms motivate an explicit proposal-validation-commit boundary between observed or proposed changes and authoritative state. 
For scoped visibility, Collaborative Memory~\cite{rezazadeh2025collaborative} complements this process through identity- and time-scoped projections, showing that shared state can remain coordinated without requiring universal visibility.
For consistency management, state generation, lost updates, and causal-order violations motivate isolation, causal ordering, and conflict-resolution mechanisms under concurrent writers~\cite{khan2026concurrency}. Event-sourced designs provide a complementary mechanism by preserving committed transitions in append-only histories that support state reconstruction, replay, and branching~\cite{nakajima2026log}. 

Overall, these studies identify the basic requirements for reliable state recording, but they do not yet provide a unified graph-native implementation.

\subsubsection{Fault Localization}

Recording state does not by itself explain why an execution has gone wrong. In a long-horizon system, a local error may propagate through dependent tasks and agents, while the visible fault appears several steps after the original deviation. \emph{Fault Localization} addresses this problem by detecting abnormal outcomes, locating the decisive error, tracing its effects through dependencies, and testing possible causes against available evidence. Runtime state supports this process by preserving dependencies, provenance, and evidence. The system treats the cause of a fault as a hypothesis and does not assume that temporal or structural links prove causality.

Structured state representations provide complementary mechanisms for localizing faults. Runtime state constrains reasoning from evidence to possible causes through explicit hypothesis-evidence dependencies and supports inspecting earlier states in detail and backtracking when evidence is insufficient~\cite{luo2026graphstates}, while MAGE~\cite{chen2026mage} represents execution as paths in a hierarchical state tree, allowing erroneous branches and nearby valid decision boundaries to be identified. These approaches illustrate how structured execution state can constrain the search space for root-cause analysis. Failure-attribution studies further identify the information needed for this process: Who \& When~\cite{zhang2025whowhen} attributes failures to both responsible agents and steps that caused the failure, MAST~\cite{cemri2025mast} distinguishes system-design, inter-agent coordination, and task-verification failures, and TraceElephant~\cite{chen2026traceelephant} considers execution traces, intermediate context, and complete inputs rather than final outputs alone. Together, they motivate preserving actors, transitions, dependencies, and validation evidence needed to formulate and test attribution hypotheses.
Diagnosis ultimately requires validating such hypotheses against externally observable evidence. TDAD~\cite{alonso2026tdad} connects code changes to affected tests through explicit code-test dependencies, while Cordon~\cite{chen2026cordon} uses typed lineage, shadow state, and semantic transaction boundaries to relate runtime actions to their external effects. These studies suggest that dependencies narrow the search for a cause, but they do not prove the cause. Recording the resulting diagnosis and its supporting evidence as part of runtime state then provides a traceable basis for subsequent recovery.

Overall, fault localization uses runtime records and external evidence to detect faults, trace their effects, test possible causes, and determine which parts of the execution remain valid for recovery.

\subsubsection{Failure Recovery}

Fault localization identifies where execution deviated, but system intelligence also needs a way to continue without discarding valid work or repeating harmful effects. 
\emph{Failure Recovery} addresses this problem by selecting an explicit recovery boundary and determining how execution can safely resume from it. The system may retract invalid states, replay recoverable computation, compensate for external effects, or branch into an alternative execution path. The runtime state layer supports these operations by preserving committed versions, dependencies, provenance, and recovery boundaries, while distinguishing reconstructable internal states from external effects that require compensation.

Existing systems implement recovery mechanisms at different levels. MAGE~\cite{chen2026mage}, ALAS~\cite{geng2025alas}, CausalFlow~\cite{bonagiri2026causalflow}, and ReflexGrad~\cite{kadu2026within} localize failures and selectively repair affected execution regions, avoiding costly global recomputation. Event sourcing~\cite{nakajima2026log}, AgentGit~\cite{li2025agentgit}, and Shepherd~\cite{yu2026shepherd} enable replay, rollback, and branching over recorded execution states, while DART~\cite{yang2026dart} further constrains restoration to semantically valid boundaries under downstream dependencies and committed effects. Together, these approaches support localized recovery while preserving unaffected progress.
In addition to the localized recovery mechanism mentioned, external effects require other recovery guidance that comes from the environment. SagaLLM~\cite{chang2025sagallm} and RAC~\cite{perera2026robust} combine checkpoints with compensation for effects that cannot be directly rolled back, while Atomix~\cite{mohammadi2026atomix} coordinates reversible and irreversible effects through transactional settlement. Aegis~\cite{song2025aegis} complements these mechanisms by improving agent-environment interactions to reduce environment-induced failures. 

Overall, effective recovery requires selective repair with explicit handling of state dependencies and external effects. This process includes recording the recovery boundary, corrective actions, and resulting state, closing the recording, diagnosis, and recovery loop.

\subsection{System Evolution}
In open-ended and long-horizon environments, execution continuously generates evidence about effective structures, coordination strategies, and failure modes~\cite{wang2026agenticeval,yang2026adaptive,yang2026ttcs}. Task organization, agent coordination, and runtime state management provide the foundations for system intelligence, but do not inherently enable improvement over time. To address this challenge, \emph{System Evolution} leverages such experience to refine its organization and operation across executions. The evolution of the system level spans three dimensions: \emph{task organization} improves objective decomposition and workflow construction; \emph{agent coordination} adapts team structures and communication patterns; and \emph{runtime state management} consolidates execution histories into reusable experience while enabling system updates to be validated, revised, or rolled back. These mechanisms turn runtime experience into sustained system-level improvement~\cite{zhou2026mem1,zhang2026memgen,ouyang2026reasoningbank,xiang2026systematic}.

\textbf{Evolution of Task Organization.}
Predefined task structures are often inadequate in open-ended environments, where intermediate outcomes and changing conditions can invalidate prior decomposition and execution plans. Task evolution addresses this limitation by refining both task structures and workflows from execution feedback. At the task-structure level, TDAG~\cite{tdag2025} dynamically decomposes complex tasks and generates specialized agents as execution unfolds. Flow~\cite{flow2025} refines subtask allocation using historical performance and prior workflow structures, while DynTaskMAS~\cite{yu2025dyntaskmas} dynamically maintains task dependencies to support adaptive scheduling and parallel execution. At the workflow level, DyFlow~\cite{dyflow2025} determines subsequent operations from intermediate outputs and real-time feedback; EvoFlow~\cite{evoflow2025} evolves heterogeneous workflow candidates through retrieval, crossover, mutation, and selection; and QualityFlow~\cite{qualityflow2025} uses intermediate quality checks to determine whether to proceed, clarify, or revert execution. Together, these approaches turn task organization from static planning into an iterative process in which execution outcomes refine subsequent decomposition and workflows. Such adaptability, however, also creates vulnerabilities: FlowSteer~\cite{flowsteer2026} shows that manipulated planning signals can steer replanning and dependency formation toward undesirable execution paths. Reliable task evolution therefore requires structural revisions to be grounded in trustworthy execution feedback.

\textbf{Evolution of Agent Coordination.}
Changing task requirements and component capabilities can render predefined coordination structures ineffective. The evolution of agent coordination addresses this mismatch by adapting both \emph{team structures} and \emph{communication patterns} through collaborative experience. For \emph{team structure evolution}, SwarmAgentic~\cite{zhang2025swarmagentic} jointly optimizes agent functionality and collaboration structures through feedback-guided population search. AgentNet~\cite{yang2025agentnet} enables decentralized specialization and reorganization by adjusting agent connectivity and task routing according to local expertise and context, while self-organizing agents~\cite{dochkina2026drop} show that specialized roles and shallow hierarchies can emerge without predefined assignments. Meta-Team~\cite{hao2026evolve} further leverages distributed execution experience to improve agent behavior, inter-agent coordination, and team organization across tasks. For \emph{communication evolution}, DyTopo~\cite{lu2026dytopo} reconstructs communication pathways at each reasoning round by matching agents' information needs and offerings. CARD~\cite{wu2026card} conditions communication structures on environmental changes in model capabilities, tools, and resources, while QueenBee Planner~\cite{tian2026queenbee} distills execution traces and evaluation outcomes into reusable design rules for improving communication in subsequent tasks. Collectively, these approaches shift agent coordination from predefined collaboration toward experience-driven evolution of both team organization and information exchange.

\textbf{Evolution of Runtime State Management.}
The evolution of runtime state management extends state from supporting execution and recovery to accumulating experience for future improvement. This involves two complementary processes: distilling execution histories into reusable knowledge and controlling state revisions to prevent erroneous experience from propagating. ReCreate~\cite{hao2026recreate} derives reusable domain patterns from interaction histories by analyzing the causes of success and failure. SkillGraph~\cite{nie2026skillgraph} distills failure cases into reasoning heuristics maintained in an evolving Skill Bank, while Swarm Skills~\cite{zhang2026swarm} extracts successful trajectories into reusable coordination skills and refines them based on effectiveness, utilization, and freshness. Beyond experience accumulation, reliable evolution requires mechanisms for validating and revising persistent state. MemTX~\cite{li2026memtx} separates tentative writes from validated belief commits and performs cascading repair when committed beliefs are retracted, limiting the propagation of invalid state. ActiveGraph~\cite{nakajima2026log} preserves event-sourced execution histories that support deterministic replay and efficient forking from prior states, enabling alternative branches to build on validated execution history. In summary, these studies make runtime state an experience substrate in which useful knowledge can be accumulated and reused, while unreliable updates can be revised, retracted, or bypassed.

Overall, system evolution enables system intelligence to improve across executions by turning runtime experience into system-level updates. Execution outcomes provide evidence for refining task organization, agent coordination, and state management, while validation and rollback mechanisms ensure that only reliable improvements persist. This establishes a closed loop between execution, experience, and evolution, allowing successful strategies to accumulate and failures to inform subsequent decisions. System intelligence thus progresses from runtime adaptation toward sustained, experience-driven evolution.

\section{Future Direction: Ontology Engineering for Next-Generation System Intelligence}
\label{sec:direction}

Graph Engineering provides a structural foundation for system intelligence by making relationships among work organization, agent coordination, and runtime state explicit, schedulable, and adaptable. However, explicit graph structures alone do not ensure that system entities and relations are defined consistently. Many existing approaches assume that goals, operations, agent capabilities, and runtime states already have clear and shared meanings. This assumption often fails in open, long-running, and industrial environments, where the same concept may be defined differently across graph views, system components, or stages of execution. Ontology Engineering~\cite{gruber1993translation,lippolis2026ontoextendframeworkrequirementdrivenscalable,ekelhart2026agento} addresses this limitation by establishing a shared, machine-interpretable model of system entities, relations, and constraints. It therefore provides the semantic foundation needed to connect, validate, reuse, and evolve graph structures, supporting the next generation of system intelligence.


\subsection{Limitation of Graph Engineering-based System Intelligence}

End-task success alone is insufficient to determine whether a system has developed System Intelligence. Performance gains may result from a stronger foundation model, longer context, additional reasoning samples, or greater computational cost rather than more effective task organization, agent coordination, or state management. Future evaluation should distinguish component-level capability from the contribution of system organization. It should assess goal formation, semantic consistency, parallel execution efficiency, heterogeneous capability allocation, collective decision quality, state consistency, failure recovery, transfer across tasks, and runtime overhead. Evaluation tasks should also include incomplete objectives, concurrent workloads, distributed information, component failures, and environmental changes to test whether the system can maintain coherent behavior under structural disturbances. Beyond end-to-end metrics, intervention studies, structural ablations, and execution-trace analysis are needed to identify the causal contributions of different system mechanisms and distinguish genuine system-level capability from gains produced by additional computation.

These challenges clarify the limits of Graph Engineering. Graph structures can explicitly organize relationships among tasks, components, and runtime states, but System Intelligence must also formulate appropriate goals, establish a shared and grounded understanding of the system, and support rigorous system-level evaluation. Ontology Engineering primarily addresses the need for shared semantics and can also provide consistent definitions for goals, roles, states, evidence, and operational constraints. It should therefore be viewed as a semantic foundation connecting Graph Engineering to broader System Intelligence rather than a complete solution to all system-level challenges.

Graph Engineering makes relationships among tasks, agents, and runtime states explicit, but explicit structures do not ensure that system components interpret them consistently. Agents may still disagree about what constitutes task completion, sufficient evidence, valid state, or authorized action. Ontology Engineering addresses this broader problem by establishing a shared, machine-interpretable model of the system. Rather than merely adding semantic annotations to graphs, it defines which entities exist, what their relations mean, which constraints must hold, and what conclusions can be derived from them.

An ontology for System Intelligence should be layered and modular. A core ontology can define concepts shared across systems, while specialized modules describe goals and values, agents and capabilities, observations and evidence, actions and states, and evaluation criteria. Domain ontologies can further extend these concepts for particular applications. Such a structure provides consistent definitions of Goals, Agents, Capabilities, Evidence, Policies, States, and Outcomes without requiring every system or domain to adopt a single monolithic model.

\subsection{Ontology Engineering}
\subsubsection{Goal Formation and Value Alignment}

For Goal Formation and Value Alignment, Ontology Engineering can represent the provenance, priority, authorization scope, completion criteria, and constraints of candidate goals. These representations allow a system to identify goal conflicts, detect unauthorized modifications, and determine what evidence is required for completion. Ontologies cannot decide which values a system should adopt, but they can make goals and normative constraints explicit and verifiable. Recent ontology-guided agent systems illustrate this shift from representing domain concepts to constraining agent reasoning~\cite{seo2026toward,wang2026agentic,ortacc2026agentology}. In LAMP, a Planner, Builder, and Verifier collaboratively access a domain-specific ontology through MCP, using explicit structured knowledge at inference time rather than relying solely on model parameters [7]. Likewise, Agentology proposes treating the ontology-defined environment, rather than the individual agent prompt, as the primary object of system design, allowing multiple specialist agents to reason over a shared and persistent semantic structure~\cite{ortacc2026agentology}. These developments point toward an ontology-centered organization of multi-agent systems in which semantic constraints are externalized from individual agents and shared across the system.


\subsubsection{Shared Semantics and World Grounding}

For Shared Semantics and World Grounding, ontologies provide common definitions and mappings across agents and systems. These concepts must also be connected to tool outputs, environmental observations, timestamps, provenance, and validation results, since semantic consistency alone does not guarantee factual correctness. Recent multi-agent ontology-enrichment systems demonstrate how this semantic layer can itself be dynamically maintained. OntoCodex coordinates decision, ontology-reading, knowledge-base, terminology, and script-generation agents to enrich an existing OWL ontology while preserving its structural constraints and grounding newly introduced concepts in curated knowledge sources~\cite{feng2026ontocodex}. CoA-Text2OWL similarly distributes ontology learning across multiple worker agents and a manager agent, demonstrating the potential of agent collaboration for constructing coherent ontologies from large textual sources [10]. Ontology-grounded tool and agent designs, such as AgentO and Ontology-to-Tools, further connect semantic concepts to executable capabilities and tool interfaces~\cite{ekelhart2026agento,zhou2026ontology}. These systems indicate that future ontology infrastructure may not remain static: agents can participate in proposing, validating, aligning, and updating the semantic model while retaining explicit provenance and human oversight.


\subsubsection{Measuring System Intelligence}

Ontology Engineering can also support the measurement of System Intelligence by standardizing the meanings of task success, failure, agent contribution, recovery, state consistency, and runtime cost. Shared representations of system configurations, execution events, evidence, interventions, and outcomes would make execution traces more comparable across systems and support structural ablation and causal analysis. Ontologies do not replace evaluation methods, but they clarify what is being measured and whether a metric concerns foundation-model capability, individual-agent performance, or system-level organization~\cite{li2026large,palantir2026ontology}. By providing a common vocabulary for system-level events and entities, ontology specifications such as Ontology SLR and Palantir Ontology could facilitate more systematic comparison of heterogeneous agent architectures~\cite{li2026large,palantir2026ontology}.

Future research should investigate how system ontologies are grounded, updated, and governed. LLMs may assist in proposing new concepts and relations, but semantic changes should undergo provenance checking, consistency validation, and impact analysis~\cite{lippolis2026ontoextendframeworkrequirementdrivenscalable,feng2026ontocodex,zhou2026ontology,ortacc2026agentology}. System ontologies must also support version control, compatibility checking, migration, and rollback as tasks and environments change. Their constraints should be connected to runtime mechanisms that enforce permission checks, evidence requirements, and valid state transitions. Recent work such as LAMP demonstrates one direction in which structured ontology knowledge is directly exposed to agents through tool interfaces, while Agentology explicitly treats the ontology as part of the operational environment within which multiple agents reason~\cite{ortacc2026agentology,zhou2026ontology}. Ontology Engineering thus defines the shared conceptual model of System Intelligence, Graph Engineering instantiates this model as task-specific structures, and runtime mechanisms enforce its operational consequences.

\section{Open Challenges and Research Opportunities}
\label{sec:challenges}

Graph Engineering provides a structural foundation for transforming LLM-based agent systems from individual intelligence into system intelligence by explicitly modeling work organization, component coordination, and state evolution. However, moving from task-specific graph structures toward general-purpose infrastructures that can operate continuously and be reused across systems introduces several unresolved challenges. These challenges concern not only how graphs are constructed, but also how they can evolve dynamically, operate reliably, scale efficiently, and interoperate across heterogeneous agent systems. More fundamentally, as the Work Organization Graph, Agent Team Graph, and State Evolution Graph become increasingly interconnected, the research focus must shift from merely constructing graph structures toward ensuring their consistent interpretation, reliable execution, and continual evolution.


\subsection{LLM-based Autonomous Ontology Construction}

Ontologies provide shared conceptualizations of entities, relations, constraints, and inference rules, enabling heterogeneous graph data to be interpreted, reused, and validated across modules and agents~\cite{gruber1993translation,noy2001ontology,suarezfigueroa2015neon}. This semantic layer is particularly important for Graph Engineering, where tasks, agents, tools, capabilities, states, events, and evidence must remain interoperable.
High-quality ontology construction therefore becomes critical: it must also ensure logical consistency, structural validity, interoperability, provenance, and maintainability.

Recent work has established a pipeline from candidate induction to constrained ontology engineering. LLMs4OL decomposes ontology learning into term typing, taxonomy discovery, and non-taxonomic relation extraction~\cite{babaei2023llms4ol}. SPIRES/OntoGPT constrains extraction with user-defined classes, slots, and vocabularies~\cite{caufield2024spires}, while NeOn-GPT combines LLM prompting with the NeOn methodology to produce structured ontology artifacts~\cite{fathallah2025neongpt,suarezfigueroa2015neon}. For reuse and consistency, BERTMap and LLMs4OM investigate language-model-based ontology matching with structural or logic-based repair~\cite{he2022bertmap,llms4om2024}; DeepOnto integrates ontology processing, reasoning, alignment, and completion with deep-learning workflows~\cite{he2024deepponto}. Emerging work such as OntoExtend further frames the problem as requirement-driven and scalable ontology lifecycle management~\cite{ontoextend2026}, in which LLM agents elicit requirements, propose ontology patches, and repair candidate changes, while formal validators and human reviewers control acceptance~\cite{zhang2025ontochat,shimizu2025accelerating}.

In short, LLM-based ontology engineering is evolving from LLM-assisted modeling toward agentic, autonomous ontology construction and continuous evolution. 
Existing systems demonstrate practical capabilities in concept induction, schema-constrained extraction, ontology matching, and completion, but unconstrained generation remains unreliable under semantic drift, logical inconsistency, and continual evolution. A robust future paradigm is therefore hybrid: LLM agents propose and explain ontology changes, whereas OWL reasoning, SHACL validation, provenance tracking, regression testing, version control, and human governance determine whether those changes are accepted.

\subsection{Graph-Native Capability Substrates}

Current Graph Engineering primarily makes the organization of tasks, agents, and runtime states explicit, while many capabilities used by an agent system are still maintained as independent collections or services~\cite{li2026goal}. Memory stores contain experiences and facts, skill libraries contain reusable procedures, and tool registries expose executable functions, but the relationships among these capabilities are often implicit~\cite{yu2026memagent,ji2026memory,wang2026mem}. As these repositories grow, capability selection becomes increasingly structural: a capability may depend on another capability, substitute for an unavailable one, compose with several others, require specific permissions, or be applicable only under particular runtime conditions.

Recent work on memory and skills already points toward graph-structured capability substrates~\cite{cao2026higmem,shen2026anchormem,fang2026memp}. A-MEM~\cite{xu2025amem} dynamically links related memories into evolving memory networks, while Zep~\cite{rasmussen2025zep} represents changing facts and their temporal relations through a temporal knowledge graph. A similar transition is emerging for reusable skills. Graph of Skills~\cite{li2026graphofskills} represents dependencies and workflow relations among skills to retrieve executable skill bundles rather than isolated entries, while SkillDAG~\cite{bai2026skilldag} further allows typed skill relations to evolve from execution evidence. These systems suggest that graphs can organize not only tasks and agents, but also the capability substrate on which they operate.

A broader direction is therefore to construct unified \emph{capability graphs} in which models, tools, skills, memories, data sources, verifiers, and execution environments are represented as typed nodes, with edges describing dependency, compatibility, composition, substitution, authorization, cost, and reliability. The main research challenge is not simply to represent each capability family as a graph, but to connect these capability graphs with task, agent, and runtime state graphs. Task decomposition should expose capability requirements, agent allocation should consider available capability subgraphs, and execution outcomes should update capability reliability and applicability. Such coupling would allow Graph Engineering to move from organizing system execution to organizing the reusable capability space from which execution is constructed~\cite{xu2026structmem,li2026timem,feng2026searl,zhang2026personaagent}.

\subsection{Self-Evolving Graph Systems}

Existing Graph Engineering methods increasingly make graph structure an optimization variable. GPTSwarm~\cite{zhuge2024gptswarm} optimizes computational graphs of language agents, while workflow and topology optimization methods such as AFlow~\cite{zhang2025aflow} and DyTopo~\cite{lu2026dytopo} adapt executable or communication structures according to task feedback. Other approaches accumulate execution experience for future improvement: ReCreate~\cite{hao2026recreate} derives reusable patterns from successful and failed trajectories, while MemTX~\cite{li2026memtx} and event-sourced agent designs~\cite{nakajima2026log} provide mechanisms for validating, revising, replaying, and forking persistent state. These developments represent early steps from fixed graph execution toward experience-driven structural adaptation.

However, runtime adaptation should be distinguished from persistent system evolution. Conditional routing, temporary worker assignment, or recovery may change one execution trajectory without changing the organization used in later tasks. A self-evolving graph system should instead transform execution evidence into persistent and reusable structural changes. This requires a closed process from execution and observation to structural credit assignment, graph modification, validation, and finally commit or rollback. Future systems must determine which task dependencies, agent relations, capability assignments, or state structures were responsible for success and failure, and whether the resulting modification generalizes beyond the current execution.

An additional challenge is that these graphs cannot evolve independently. Modifying a task graph may change the capabilities required from the agent team, while replacing an agent may invalidate communication relations, permissions, or runtime assumptions. Future research should therefore study \emph{cross-graph evolution}, where changes to task, agent, capability, and state graphs are coordinated under shared constraints. Structural evolution must also remain governable through provenance, versioning, validation, replay, and rollback. The long-term objective is not unrestricted self-modification, but systems that can accumulate useful organizational experience while preventing unreliable structural changes from propagating across executions.

\subsection{Graph-Native Agent Operating Systems}

The growing complexity of agent systems also raises an infrastructure question. Current engineering stacks separate model serving, harnesses, workflow engines, memory systems, multi-agent frameworks, and state stores, each using different abstractions for tasks, tools, messages, agents, events, and execution state. Protocols such as MCP~\cite{anthropic2024mcp} improve access to external capabilities, while graph-oriented frameworks such as LangGraph~\cite{langgraph2026} provide explicit workflow and state representations. AIOS~\cite{mei2025aios} takes a complementary operating-system view by providing scheduling, context, memory, storage, tool, and access-control services for LLM agents. However, these mechanisms do not yet provide a common structural substrate for organizing complete agent systems. AIOS itself is an important precedent: its kernel explicitly separates agent applications from scheduling, memory, storage, tools, and access control, showing why these concerns increasingly resemble operating-system services rather than application-specific logic~\cite{mei2025aios}.

A future \emph{graph-native agent operating system} could make tasks, agents, capabilities, and runtime states first-class system objects represented through typed and versioned graphs. Instead of each framework separately implementing workflow scheduling, resource allocation, persistent state, communication, and recovery, a shared runtime could provide graph scheduling, capability discovery, state storage, event and provenance logging, structural transactions, permission enforcement, checkpointing, replay, rollback, and graph-level observability. Ontology Engineering would define the types, relations, and constraints of these objects, while the graph runtime would enforce their operational semantics.

Such an infrastructure would also provide the foundation required for safe system evolution. Execution traces could be linked directly to the graph structures that produced them, candidate structural changes could be evaluated against historical or counterfactual executions, and validated improvements could be committed as new graph versions. Graph Engineering would then evolve from a method for designing individual workflows or multi-agent topologies into a reusable system substrate for constructing, executing, observing, and continuously improving agent systems. This progression from shared semantics, to graph-structured capabilities, to controlled structural evolution, and finally to graph-native runtime infrastructure represents a possible path toward scalable and persistent System Intelligence.


\subsection{Privacy and Ethics}

System intelligence introduces broader privacy and ethical risks because it coordinates multiple agents, tools, memories, and shared states over long horizons. Compared with a single-agent setting, sensitive information may be replicated across components, propagated through workflows, and preserved in persistent state, increasing the risk of unauthorized access, cross-task leakage, and unintended inference of private attributes from execution traces. Moreover, as decisions are distributed across interacting components, accountability becomes harder to assign when biased evidence, faulty reasoning, or adversarial inputs are amplified through the system. Future system-intelligent agents therefore require privacy-preserving state management, scoped permissions, provenance-aware logging, and strong human oversight to ensure that autonomy does not come at the cost of user privacy, fairness, or controllability.

\section{Benchmarks, Datasets, and Evaluation}
\label{sec:benchmarks}

Evaluation should follow the unit of intelligence being studied. Model Intelligence concerns capabilities expressed within bounded model interactions. Individual Intelligence concerns whether a single autonomous agent can combine reasoning with external capabilities and environmental feedback over a sustained trajectory. System Intelligence further concerns whether multiple intelligent components and their relations can be organized, coordinated, maintained, and improved as a coherent system. We therefore organize evaluation resources around these three levels rather than by task domain or graph type.

We distinguish three forms of evaluation resource. A \emph{benchmark} defines tasks, an evaluation protocol, and scoring rules. A \emph{dataset} provides reusable instances, annotations, graphs, interaction records, or execution traces. An \emph{environment} exposes executable state that an agent or agent system can observe and modify. These forms are not mutually exclusive. Table~\ref{tab:intelligence-evaluation-resources} uses B, D, and E to denote benchmark, dataset, and executable environment, respectively.

\providecommand{\BenchPaper}[1]{\href{#1}{\textcolor{cyan!60!black}{\faFilePdf}}}
\providecommand{\BenchCode}[1]{\href{#1}{\textcolor{black}{\faGithub}}}
\providecommand{\BenchData}[1]{\href{#1}{\textcolor{orange!85!black}{\faDatabase}}}

\begin{table*}[!t]
\centering
\caption{Representative benchmarks, datasets, and executable environments across Model, Individual, and System Intelligence. Type: B = benchmark or evaluation protocol; D = released dataset, annotations, or traces; E = executable or interactive environment. Focus denotes the principal capability or structural property evaluated.}
\label{tab:intelligence-evaluation-resources}

\fontsize{5.35}{5.95}\selectfont
\setlength{\tabcolsep}{1.15pt}
\renewcommand{\arraystretch}{0.96}

\begin{tabular*}{\textwidth}{
@{\extracolsep{\fill}}
>{\raggedright\arraybackslash}p{2.75cm}
>{\centering\arraybackslash}p{0.62cm}
>{\raggedright\arraybackslash}p{2.05cm}
>{\raggedright\arraybackslash}p{1.18cm}
>{\raggedright\arraybackslash}p{8.05cm}
>{\centering\arraybackslash}p{0.52cm}
@{}
}
\toprule
\textbf{Name} &
\textbf{Type} &
\textbf{Primary Unit} &
\textbf{Focus} &
\textbf{Evaluation} &
\textbf{Link} \\
\midrule

\multicolumn{6}{c}{\textit{Model Intelligence}} \\
\midrule

MMLU/MMLU-Pro~\cite{hendrycks2021mmlu,wang2024mmlupro}
& B/D
& QA instances
& Knowledge
& Broad knowledge and problem solving; MMLU-Pro increases reasoning difficulty and prompt robustness.
& \BenchPaper{https://arxiv.org/abs/2406.01574} \\

GPQA~\cite{rein2024gpqa}
& B/D
& Expert science QA
& Reasoning
& Graduate-level scientific knowledge and difficult multi-step reasoning.
& \BenchPaper{https://arxiv.org/abs/2311.12022} \\

NPPC~\cite{yang2026nppc}
& B/D
& NP instances
& Reasoning
& Scalable, automatically verifiable reasoning over NP-complete problems.
& \BenchPaper{https://arxiv.org/abs/2504.11239} \\

OlymMATH~\cite{sun2026olymmath}
& B/D
& Math problems
& Reasoning
& Olympiad-level reasoning with objective and formal verification.
& \BenchPaper{https://aclanthology.org/2026.acl-long.792/} \\

IFEval~\cite{zhou2023ifeval}
& B/D
& Instruction pairs
& Following
& Verifiable instruction following under objectively checkable constraints.
& \BenchPaper{https://arxiv.org/abs/2311.07911} \\

EvolIF~\cite{jia2026evolif}
& B/D
& Multi-turn dialogues
& Following
& Evolving instruction following, constraint tracking, and failure recovery.
& \BenchPaper{https://aclanthology.org/2026.acl-long.433/} \\

HumanEval/EvalPlus~\cite{chen2021codex,liu2023evalplus}
& B/D
& Coding problems
& Coding
& Executable functional correctness with strengthened test coverage.
& \BenchPaper{https://arxiv.org/abs/2305.01210} \\

MMMU~\cite{yue2024mmmu}
& B/D
& Multimodal QA
& Multimodal
& Expert-level multimodal understanding and reasoning across disciplines.
& \BenchPaper{https://openaccess.thecvf.com/content/CVPR2024/html/Yue_MMMU_A_Massive_Multi-discipline_Multimodal_Understanding_and_Reasoning_Benchmark_for_CVPR_2024_paper.html} \\

OMHBench~\cite{kim2026omhbench}
& B/D
& Omni-modal QA
& Multimodal
& Grounded multi-hop reasoning across text, vision, and speech.
& \BenchPaper{https://aclanthology.org/2026.findings-acl.911/} \\

LiveBench~\cite{white2025livebench}
& B/D
& Refreshable tasks
& General
& Frequently refreshed capability evaluation with objective scoring to reduce contamination.
& \BenchPaper{https://arxiv.org/abs/2406.19314} \\

GraphRAG-Bench~\cite{xiang2026graphragbench}
& B/D
& RAG tasks
& Retrieval
& Graph construction, retrieval, reasoning, and generation in GraphRAG.
& \BenchPaper{https://proceedings.iclr.cc/paper_files/paper/2026/hash/6c9e01d6cefbbf4cdd265032550e767f-Abstract-Conference.html} \\

\midrule
\multicolumn{6}{c}{\textit{Individual Intelligence}} \\
\midrule

$A^2E$~\cite{arxiv:Wang_2026}
& B/D
& Harness executions
& Harness
& End-to-end agent harness auditing; execution efficiency, tool use, task planning, and error recovery.
& \BenchCode{https://github.com/datamllab/A2E/tree/main} \\

AgentBench~\cite{liu2024agentbench}
& B/D/E
& Agent trajectories
& General
& Reasoning and decision making across multiple interactive environments.
& \BenchPaper{https://arxiv.org/abs/2308.03688} \\

GAIA~\cite{mialon2024gaia}
& B/D
& Assistant tasks
& General
& Integrated reasoning, browsing, multimodal understanding, and tool use.
& \BenchPaper{https://arxiv.org/abs/2311.12983} \\

AgencyBench~\cite{li2026agencybench}
& B/D/E
& Long-horizon tasks
& General
& Long-horizon real-world autonomy with tools and extended context.
& \BenchPaper{https://aclanthology.org/2026.acl-long.337/} \\

AgentGym2~\cite{xi2026agentgym2}
& B/D/E
& Real-world tasks
& General
& Tool discovery and composition under noisy and underspecified environments.
& \BenchPaper{https://aclanthology.org/2026.acl-long.2058/} \\

WebArena~\cite{zhou2024webarena}
& B/D/E
& Web trajectories
& Web
& Long-horizon interaction with realistic websites and execution-based evaluation.
& \BenchPaper{https://arxiv.org/abs/2307.13854} \\

OSWorld~\cite{xie2024osworld}
& B/D/E
& Computer trajectories
& Computer
& Open-ended interaction with real desktop applications and operating systems.
& \BenchPaper{https://arxiv.org/abs/2404.07972} \\

Terminal-Bench 2.0~\cite{merrill2026terminalbench}
& B/D/E
& Terminal tasks
& Terminal
& Realistic command-line tasks with executable verification.
& \BenchPaper{https://arxiv.org/abs/2601.11868} \\

SWE-bench/Pro~\cite{jimenez2024swebench,deng2025swebenchpro}
& B/D/E
& Repository tasks
& Software
& Repository-level issue resolution with executable verification and harder long-horizon tasks.
& \BenchPaper{https://arxiv.org/abs/2509.16941} \\

LongCLI-Bench~\cite{feng2026longcli}
& B/D/E
& CLI tasks
& Software
& Long-horizon command-line software engineering with step-level evaluation.
& \BenchPaper{https://aclanthology.org/2026.findings-acl.1497/} \\

AppWorld~\cite{trivedi2024appworld}
& B/D/E
& Application state
& Tools
& API use, code generation, application-state transitions, and task completion.
& \BenchData{https://appworld.dev/} \\

$\tau$-bench/$\tau^2$-bench~\cite{yao2024taubench,barres2025tau2}
& B/D/E
& Agent-user tasks
& Tools
& Policy following, tool use, and user-agent coordination in shared environments.
& \BenchPaper{https://arxiv.org/abs/2506.07982} \\

ToolSandbox~\cite{lu2024toolsandbox}
& B/D/E
& Stateful dialogues
& Tools
& Stateful tool execution, dependencies, intermediate milestones, and recovery.
& \BenchCode{https://github.com/apple/ToolSandbox} \\

AgentDojo~\cite{debenedetti2024agentdojo}
& B/D/E
& Adversarial episodes
& Security
& Agent utility and robustness under prompt injection attacks.
& \BenchCode{https://github.com/ethz-spylab/agentdojo} \\

Harness-Bench~\cite{yao2026harnessbench}
& B/D/E
& Sandboxed workflows
& Harness
& Effects of context, tools, state, constraints, permissions, tracing, and recovery.
& \BenchPaper{https://arxiv.org/abs/2605.27922} \\

HarnessOpt-Bench~\cite{ursekar2026harnessopt}
& B/E
& Harness variants
& Harness
& Evaluation-guided harness optimization under fixed budgets.
& \BenchPaper{https://arxiv.org/abs/2608.06301} \\

Skill-Use~\cite{han2026skilluse}
& B/D/E
& Skill tasks
& Skills
& Skill triggering, procedural compliance, capability boundaries, and harness dependence.
& \BenchPaper{https://arxiv.org/abs/2608.04828} \\

LongMemEval~\cite{wu2025longmemeval}
& B/D
& Multi-session QA
& Memory
& Information extraction, temporal reasoning, updates, and long-term interactive memory.
& \BenchPaper{https://arxiv.org/abs/2410.10813} \\

MemoryAgentBench~\cite{hu2026memoryagentbench}
& B/D
& Multi-turn memory
& Memory
& Retrieval, test-time learning, long-range understanding, and selective forgetting.
& \BenchCode{https://github.com/HUST-AI-HYZ/MemoryAgentBench} \\

MemoryArena~\cite{he2026memoryarena}
& B/D/E
& Multi-session tasks
& Memory
& Acquisition and reuse of experience across interdependent sessions.
& \BenchPaper{https://arxiv.org/abs/2602.16313} \\

GateMem~\cite{ren2026gatemem}
& B/D
& Memory episodes
& Memory
& Access control, deletion, selective forgetting, and memory governance.
& \BenchData{https://rzhub.github.io/GateMem/project.html} \\

MemSyco-Bench~\cite{xiang2026memsyco}
& B/D
& Memory decisions
& Memory
& Appropriate use of retrieved memory under factual, scope, conflict, update, and personalization conditions.
& \BenchPaper{https://arxiv.org/abs/2607.01071} \\

Mem2ActBench~\cite{shen2026mem2act}
& B/D
& Memory-tool chains
& Memory
& Contribution of retained memory to subsequent tool actions.
& \BenchCode{https://github.com/Cantaloupe-M/Mem2ActBench} \\

LongDS-Bench~\cite{xu2026longds}
& B/D/E
& Long trajectories
& Long-term
& State maintenance, restoration, adaptation, and rollback over long executions.
& \BenchPaper{https://arxiv.org/abs/2605.30434} \\

EvoMemBench~\cite{wang2026evomembench}
& B/D
& Memory episodes
& Evolution
& Memory evolution and selective retention within and across episodes.
& \BenchCode{https://github.com/DSAIL-Memory/EvoMemBench} \\

Trainee-Bench~\cite{fu2026traineebench}
& B/D/E
& Workplace streams
& Evolution
& Scheduling, exploration, and continual learning in dynamic workplaces.
& \BenchPaper{https://arxiv.org/abs/2601.08173} \\

SEA-Eval~\cite{jiang2026seaeval}
& B/D/E
& Task streams
& Evolution
& Cross-task evolutionary gain, stability, and execution efficiency.
& \BenchPaper{https://arxiv.org/abs/2604.08988} \\

Evo-Bench~\cite{huang2026evobench}
& B/D
& Harness evolution
& Evolution
& Autonomous harness improvement and cross-domain transfer.
& \BenchPaper{https://arxiv.org/abs/2608.09096} \\

OpenClawBench~\cite{liu2026openclawbench}
& B/D
& Execution traces
& Failure
& Process-side anomalies, robustness, and failures in real agent trajectories.
& \BenchPaper{https://arxiv.org/abs/2605.29253} \\

BenchTrace~\cite{huang2026benchtrace}
& B/D
& Repeated episodes
& Reflection
& Whether reflection on failures improves behavior in subsequent executions.
& \BenchPaper{https://arxiv.org/abs/2605.29225} \\

TheAgentCompany~\cite{xu2025agentcompany}
& B/D/E
& Workplace episodes
& Long-term
& Long-horizon workplace tasks spanning browsing, coding, and communication.
& \BenchPaper{https://arxiv.org/abs/2412.14161} \\

\midrule
\multicolumn{6}{c}{\textit{System Intelligence}} \\
\midrule

TaskBench~\cite{shen2024taskbench}
& B/D
& Tool graphs
& Work
& Task decomposition, tool selection, grounding, and explicit tool-graph construction.
& \BenchPaper{https://arxiv.org/abs/2311.18760} \\

WorFBench~\cite{qiao2025worfbench}
& B/D
& Workflow graphs
& Work
& Workflow generation with sequence-level and graph-level structure matching.
& \BenchPaper{https://arxiv.org/abs/2410.07869} \\

FlowBench~\cite{xiao2024flowbench}
& B/D
& Workflow pairs
& Work
& Workflow-guided planning across heterogeneous workflow representations.
& \BenchPaper{https://aclanthology.org/2024.findings-emnlp.638/} \\

ComfyBench~\cite{xue2025comfybench}
& B/D/E
& Executable workflows
& Work
& Construction and execution of explicit node-edge workflows.
& \BenchPaper{https://openaccess.thecvf.com/content/CVPR2025/html/Xue_ComfyBench_Benchmarking_LLM-based_Agents_in_ComfyUI_for_Autonomously_Designing_Collaborative_CVPR_2025_paper.html} \\

TPS-Bench~\cite{xu2026tpsbench}
& B/D/E
& Scheduling tasks
& Work
& Dependency-aware planning, parallel scheduling, throughput, and execution efficiency.
& \BenchPaper{https://aclanthology.org/2026.acl-long.1614/} \\

JourneyBench~\cite{balaji2026journeybench}
& B/D/E
& Policy workflows
& Work
& Policy-constrained service workflows and business-rule adherence.
& \BenchPaper{https://aclanthology.org/2026.eacl-industry.15/} \\

ETOM~\cite{dong2026etom}
& B/D/E
& Tool hierarchies
& Work
& Hierarchical orchestration, server selection, and out-of-scope robustness.
& \BenchPaper{https://aclanthology.org/2026.findings-eacl.75/} \\

LLM-Coordination~\cite{agashe2023coordination}
& B/D/E
& Coordination games
& Team
& Joint planning, theory of mind, sustained coordination, and partner robustness.
& \BenchPaper{https://aclanthology.org/2025.findings-naacl.448/} \\

VillagerBench~\cite{dong2024villageragent}
& B/D/E
& Minecraft tasks
& Team
& Workload distribution, task dependencies, adaptation, and synchronized execution.
& \BenchPaper{https://aclanthology.org/2024.findings-acl.964/} \\

MultiAgentBench~\cite{zhu2025multiagentbench}
& B/D/E
& Multi-agent episodes
& Team
& Collaboration, competition, milestones, and topology-sensitive coordination.
& \BenchPaper{https://aclanthology.org/2025.acl-long.421/} \\

AgentsNet~\cite{groetschla2025agentsnet}
& B/D/E
& Networked tasks
& Evolution
& Self-organization, adaptive communication, and network scaling.
& \BenchPaper{https://arxiv.org/abs/2507.08616} \\

DBS~\cite{wan2026dawn}
& B/D
& Workflow graphs
& Evolution
& Adaptive workflow synthesis under distributed heterogeneity and privacy.
& \BenchPaper{https://ojs.aaai.org/index.php/AAAI/article/view/39812} \\

SILO-BENCH~\cite{zhang2026silobench}
& B/D/E
& Distributed tasks
& Team
& Role-free coordination under information silos and agent scaling.
& \BenchPaper{https://aclanthology.org/2026.acl-long.1354/} \\

CoLLAB~\cite{mahmud2025collab}
& B/D
& Coordination tasks
& Team
& Constraint-based coordination and structural credit assignment.
& \BenchPaper{https://openreview.net/forum?id=372FjQy1cF} \\

Collab-Overcooked~\cite{sun2025collabovercooked}
& B/D/E
& Collaboration games
& Team
& Process-oriented collaboration quality beyond final task success.
& \BenchPaper{https://aclanthology.org/2025.emnlp-main.249/} \\

MAS-BENCH~\cite{yang2026masbench}
& B/D/E
& Distributed sorting
& Team
& Shared-state consistency, protocol alignment, termination, and agent scaling.
& \BenchPaper{https://aclanthology.org/2026.findings-acl.1698/} \\

DPBench~\cite{hasan2026dpbench}
& B/D/E
& Contention games
& Team
& Sequential and simultaneous coordination under shared-resource contention.
& \BenchCode{https://github.com/najmulhasan-code/dpbench} \\

CalBench~\cite{zou2026calbench}
& B/D/E
& Decentralized tasks
& Team
& Coordination, communication efficiency, fairness, and privacy under private information.
& \BenchPaper{https://arxiv.org/abs/2605.09823} \\

TAMAS~\cite{kavathekar2026tamas}
& B/D/E
& Adversarial episodes
& Team
& Robustness and safety under adversarial multi-agent interaction.
& \BenchPaper{https://aclanthology.org/2026.acl-long.1442/} \\

SyncBench~\cite{guo2025syncmind}
& B/D/E
& Recovery instances
& State
& Belief-world consistency, diagnosis, resource awareness, and recovery.
& \BenchPaper{https://proceedings.mlr.press/v267/guo25l.html} \\

MAST~\cite{cemri2025mast}
& D
& Failure annotations
& State
& Multi-agent failure modes in system design, alignment, and verification.
& \BenchPaper{https://arxiv.org/abs/2503.13657} \\

Who\&When/Pro~\cite{zhang2025whowhen,liu2026pro}
& B/D
& Failed trajectories
& State
& Agent- and step-level failure attribution under controlled failures.
& \BenchPaper{https://arxiv.org/abs/2607.09996} \\

TraceElephant~\cite{chen2026traceelephant}
& B/D/E
& Execution traces
& State
& Failure attribution under complete execution observability.
& \BenchPaper{https://aclanthology.org/2026.acl-long.912/} \\

MP-Bench~\cite{in2026mpbench}
& B/D
& Failure cases
& State
& Multi-perspective evaluation when failures admit several plausible attributions.
& \BenchPaper{https://arxiv.org/abs/2603.25001} \\

R2Act~\cite{qi2026r2act}
& B/D/E
& Incident states
& State
& Diagnosis-to-action reasoning, admissible recovery, and recovery validity.
& \BenchPaper{https://arxiv.org/abs/2607.04623} \\

MASEval~\cite{emde2026maseval}
& B/E
& System variants
& Evolution
& Topology, orchestration, framework, and runtime design comparison.
& \BenchPaper{https://aclanthology.org/2026.acl-demo.34/} \\

MAS-PromptBench~\cite{bai2026maspromptbench}
& B/E
& MAS configurations
& Evolution
& Optimization across workflow topologies, protocols, and team sizes.
& \BenchPaper{https://arxiv.org/abs/2606.23664} \\

BenchAgent~\cite{fu2026benchagent}
& B/E
& Agent workflows
& Evolution
& Controlled comparison of single, fixed multi-agent, and evolving agent workflows.
& \BenchPaper{https://arxiv.org/abs/2606.05670} \\

\bottomrule
\end{tabular*}
\end{table*}

\subsection{Model Intelligence}

Model Intelligence evaluation focuses on capabilities expressed within bounded model interactions. Representative benchmarks cover broad knowledge and reasoning, instruction following, executable code generation, multimodal understanding, and retrieval-augmented reasoning~\cite{hendrycks2021mmlu,wang2024mmlupro,rein2024gpqa,zhou2023ifeval,chen2021codex,liu2023evalplus,yue2024mmmu,xiang2026graphragbench}. Recent resources also address the rapid saturation of static evaluation: LiveBench refreshes questions to reduce contamination~\cite{white2025livebench}, while NPPC generates automatically verifiable NP-complete problem instances with scalable difficulty~\cite{yang2026nppc}. The primary evaluation unit at this level remains the model output; persistent interaction and environment state are largely outside the evaluation target.

\subsection{Individual Intelligence}

Individual Intelligence shifts the evaluation unit from outputs to trajectories. AgentBench and GAIA evaluate general agent capabilities, while WebArena, OSWorld, Terminal-Bench, SWE-bench, AppWorld, LongCLI-Bench, and TheAgentCompany test sustained interaction with web, computer, terminal, software, API, and workplace environments~\cite{liu2024agentbench,mialon2024gaia,zhou2024webarena,xie2024osworld,merrill2026terminalbench,jimenez2024swebench,deng2025swebenchpro,trivedi2024appworld,feng2026longcli,xu2025agentcompany}. More recent benchmarks push this setting toward longer and less idealized execution: AgencyBench evaluates extended real-world tasks~\cite{li2026agencybench}, while AgentGym2 evaluates end-to-end execution, tool discovery, tool composition, and robustness to noisy or underspecified information~\cite{xi2026agentgym2}. Tool and Harness resources further evaluate whether an agent can reliably access, govern, and improve external capabilities~\cite{yao2024taubench,barres2025tau2,lu2024toolsandbox,debenedetti2024agentdojo,yao2026harnessbench,han2026skilluse,ursekar2026harnessopt}.

Long-horizon evaluation additionally examines whether information and experience remain useful across extended or repeated executions. Existing resources cover long-term memory, memory governance, memory-to-action transfer, and reliable use of retrieved memories~\cite{wu2025longmemeval,hu2026memoryagentbench,he2026memoryarena,ren2026gatemem,bei2026mem,shen2026mem2act,xiang2026memsyco}. Recent benchmarks increasingly move beyond isolated episodes toward explicit adaptation and evolution: $A^2E$ provides an end-to-end evaluation engine for agent harnesses, capturing standardized execution traces and assessing harness capabilities in execution efficiency, tool use, task planning, and error recovery~\cite{arxiv:Wang_2026}; HarnessOpt-Bench directly evaluates evaluation-guided improvement of agent harnesses under fixed search and evaluation budgets~\cite{ursekar2026harnessopt}; Trainee-Bench evaluates scheduling, exploration, and continual learning in dynamic workplace streams~\cite{fu2026traineebench}; SEA-Eval measures evolutionary gain and stability across sequential tasks~\cite{jiang2026seaeval}; and Evo-Bench evaluates whether models can improve their own agent harnesses~\cite{huang2026evobench}. Other resources examine persistent state, evolving memory, process anomalies, and reflection across executions~\cite{xu2026longds,wang2026evomembench,liu2026openclawbench,huang2026benchtrace}. These benchmarks evaluate increasingly persistent and adaptive agents, but responsibility for task organization and execution remains centered on one agent or one local runtime.

\subsection{System Intelligence}

System Intelligence expands evaluation from an individual trajectory to the organization of multiple components and their relations. Existing resources provide partial probes of this broader objective. Work-oriented benchmarks evaluate decomposition, workflow structure, dependency-aware scheduling, and hierarchical orchestration~\cite{shen2024taskbench,qiao2025worfbench,xiao2024flowbench,xue2025comfybench,xu2026tpsbench,balaji2026journeybench,dong2026etom}. Coordination benchmarks evaluate collaboration, communication topology, distributed information, resource contention, scalability, privacy, and adversarial robustness~\cite{agashe2023coordination,dong2024villageragent,zhu2025multiagentbench,zhang2026silobench,mahmud2025collab,sun2025collabovercooked,yang2026masbench,hasan2026dpbench,zou2026calbench,kavathekar2026tamas}. State-oriented resources further expose failure attribution, consistency, diagnosis, and recovery as explicit system-level evaluation targets~\cite{guo2025syncmind,cemri2025mast,zhang2025whowhen,liu2026pro,chen2026traceelephant,in2026mpbench,qi2026r2act}.

Evaluation of system adaptation and evolution is also beginning to emerge. AgentsNet examines self-organization and scaling of networked agents~\cite{groetschla2025agentsnet}, while DBS studies adaptive workflow synthesis under distributed heterogeneity and privacy~\cite{wan2026dawn}. MASEval treats topology, orchestration, framework, and runtime design as system-level evaluation variables~\cite{emde2026maseval}, and MAS-PromptBench evaluates optimization across different multi-agent configurations~\cite{bai2026maspromptbench}. BenchAgent further contrasts single-agent, fixed multi-agent, and evolving workflows under controlled protocols~\cite{fu2026benchagent}. Despite this progress, persistent system evolution remains comparatively underexplored: current resources rarely evaluate whether runtime evidence produces durable and transferable improvements to work organization, team structure, and runtime management across repeated executions.

\subsection{Evaluation Principles and Open Challenges}

Across all three levels, evaluation should report \emph{effectiveness}, \emph{efficiency}, and \emph{robustness}. Graph-engineered systems additionally require \emph{structural fidelity}, \emph{operational correctness}, and \emph{evolution and governance}. These dimensions distinguish whether a system succeeds from whether its underlying structure is valid, its graph operations are executed correctly, and its structural changes remain traceable and controllable.

Three gaps are especially important. First, system-level improvements must be separated from gains caused by stronger models, larger contexts, additional tools, retries, or compute. Second, current resources remain fragmented across work organization, coordination, runtime state, and evolution, making cross-structure effects difficult to measure. Third, structural credit assignment and dynamic system-level evaluation remain weak. Future benchmarks should therefore provide matched execution budgets, versioned graph artifacts, complete traces and state snapshots, controlled structural perturbations, and repeated evaluations across tasks and time.

\section{Open-Source Libraries and Engineering Ecosystem}
\label{sec:open-source-libraries}

Open-source libraries translate the evolution from Model Intelligence to Individual Intelligence and System Intelligence into executable engineering stacks. Modern libraries often span several levels: a model-serving engine may also serve as the rollout backend of reinforcement learning, while an agent framework may support both a single tool-using agent and a multi-agent workflow. We therefore organize libraries by their \emph{primary engineering target} rather than by exclusive functionality. Model Intelligence libraries primarily construct, post-train, or execute model capabilities; Individual Intelligence libraries provide the persistent capabilities and control required by an autonomous agent; and System Intelligence libraries organize multiple intelligent components, their relations, shared execution structures, and runtime state.

We include reusable projects whose source and technical documentation are publicly available and whose abstractions directly affect the construction or execution of intelligent systems. Table~\ref{tab:ge-open-source-libraries} summarizes representative systems. The \emph{Focus} column records their main engineering concerns but is intentionally non-exclusive. Generic machine learning utilities, graph databases, workflow schedulers, and domain-specific agent applications are omitted unless they expose a reusable abstraction that is directly relevant to the intelligence stack. Source-available systems are retained when they have substantial engineering relevance, but their licensing status is stated explicitly.

\providecommand{\LibCode}[1]{\href{#1}{\textcolor{black}{\faGithub}}}

\begin{table*}[!t]
\centering
\caption{Representative open-source projects and engineering systems across Model, Individual, and System Intelligence. Libraries are grouped by their primary engineering target rather than exclusive functionality. Focus summarizes the main engineering concerns exposed by each system.}
\label{tab:ge-open-source-libraries}

\fontsize{5.25}{5.90}\selectfont
\setlength{\tabcolsep}{1.25pt}
\renewcommand{\arraystretch}{0.98}

\begin{tabular*}{\textwidth}{
@{\extracolsep{\fill}}
>{\raggedright\arraybackslash}p{2.20cm}
>{\raggedright\arraybackslash}p{1.50cm}
>{\raggedright\arraybackslash}p{3.20cm}
>{\raggedright\arraybackslash}p{5.25cm}
>{\raggedright\arraybackslash}p{2.05cm}
>{\centering\arraybackslash}p{0.62cm}
@{}
}
\toprule
\textbf{Project/System} &
\textbf{Focus} &
\textbf{Abstraction} &
\textbf{Engineering Paradigm} &
\textbf{License} &
\textbf{Link} \\
\midrule

\multicolumn{6}{c}{\textit{Model Intelligence}} \\
\midrule

Transformers~\cite{wolf2020transformers}
& Model interface
& Unified model definitions, configurations, tokenizers, and generation APIs
& Model loading, training, generation, multimodal models, and integration with downstream training and inference stacks
& Python; Apache-2.0
& \LibCode{https://github.com/huggingface/transformers} \\

Megatron Core~\cite{shoeybi2019megatron,nvidia2026megatron}
& Pretraining
& Distributed transformer training building blocks
& Tensor, pipeline, data, expert, and context parallelism; mixed precision and scalable distributed training
& Python; Apache-2.0
& \LibCode{https://github.com/NVIDIA/Megatron-LM} \\

LLaMA-Factory~\cite{zheng2024llamafactory}
& Post-training
& Unified fine-tuning and post-training recipes
& Continued pretraining, SFT, preference optimization, reward modeling, PPO, LoRA, and quantized fine-tuning
& Python; Apache-2.0
& \LibCode{https://github.com/hiyouga/LlamaFactory} \\

verl~\cite{sheng2025hybridflow}
& RL post-train
& Distributed RL post-training dataflow
& PPO, GRPO and related algorithms; integration with FSDP/Megatron for training and vLLM/SGLang for rollout generation
& Python; Apache-2.0
& \LibCode{https://github.com/verl-project/verl} \\

slime~\cite{slime2025}
& RL scaling
& Training--rollout--data-buffer loop
& Megatron training, SGLang rollout, custom rewards, verifiers, tool interaction, sandboxes, and asynchronous agentic data generation
& Python; Apache-2.0
& \LibCode{https://github.com/THUDM/slime} \\

vLLM~\cite{kwon2023vllm}
& Serving
& PagedAttention-based inference engine
& High-throughput batched inference, continuous serving, efficient KV-cache management, and model-serving APIs
& Python/CUDA; Apache-2.0
& \LibCode{https://github.com/vllm-project/vllm} \\

SGLang~\cite{zheng2023sglang}
& Serving/rollout
& Structured generation frontend and high-performance runtime
& Prefix-cache-aware execution, structured outputs, parallel inference, distributed serving, and rollout integration
& Python/CUDA; Apache-2.0
& \LibCode{https://github.com/sgl-project/sglang} \\

\midrule
\multicolumn{6}{c}{\textit{Individual Intelligence}} \\
\midrule

LangChain~\cite{langchain2026}
& Harness/Loop
& Agent loop over models, tools, middleware, and state
& Dynamic tools, middleware, tool retries, context control, structured output, state persistence, and human intervention
& Python; MIT
& \LibCode{https://github.com/langchain-ai/langchain} \\

OpenAI Agents SDK~\cite{openaiagentssdk2026}
& Harness/Loop/Team
& Agent runner with tools, guardrails, sessions, and delegation
& Tool execution, agents-as-tools, handoffs, guardrails, sessions, HITL, tracing, and multi-agent composition
& Python; MIT
& \LibCode{https://github.com/openai/openai-agents-python} \\

Claude Agent SDK~\cite{anthropicagentsdk2026}
& Harness/Exec.
& Programmable Claude Code agent runtime
& Filesystem and shell tools, permission control, MCP tools, hooks, sessions, custom tools, and programmatic subagents
& Python; MIT
& \LibCode{https://github.com/anthropics/claude-agent-sdk-python} \\

Pydantic AI~\cite{pydanticai2026}
& Harness/State
& Typed agents, capabilities, and \texttt{pydantic-graph}
& Typed tools and outputs, validation, MCP, HITL approval, graph/state-machine control, and durable execution integrations
& Python; MIT
& \LibCode{https://github.com/pydantic/pydantic-ai} \\

LlamaIndex Workflows~\cite{llamaindexworkflows2026}
& Context/Workflow
& Event-driven asynchronous workflow of typed steps and events
& Retrieval-oriented agents, event routing, branching, loops, parallel steps, persistence, recovery, and HITL
& Python; MIT
& \LibCode{https://github.com/run-llama/workflows-py} \\

Haystack~\cite{haystack2026}
& Context/Workflow
& Modular pipelines and agent workflows
& Retrieval, routing, memory, tools, conditional branches, loops, component composition, tracing, and deployment
& Python; Apache-2.0
& \LibCode{https://github.com/deepset-ai/haystack} \\

Apache Burr~\cite{burr2026}
& Loop/State
& Action graph interpreted as a persistent state machine
& Explicit transitions and state updates, persistence, resumability, streaming, HITL, telemetry, and trace inspection
& Python; Apache-2.0
& \LibCode{https://github.com/apache/burr} \\

Letta Agent SDK~\cite{lettaagentsdk2026}
& Memory/State
& Stateful agent backed by a persistent agent harness
& Persistent memory, sessions, skills, subagents, local or remote execution, and long-lived personalized agent state
& TypeScript; Apache-2.0
& \LibCode{https://github.com/letta-ai/letta-agent-sdk} \\

Graphiti~\cite{rasmussen2025zep}
& Memory/Graph
& Temporal context graph of entities, episodes, facts, and provenance
& Incremental graph updates, temporal validity, changing facts, source provenance, ontology support, and historical retrieval
& Python; Apache-2.0
& \LibCode{https://github.com/getzep/graphiti} \\

MCP Python SDK~\cite{mcp2026}
& Capability I/O
& Standard client/server interface for resources, tools, and prompts
& Capability discovery, tool execution, context resources, prompts, lifecycle management, authentication, and interoperable transports
& Python; MIT
& \LibCode{https://github.com/modelcontextprotocol/python-sdk} \\

Langflow~\cite{langflow2026}
& Visual workflow
& Visual node-edge canvas for agents, models, tools, and data
& Visual composition, reusable components, agent/tool integration, MCP exposure, execution inspection, and deployable flows
& Python/TS; MIT
& \LibCode{https://github.com/langflow-ai/langflow} \\

Dify$^{b}$~\cite{dify2026}
& Visual workflow
& Visual Workflow/Chatflow graph and application runtime
& RAG, agents, tools, branching, loops, variables, triggers, HITL, node-level traces, and workflow versions
& Python/TS; source-available
& \LibCode{https://github.com/langgenius/dify} \\

\midrule
\multicolumn{6}{c}{\textit{System Intelligence}} \\
\midrule

LangGraph~\cite{langgraph2026}
& Work/Team/State
& Typed \texttt{StateGraph} of nodes, edges, reducers, and subgraphs
& Conditional and cyclic routing, parallel fan-out, multi-agent composition, durable execution, checkpoints, interrupts, replay, and state inspection
& Python/TS; MIT
& \LibCode{https://github.com/langchain-ai/langgraph} \\

Microsoft Agent Framework~\cite{microsoftagentframework2026}
& Work/Team/State
& Graph-based workflows of agents and deterministic executors
& Sequential, concurrent, handoff, and group collaboration; checkpoints, time travel, HITL, middleware, streaming, and tracing
& Python/.NET; MIT
& \LibCode{https://github.com/microsoft/agent-framework} \\

Google ADK~\cite{googleadk2026}
& Work/Team/State
& Workflow graphs combining agents and executable nodes
& Sequential, parallel, loop, graph, dynamic, and collaborative workflows; routing, session state, evaluation, and deployment
& Python; Apache-2.0
& \LibCode{https://github.com/google/adk-python} \\

AutoGen/GraphFlow$^{a}$~\cite{wu2024autogen,autogen2026}
& Work/Team
& Event-driven agents and explicit multi-agent interaction patterns
& Message passing, group chat, distributed runtime, tool execution, GraphFlow-style directed interaction, logging, and inspection
& Python; MIT
& \LibCode{https://github.com/microsoft/autogen} \\

AG2~\cite{ag22026}
& Team/Harness
& Protocol-driven agents and multi-agent orchestration
& Tools, HITL, agent cooperation, multi-agent conversation patterns, knowledge, compaction, and extensible agent protocols
& Python; Apache-2.0/MIT
& \LibCode{https://github.com/ag2ai/ag2} \\

CrewAI~\cite{crewai2026}
& Work/Team/State
& Role/task-based Crews plus event-driven Flows
& Role specialization, task ownership, sequential and hierarchical processes, event routing, shared state, persistence, callbacks, and tracing
& Python; MIT
& \LibCode{https://github.com/crewAIInc/crewAI} \\

CAMEL~\cite{li2023camel}
& Work/Team/Evol.
& Workforce hierarchy and task-dependency structure
& Task decomposition, capability-based assignment, parallel workers, dependencies, role interaction, failure handling, shared memory, and workforce state
& Python; Apache-2.0
& \LibCode{https://github.com/camel-ai/camel} \\

Mastra$^{c}$~\cite{mastra2026}
& Work/Team/State
& Agents plus graph-based workflow engine
& Sequential, branch, and parallel flows; agent composition, memory, HITL, suspend/resume, storage-backed state, MCP, evaluation, and observability
& TypeScript; Apache core
& \LibCode{https://github.com/mastra-ai/mastra} \\

GPTSwarm~\cite{zhuge2024gptswarm}
& Team/Evolution
& Optimizable computational graph of LLM operations and agents
& Agent-graph construction, composite swarm graphs, node and prompt optimization, inter-agent edge creation or pruning, cost tracking, and graph optimization
& Python; MIT
& \LibCode{https://github.com/metauto-ai/GPTSwarm} \\

\bottomrule
\end{tabular*}

\vspace{1pt}
\begin{minipage}{0.985\textwidth}
\fontsize{5.10}{5.75}\selectfont
$^{a}$AutoGen is in maintenance mode and is retained because of its historical influence on multi-agent programming; Microsoft recommends Agent Framework for new projects.
$^{b}$Dify uses a modified Apache-2.0 license with additional deployment and branding restrictions and is included as a source-available ecosystem reference.
$^{c}$Mastra's core is Apache-2.0, while code under its enterprise directories is governed by a separate enterprise license.
\end{minipage}
\end{table*}

\subsection{Model Intelligence}

At the Model Intelligence level, open-source infrastructure determines how model capabilities are constructed, refined, and exposed to higher layers. Transformers provides a common model-definition and execution interface, while Megatron Core addresses large-scale distributed pretraining~\cite{wolf2020transformers,shoeybi2019megatron,nvidia2026megatron}. LLaMA-Factory packages supervised and preference-oriented post-training into a unified toolkit, whereas verl and slime focus on scalable reinforcement learning pipelines that connect training, rollout generation, reward computation, and increasingly agentic environment interaction~\cite{zheng2024llamafactory,sheng2025hybridflow,slime2025}. vLLM and SGLang provide the inference and rollout substrate on which both interactive agents and modern post-training systems depend~\cite{kwon2023vllm,zheng2023sglang}. These libraries primarily engineer model parameters and model execution rather than persistent agent behavior or system organization.

\subsection{Individual Intelligence}

At the Individual Intelligence level, the main engineering object shifts from model parameters to the runtime surrounding a model. LangChain, OpenAI Agents SDK, Claude Agent SDK, and Pydantic AI expose variants of the model--tool loop together with middleware, permissions, validation, sessions, state, and human control~\cite{langchain2026,openaiagentssdk2026,anthropicagentsdk2026,pydanticai2026}. LlamaIndex Workflows, Haystack, and Burr provide more explicit control over context construction, workflow transitions, event routing, and persistent execution~\cite{llamaindexworkflows2026,haystack2026,burr2026}. Their abstractions closely match the progression from Context Engineering to Harness Engineering and Loop Engineering: external capabilities are made accessible to the model and then organized into persistent, observable execution processes.

Persistent information is increasingly treated as another runtime capability. Letta maintains long-lived agent state and memory, while Graphiti represents changing contextual knowledge as a temporal graph with provenance~\cite{lettaagentsdk2026,rasmussen2025zep}. MCP addresses a complementary problem by standardizing how tools, resources, and prompts are exposed across agent runtimes~\cite{mcp2026}. Langflow and Dify lower the implementation barrier through visual workflow composition~\cite{langflow2026,dify2026}. Several of these systems can also compose multiple agents, but their primary abstractions remain centered on building and operating an agent or agent application rather than explicitly engineering system-level organization.

\subsection{System Intelligence}

System Intelligence libraries make relationships among tasks, agents, executors, and shared state explicit engineering objects. LangGraph, Microsoft Agent Framework, and Google ADK provide graph-oriented execution models in which agents and deterministic operations can be composed through conditional, concurrent, cyclic, or collaborative structures~\cite{langgraph2026,microsoftagentframework2026,googleadk2026}. AutoGen established an influential multi-agent programming model based on message-passing agents and flexible conversation patterns, although it is now maintained primarily for existing users~\cite{wu2024autogen,autogen2026}. AG2 represents a community continuation of this lineage with its own agent protocol and multi-agent abstractions~\cite{ag22026}.

Other systems place stronger emphasis on organizational semantics. CrewAI separates role-oriented Crews from event-driven Flows, while CAMEL's Workforce couples task decomposition with worker assignment and hierarchical coordination~\cite{crewai2026,li2023camel}. Mastra combines agents with an explicit workflow engine and persistent execution state~\cite{mastra2026}. GPTSwarm is particularly relevant to Graph Engineering because it treats graph connectivity itself as an optimization variable: both node-level prompts and inter-agent edges can be modified to improve the resulting system~\cite{zhuge2024gptswarm}. Nevertheless, most production-oriented frameworks still operate within developer-defined organizational templates. Dynamic routing is common, but persistent creation, removal, or rewiring of system structure from accumulated execution evidence remains rare.

\subsection{Open Challenges in the Engineering Ecosystem}

The ecosystem shows a clear progression from model infrastructure to persistent agent runtimes and multi-component orchestration, but the boundaries between these layers remain fragmented. Model training and serving systems expose different execution semantics from agent runtimes; agent frameworks use incompatible representations of tools, messages, workflows, events, and state; and multi-agent systems rarely share a common representation of task dependencies, capabilities, authority, communication, and runtime state. Protocols such as MCP improve capability interoperability, but they do not provide a common representation for executable system organization.

A second limitation is that current dynamism is primarily \emph{within} predefined structures. Conditional edges, routing, parallel fan-out, worker assignment, and recovery can change an execution path without changing the persistent organization that governs future executions. GPTSwarm and a small number of research-oriented systems expose topology optimization, but systematic cross-run evolution remains uncommon. This creates a gap between current orchestration frameworks and the RSI view of Graph Engineering, where runtime evidence should be abstracted into reusable structural changes.

Finally, state remains divided among model checkpoints, agent memories, workflow snapshots, message histories, event logs, and temporal knowledge stores. Existing observability tools can reconstruct what executed, but they seldom capture typed causal relations between observations, decisions, structural mutations, failures, recovery actions, and later system improvements. A more complete Graph Engineering substrate should therefore support typed and versioned work, team, and runtime structures; safe structural transactions and validators; persistent provenance; replay and rollback; graph-level tracing and counterfactual comparison; and controlled mechanisms for retaining successful structural changes across executions.

\section{Applications of Graph Engineering}
\label{sec:applications}

Applications provide a complementary view of Graph Engineering. Unlike benchmarks and open-source libraries, which can be organized naturally by intelligence level, applications are better distinguished by the domains in which structural decisions affect real work. We therefore organize this section by application domain while using intelligence level and Graph Engineering focus as cross-domain descriptors.

We include both research prototypes and deployed agent systems when task organization, agent relations, or runtime state have operational consequences. The term \emph{Graph Engineering} need not be used explicitly by the original system. A workflow, team, dependency structure, or persistent environment is relevant when changing that structure changes how the system executes. Systems that use a knowledge graph only as an external retrieval source are not included unless the graph also affects task organization, agent coordination, or runtime behavior.

Table~\ref{tab:ge-applications} uses \textsc{I} and \textsc{S} to denote Individual and System Intelligence. Some systems span both levels as they evolve from a single persistent agent toward parallel or multi-agent execution. The \emph{Focus} column maps each application to the current Graph Engineering directions of Work Organization, Agent Team, Runtime State, and System Evolution. These assignments are non-exclusive and reflect our interpretation of the operational structure exposed by each system. Pure Model Intelligence applications are omitted because model capability alone does not constitute Graph Engineering without persistent agent execution or system structure.

\providecommand{\AppPaper}[1]{\href{#1}{\textcolor{cyan!60!black}{\faFilePdf}}}
\providecommand{\AppCode}[1]{\href{#1}{\textcolor{black}{\faGithub}}}
\providecommand{\AppWeb}[1]{\href{#1}{\textcolor{blue!65!black}{\faGlobe}}}

\begin{table*}[!t]
\centering
\caption{Representative applications of System Intelligence. Level denotes the primary intelligence level: \textsc{I} = Individual Intelligence, \textsc{S} = System Intelligence, and \textsc{I/S} = systems spanning both. Focus summarizes the principal Graph Engineering concerns: Work Organization, Agent Team, Runtime State, and System Evolution.}
\label{tab:ge-applications}

\fontsize{5.05}{5.70}\selectfont
\setlength{\tabcolsep}{1.15pt}
\renewcommand{\arraystretch}{0.97}

\begin{tabular*}{\textwidth}{
@{\extracolsep{\fill}}
>{\raggedright\arraybackslash}p{1.85cm}
>{\raggedright\arraybackslash}p{2.15cm}
>{\centering\arraybackslash}p{1.05cm}
>{\raggedright\arraybackslash}p{1.45cm}
>{\raggedright\arraybackslash}p{4.55cm}
>{\raggedright\arraybackslash}p{3.60cm}
>{\centering\arraybackslash}p{0.52cm}
@{}
}
\toprule
\textbf{System} &
\textbf{Application Domain} &
\textbf{Level} &
\textbf{Focus} &
\textbf{Structural role} &
\textbf{Evidence or artifact} &
\textbf{Link} \\
\midrule

\multicolumn{7}{c}{\textit{Software Engineering and IT Operations}} \\
\midrule

MetaGPT~\cite{hong2024metagpt}
& End-to-end software production
& S
& Work/Team
& SOP-derived stages assign requirements, architecture, implementation, and review to specialized roles with structured intermediate artifacts
& Collaborative software generation with executable projects and role-specific artifacts
& \AppPaper{https://proceedings.iclr.cc/paper_files/paper/2024/hash/6507b115562bb0a305f1958ccc87355a-Abstract-Conference.html} \\

SWE-agent~\cite{yang2024sweagent}
& Repository issue resolution
& I
& Work/State
& Agent-computer interface constrains repository navigation, editing, commands, and test feedback within an iterative execution trajectory
& Patch resolution on real repository issues with observable action and test trajectories
& \AppPaper{https://proceedings.neurips.cc/paper_files/paper/2024/hash/5a7c947568c1b1328ccc5230172e1e7c-Abstract-Conference.html} \\

OpenHands~\cite{wang2025openhands}
& General software development
& I/S
& Team/State
& Event-stream runtime connects code, shell, browser, observations, and delegation while preserving execution history
& Reproducible platform and evaluation across software-engineering tasks
& \AppPaper{https://openreview.net/forum?id=OJd3ayDDoF} \\

Codex~\cite{openai2026codexapp}
& Production software engineering
& I/S
& Work/Team/State
& Multiple coding agents execute parallel tasks in isolated worktrees with project threads, skills, review, and background automation
& Deployed coding system supporting parallel long-running engineering work
& \AppWeb{https://openai.com/index/introducing-the-codex-app/} \\

Claude Code~\cite{anthropic2026claudecode}
& Repository-scale coding
& I/S
& Work/Team/State
& Agent loop combines repository tools, checkpoints, hooks, background tasks, subagents, and parallel agent teams
& Long-running coding workflows and demonstrated parallel agent-team software development
& \AppWeb{https://www.anthropic.com/product/claude-code} \\

OpenCode~\cite{opencode2026}
& Open coding agent
& I/S
& Work/Team
& Primary agents delegate specialized work to configurable subagents with separate permissions, tools, and child sessions
& Open implementation supporting planning, coding, review, research, and parallel delegated tasks
& \AppWeb{https://opencode.ai/docs/agents/} \\

Cline~\cite{cline2026}
& Parallel software development
& S
& Work/Team/State
& Dependency-linked tasks execute in isolated worktrees; persistent teams use a shared task board, mailbox, and mission log
& Parallel coding tasks, cross-session team state, automated dependency chains, and reviewable diffs
& \AppWeb{https://docs.cline.bot/sdk/guides/multi-agent-teams} \\

Project ALICE~\cite{ibm2025alice}
& Cloud incident localization
& S
& Work/Team/State
& Specialist agents collect telemetry, construct service and code dependency evidence, and localize operational faults
& Incident investigation artifacts and validation on ITBench scenarios
& \AppWeb{https://research.ibm.com/blog/project-alice-software-bugs-agents} \\

\midrule
\multicolumn{7}{c}{\textit{Scientific Discovery and Laboratory Automation}} \\
\midrule

SciAgents~\cite{ghafarollahi2025sciagents}
& Materials discovery
& S
& Work/Team
& Ontological knowledge structures ground specialized agents that generate, criticize, and refine scientific hypotheses
& Generated hypotheses, mechanisms, design principles, and materials proposals
& \AppPaper{https://doi.org/10.1002/adma.202413523} \\

The AI Scientist~\cite{lu2024aiscientist}
& Automated ML research
& I
& Work/State
& A long-running research workflow links ideation, implementation, experiments, visualization, writing, and simulated review
& End-to-end generated experiments and manuscripts across multiple ML subfields
& \AppCode{https://github.com/SakanaAI/AI-Scientist} \\

Virtual Lab~\cite{swanson2025virtuallab}
& Nanobody design
& S
& Work/Team/State
& A principal-investigator agent organizes specialist scientist agents and external computational tools through research meetings
& Experimentally validated SARS-CoV-2 nanobody designs with human oversight
& \AppPaper{https://doi.org/10.1038/s41586-025-09442-9} \\

Co-Scientist~\cite{gottweis2026coscientist}
& Scientific hypothesis generation
& S
& Work/Team/State
& Supervisor-managed specialized agents asynchronously generate, critique, rank, and refine hypotheses under a structured research objective
& Experimentally validated biomedical hypotheses and test-time scaling of hypothesis quality
& \AppPaper{https://doi.org/10.1038/s41586-026-10644-y} \\

Robin~\cite{ghareeb2026robin}
& Experimental biological discovery
& S
& Work/Team/State
& Literature and data-analysis agents connect hypothesis generation, experiment proposals, laboratory results, analysis, and revised hypotheses
& Lab-in-the-loop discovery and experimental validation of therapeutic candidates
& \AppPaper{https://doi.org/10.1038/s41586-026-10652-y} \\

\midrule
\multicolumn{7}{c}{\textit{Healthcare and Clinical Decision Support}} \\
\midrule

DeepRare~\cite{zhao2026deeprare}
& Rare-disease diagnosis
& S
& Work/Team/State
& A central host coordinates specialized phenotype, genotype, retrieval, and analysis agents while maintaining accumulated diagnostic evidence
& Evaluation across heterogeneous clinical datasets with traceable evidence-supported reasoning
& \AppPaper{https://doi.org/10.1038/s41586-025-10097-9} \\

AMIE~\cite{lievin2026amie}
& Longitudinal disease management
& S
& Work/Team/State
& Dialogue and management-reasoning agents share patient history across visits and ground evolving care plans in clinical guidelines
& Multi-visit virtual OSCE evaluation against primary-care physicians
& \AppPaper{https://doi.org/10.1038/s41586-026-10764-5} \\

CARE-AD~\cite{li2025caread}
& Longitudinal Alzheimer risk
& S
& Work/Team/State
& Specialized assessments aggregate multimodal evidence across clinical time points into coordinated longitudinal predictions
& Retrospective prediction across multiple horizons from longitudinal EHR notes
& \AppPaper{https://doi.org/10.1038/s41746-025-01940-4} \\

MAP~\cite{chen2026map}
& Inpatient clinical pathways
& S
& Work/Team/State
& Triage, diagnosis, and treatment agents encode staged responsibilities along a clinical pathway
& Evaluation of multi-agent enhancement across inpatient pathway decisions
& \AppPaper{https://doi.org/10.1038/s44401-026-00085-0} \\

\midrule
\multicolumn{7}{c}{\textit{Enterprise Workflows and Digital Organizations}} \\
\midrule

WorkTeam~\cite{liu2025workteam}
& Natural-language-to-workflow
& S
& Work/Team
& Supervisor, orchestrator, and filler agents jointly transform natural-language requirements into executable workflows
& Workflow generation over 3,695 real-world enterprise samples
& \AppPaper{https://aclanthology.org/2025.naacl-industry.3/} \\

SOAN~\cite{xiong2025soan}
& Nested workflow automation
& S
& Work/Team
& Reusable structural units are incrementally encapsulated as agents in a formalized hierarchical network
& Improved adaptability, fault tolerance, and execution efficiency on complex workflows
& \AppPaper{https://arxiv.org/abs/2508.13732} \\

FinRobot-ERP~\cite{yang2025finrobot}
& Financial ERP processes
& S
& Work/Team/State
& Business-process structures coordinate specialist agents and insert operational controls around consequential transactions
& Case studies in wire transfers and employee reimbursement
& \AppPaper{https://arxiv.org/abs/2506.01423} \\

Agent-Ops~\cite{singh2026agentops}
& E-commerce SOP automation
& S
& Work/Team/State
& SOP grooming, web execution, and document verification are assigned to cooperating components in an auditable operational chain
& Production deployment across seven SOP categories with more than 1,000 account managers
& \AppPaper{https://aclanthology.org/2026.acl-industry.29/} \\

Gemini Enterprise Agentic RAG~\cite{google2026agenticrag}
& Enterprise knowledge workflows
& S
& Work/Team/State
& Root, planning, query-rewriting, retrieval, sufficient-context, and synthesis agents iteratively coordinate multi-source information gathering
& Cross-corpus enterprise retrieval with iterative sufficiency checking and public-preview deployment
& \AppWeb{https://research.google/blog/unlocking-dependable-responses-with-gemini-enterprise-agent-platforms-agentic-rag/} \\

\midrule
\multicolumn{7}{c}{\textit{General-Purpose Digital Agents and Personal Automation}} \\
\midrule

OpenClaw~\cite{openclaw2026}
& Persistent digital assistance
& I/S
& Team/State
& A gateway manages isolated agent identities, workspaces, authentication, session stores, skills, and channel-to-agent routing
& Persistent agents operating across communication channels with independent state and workspace boundaries
& \AppCode{https://github.com/openclaw/openclaw} \\

Hermes Agent~\cite{nous2026hermes}
& General task execution
& I/S
& Work/State
& A persistent agent combines tools, delegation, memory, reusable skills, scheduled execution, and multi-platform access
& Cross-session memory and agent-generated skills that retain procedures learned during previous tasks
& \AppCode{https://github.com/NousResearch/hermes-agent} \\

\midrule
\multicolumn{7}{c}{\textit{Social and Economic Simulation}} \\
\midrule

AgentSociety~\cite{zhang2025agentsociety}
& Large-scale social simulation
& S
& Team/State
& Large agent populations interact through a realistic shared environment and parallelized social processes
& Simulations of up to 30,000 agents and intervention-based social experiments
& \AppPaper{https://aclanthology.org/2025.acl-industry.94/} \\

EconAgent~\cite{li2024econagent}
& Macroeconomic simulation
& S
& Team/State
& Heterogeneous households repeatedly interact with labor and consumption markets while memory incorporates prior personal and market experience
& Multi-period macroeconomic dynamics compared with rule-based and learned agents
& \AppPaper{https://aclanthology.org/2024.acl-long.829/} \\

SRAP-Agent~\cite{ji2024srap}
& Public-housing allocation
& S
& Work/Team/State
& Applicant agents, allocation rules, scarce resources, and outcomes form an explicit policy simulation and optimization process
& Policy simulation and optimization for efficiency and equity
& \AppPaper{https://aclanthology.org/2024.findings-emnlp.15/} \\

TwinMarket~\cite{yang2025twinmarket}
& Financial-market simulation
& S
& Team/State
& Social and trading interactions connect heterogeneous agent decisions to a shared market environment and collective feedback
& Emergent group behavior, bubbles, and recessions in simulated financial markets
& \AppPaper{https://proceedings.neurips.cc/paper_files/paper/2025/hash/5bf234ecf83cd77bc5b77a24ba9338b0-Abstract-Conference.html} \\

\bottomrule
\end{tabular*}
\end{table*}

\subsection{Software Engineering and IT Operations}

Software engineering is one of the clearest domains in which the progression from Individual to System Intelligence is already visible. Early multi-agent systems such as MetaGPT and ChatDev structured software development around predefined stages and specialist roles~\cite{hong2024metagpt,qian2024chatdev}, while SWE-agent showed that the interface between an agent and a repository can itself strongly shape execution~\cite{yang2024sweagent}. OpenHands broadened this interaction model through a persistent event stream connecting code, shell, browser, and delegation~\cite{wang2025openhands}.

Recent coding systems increasingly make parallel agent work an explicit engineering object. Codex supports concurrent agents operating in isolated worktrees, while Claude Code combines subagents, checkpoints, hooks, background execution, and agent teams~\cite{openai2026codexapp,anthropic2026claudecode}. OpenCode exposes configurable primary agents and subagents, whereas Cline represents tasks and dependencies on a shared board and persists team state across sessions~\cite{opencode2026,cline2026}. These systems shift software engineering from managing one agent trajectory toward managing concurrent work, isolated branches, dependencies, test feedback, and merge decisions. Project ALICE extends the same structural view to IT operations by coordinating specialist agents over telemetry and software dependency evidence~\cite{ibm2025alice}. The remaining challenge is to connect planning, code dependencies, ownership, external side effects, testing, and recovery in a versioned structure that can explain not only whether a patch succeeded but why a particular organization of work succeeded.

\subsection{Scientific Discovery and Laboratory Automation}

Scientific discovery is naturally structured by dependencies among hypotheses, evidence, tools, experiments, and researchers. SciAgents uses an ontological knowledge structure to ground coordinated scientific agents, while the AI Scientist organizes ideation, implementation, experimentation, writing, and review into a long-running research process~\cite{ghafarollahi2025sciagents,lu2024aiscientist}. The Virtual Lab makes team structure explicit through a principal-investigator agent and specialist scientist agents, and importantly connects their computational work to physical experimental validation~\cite{swanson2025virtuallab}.

More recent systems move toward closed scientific feedback loops. Co-Scientist assigns generation, critique, ranking, and refinement to specialized agents managed by an asynchronous supervisor~\cite{gottweis2026coscientist}. Robin combines literature-search and data-analysis agents with laboratory results so that experimental evidence can directly update subsequent hypotheses~\cite{ghareeb2026robin}. These systems demonstrate increasingly sophisticated Work, Team, and Runtime State structures, but iterative hypothesis refinement should not be confused with persistent evolution of the agent organization itself. For Graph Engineering, the stronger requirement is to preserve hypotheses, negative results, data lineage, experimental interventions, and causal dependencies while allowing evidence to influence future system structure in a reproducible manner.

\subsection{Healthcare and Clinical Decision Support}

Healthcare exposes the need to jointly engineer specialization, longitudinal state, authority, and evidence provenance. Earlier multi-agent consultation systems such as MAC showed that several doctor agents and a supervisor can reproduce aspects of multidisciplinary diagnosis~\cite{chen2025macdiagnosis}. DeepRare provides a more explicit systems architecture in which a central host coordinates specialized phenotype, genotype, retrieval, and analysis agents while accumulating traceable diagnostic evidence~\cite{zhao2026deeprare}. CARE-AD and MAP similarly organize specialist reasoning across longitudinal evidence and staged clinical responsibilities~\cite{li2025caread,chen2026map}.

AMIE extends this problem from one diagnostic episode to disease management over multiple visits. Its dialogue agent maintains conversational state while a management-reasoning agent synthesizes longitudinal patient information and clinical guidelines into evolving care plans~\cite{lievin2026amie}. This illustrates why Runtime State in healthcare is more than conversation memory: previous symptoms, treatments, responses, investigations, and recommendations change the validity of later actions. Clinical Graph Engineering must therefore preserve provenance, uncertainty, access boundaries, and human authorization together with the task and agent structures. Graph organization can improve coordination and traceability, but it does not establish clinical correctness by itself.

\subsection{Enterprise Workflows and Digital Organizations}

Enterprise applications make structural constraints concrete because actions are governed by business processes, organizational roles, permissions, and consequential updates to external systems. WorkTeam assigns workflow construction to supervisor, orchestrator, and filler agents, while SOAN builds hierarchical networks by encapsulating reusable workflow structures as agents~\cite{liu2025workteam,xiong2025soan}. FinRobot-ERP connects specialist agents to business-process models for financial operations, and Agent-Ops combines SOP refinement, web execution, and document verification in a production-oriented multi-agent pipeline~\cite{yang2025finrobot,singh2026agentops}.

Enterprise knowledge access is undergoing a similar transition. Gemini Enterprise Agentic RAG decomposes multi-source retrieval into orchestration, planning, query rewriting, search, context sufficiency checking, and synthesis, and uses feedback to continue retrieval when required information remains missing~\cite{google2026agenticrag}. Across these systems, task completion alone is insufficient. A structurally valid enterprise agent must also respect permissions, separation of duties, policy constraints, transaction boundaries, and rollback obligations. This distinction between a planned workflow and committed external state makes enterprise automation a particularly important setting for Runtime State Management and governed Graph Engineering.

\subsection{General-Purpose Digital Agents and Personal Automation}

A newer application class consists of persistent digital agents that are not confined to a single professional domain. OpenClaw uses a gateway to maintain separate agent identities, workspaces, authentication profiles, sessions, and channel bindings, allowing persistent agents to operate across communication surfaces while retaining explicit state boundaries~\cite{openclaw2026}. Hermes Agent similarly combines system tools, delegation, scheduled execution, persistent memory, and reusable skills across sessions and platforms~\cite{nous2026hermes}. In these systems, the agent is no longer instantiated only for one task; it becomes a persistent computational entity with accumulated state and continuing access to external capabilities.

This persistence also exposes an important boundary of current Graph Engineering. Hermes can convert successful procedures into reusable skills and revise them after later experience, while OpenClaw can maintain several isolated agents and route interactions between users, channels, and agent identities. These mechanisms provide cross-run adaptation and persistent organization, but they do not yet amount to general structural self-evolution. The broader challenge is to determine which experiences should alter future work structures, capability assignments, or agent relations, and how such changes can be validated, versioned, and reversed.

\subsection{Social and Economic Simulation}

Social and economic simulation moves the graph from an internal execution mechanism to part of the phenomenon being studied. AgentSociety supports large populations of interacting agents in realistic parallel environments~\cite{zhang2025agentsociety}; EconAgent models heterogeneous households whose repeated work and consumption decisions interact with macroeconomic state~\cite{li2024econagent}. SRAP-Agent connects applicant decisions, allocation rules, scarce resources, and policy outcomes~\cite{ji2024srap}, while TwinMarket couples individual social and trading behavior to shared market feedback and emergent financial dynamics~\cite{yang2025twinmarket}. Here, Agent Team structure determines who interacts with whom, while Runtime State records how local decisions alter the environment faced by later agents.

The same structure that enables simulation also creates an epistemic risk. Emergent behavior in an agent society depends on model choice, persona construction, interaction topology, memory, prompting, and environment rules. A graph-engineered simulator can make these assumptions explicit and support topology interventions, replay, and controlled ablations, but simulated emergence should not be interpreted as evidence of real-world causality without calibration against observations and explicit uncertainty analysis.

\subsection{Cross-Domain Findings}

Across application domains, the maturity of Graph Engineering is uneven. Work Organization and Agent Team Engineering are already common: applications routinely decompose objectives, assign specialized roles, schedule parallel work, and define communication or dependency structures. Explicit Runtime State Management is also becoming more visible through checkpoints, longitudinal patient records, shared task boards, event streams, experimental evidence, and evolving environments. Persistent System Evolution, however, remains rare. Most systems adapt execution within a predefined organizational structure rather than permanently revising that structure from accumulated evidence.

A second trend is the practical transition from Individual to System Intelligence. Software agents provide the clearest example: systems that initially centered on one coding trajectory now expose subagents, parallel worktrees, persistent task boards, agent teams, and supervisory interfaces. Similar changes are appearing in scientific discovery, enterprise workflows, and persistent digital assistants. Nevertheless, additional agents do not by themselves produce System Intelligence. The value of the system depends on how work is decomposed, how responsibilities are assigned, how state is shared, and how failures are diagnosed and recovered.

The application evidence therefore supports a narrower distinction between being \emph{graph-structured} and being \emph{graph-engineered}. Contemporary systems increasingly execute through explicit work, team, and state structures, but these structures are still usually selected manually or fixed before execution. Advancing toward full Graph Engineering requires structural objectives, graph-level observability, controlled mutation, cross-structure consistency, and evidence that successful structural changes persist and transfer across tasks and time.

\section{Conclusion}

Large language models have rapidly evolved from standalone generators into individual agents capable of sustained interaction, tool use, and iterative execution. Yet, as tasks become more heterogeneous, interdependent, and long-horizon, the limitations of individual intelligence become increasingly clear: a single agent loop struggles to support parallel work, specialized expertise, independent verification, and persistent state. This survey argues that the next frontier is \emph{System Intelligence}, the ability of an agent system to organize complex objectives, coordinate heterogeneous components, and maintain coherent runtime state across the task lifecycle.

To support this transition, we introduce \emph{Graph Engineering} as a structure-centered engineering paradigm that uses graph abstractions to make system relations explicit, operational, and adaptable. We organize the literature around three complementary graph views: work organization, agent coordination, and runtime state management. Together, these views show how graphs can be used not only to represent tasks, agents, and states, but also to schedule work, bind capabilities, trace execution, localize failures, and enable controlled evolution. Across the surveyed methods, a common lesson emerges: system-level intelligence depends less on simply adding more models or agents, and more on explicitly organizing the relations among work, actors, and state.

Despite rapid progress, Graph Engineering remains an emerging field with important open challenges, including semantic alignment, graph governance, evaluation, privacy, and safe self-improvement. We hope this survey provides a useful foundation for understanding how graph-based abstractions can support the design of more scalable, controllable, and evolvable agent systems, and for guiding future work toward graph-native infrastructure for system intelligence.

\bibliographystyle{abbrvnat}

\bibliography{reference}
\clearpage
\section{Appendix}

\subsection{Comparison with Related Surveys}
\label{sec:existing-surveys}

Recent surveys examine LLM-based agents from several complementary perspectives. Broad agent surveys organize the field around planning and reasoning, memory, tool use, interaction with environments, and multi-agent coordination \cite{lee2026llmagents}. Graph--agent surveys focus more specifically on how graph structures enhance planning, execution, memory, tool use, reasoning, and multi-agent interaction \cite{bei2025graphsmeetagents,liu2025graphaugmented,shao2026augmenting}. These studies provide comprehensive accounts of agent capabilities and graph-enhanced agent functions, but their primary objective is to characterize or improve the capabilities of individual agents and multi-agent architectures rather than to organize the full engineering progression from models to system-level intelligence.

A second line of work focuses on the engineering infrastructure that turns foundation models into operational agents. Recent surveys trace the transition from prompting and context construction toward workflow and harness engineering, emphasizing that agent performance depends on both model capability and the surrounding execution infrastructure \cite{guo2026qatask}. Harness-centered surveys further formalize execution loops, tool interfaces, context management, persistent state, lifecycle control, and evaluation as first-class runtime components \cite{meng2026agentharness}. Related studies examine externalized capabilities and code-centered harnesses \cite{zhou2026externalization,ning2026codeharness}, while workflow surveys study planning, orchestration structures, executable workflows, and their optimization \cite{yu2025agentworkflow,yue2026runtime}. These surveys substantially overlap with the transition from Model Intelligence to Individual Intelligence and with parts of task organization, but they generally take the individual agent runtime or executable workflow as their primary engineering object.

A third line of work moves toward system-level organization and evolution. Multi-agent orchestration surveys study task decomposition and allocation, coordination topology, communication protocols, state management, control-flow sequencing, failure recovery, and dynamic orchestration \cite{zhu2026orchestration}. The LIFE survey connects individual capability, multi-agent collaboration, failure attribution, and autonomous self-evolution \cite{qi2026beyondindividual}, while broader surveys of self-evolving agents organize how different agent components are persistently improved through experience and feedback \cite{gao2026selfevolving}. Most closely related to our structural perspective, recent work formulates self-evolving agents as dynamic graph transformations, representing memories, tools, skills, workflows, and inter-agent relations as typed graph objects whose structures can evolve over time \cite{xu2026dynamicgraph}. Its primary question is how agent evolution can be modeled and governed through dynamic graph transformation. Our survey instead begins from the organization of system intelligence and treats Task Organization, Agent Coordination, and Runtime State Management as explicit and interconnected system-level structures, with System Evolution describing how execution experience persistently improves these structures. Beyond structural organization and evolution, we further consider Ontology Engineering as a semantic foundation for defining shared system entities, relations, and constraints.

Table~\ref{tab:survey-coverage-comparison} compares representative surveys according to their substantive coverage of major topics in agent and system engineering. We focus on Harness, Loop, Planning, Workflow, multi-agent systems, Runtime State, Self-Evolution, and Ontology because these dimensions more directly distinguish how existing surveys treat the construction, execution, organization, and evolution of agent systems. Model-level topics such as foundation models, prompting, and context are not separately tabulated because they are widely used as general background across agent surveys and therefore provide limited discriminative value. Component-specific surveys devoted only to memory, tools, GraphRAG, evaluation, or individual application domains are not tabulated, although they remain important background references.

\begin{table*}[t]
\caption{Topic coverage of representative surveys related to LLM agents and agent systems.}
\label{tab:survey-coverage-comparison}

\centering
\scriptsize
\setlength{\tabcolsep}{2.6pt}
\renewcommand{\arraystretch}{1.06}

\begin{tabular*}{\textwidth}{
@{\extracolsep{\fill}}
>{\raggedright\arraybackslash}p{0.255\textwidth}
*{8}{c}
@{}
}
\toprule

\textbf{Survey / study}
&
\textbf{Harness}
&
\textbf{Loop}
&
\textbf{Planning}
&
\textbf{Workflow}
&
\textbf{MAS}
&
\textbf{State}
&
\textbf{Self-Evolution}
&
\textbf{Ontology}
\\

\midrule

LLM Agents'26 \cite{lee2026llmagents}
& \surveypart
& \surveynone
& \surveyfull
& \surveynone
& \surveyfull
& \surveynone
& \surveynone
& \surveynone \\

Graphs Meet Agents'25 \cite{bei2025graphsmeetagents}
& \surveypart
& \surveynone
& \surveyfull
& \surveypart
& \surveyfull
& \surveynone
& \surveynone
& \surveynone \\

Graph-Aug. Agents'25 \cite{liu2025graphaugmented}
& \surveypart
& \surveynone
& \surveyfull
& \surveypart
& \surveyfull
& \surveynone
& \surveypart
& \surveynone \\

Agent Intelligence + Graphs'26 \cite{shao2026augmenting}
& \surveypart
& \surveypart
& \surveypart
& \surveypart
& \surveyfull
& \surveynone
& \surveynone
& \surveynone \\

\addlinespace[0.7pt]

QA-to-Task Completion'26 \cite{guo2026qatask}
& \surveyfull
& \surveyfull
& \surveypart
& \surveyfull
& \surveypart
& \surveyfull
& \surveyfull
& \surveynone \\

Agent Harness'26 \cite{meng2026agentharness}
& \surveyfull
& \surveyfull
& \surveyfull
& \surveypart
& \surveyfull
& \surveyfull
& \surveypart
& \surveynone \\

Runtime Graphs'26 \cite{yue2026runtime}
& \surveypart
& \surveypart
& \surveypart
& \surveyfull
& \surveypart
& \surveypart
& \surveypart
& \surveynone \\

\addlinespace[0.7pt]

Multi-Agent Orchestration'26 \cite{zhu2026orchestration}
& \surveypart
& \surveypart
& \surveyfull
& \surveyfull
& \surveyfull
& \surveyfull
& \surveynone
& \surveynone \\

Beyond Individual'26 \cite{qi2026beyondindividual}
& \surveypart
& \surveypart
& \surveypart
& \surveypart
& \surveyfull
& \surveypart
& \surveyfull
& \surveynone \\

Dynamic Graph Transform.'26 \cite{xu2026dynamicgraph}
& \surveypart
& \surveynone
& \surveypart
& \surveyfull
& \surveyfull
& \surveyfull
& \surveyfull
& \surveypart \\

\midrule

\rowcolor{blue!10}
\textbf{Ours}
& \surveyfull
& \surveyfull
& \surveyfull
& \surveyfull
& \surveyfull
& \surveyfull
& \surveyfull
& \surveyfull \\

\bottomrule
\end{tabular*}

\vspace{1.5pt}

\parbox{\textwidth}{\footnotesize
\surveyfull: primary organizing axis or dedicated taxonomy;
\surveypart: substantive secondary coverage;
\surveynone: absent, incidental, or only briefly mentioned as background.
\textbf{Harness}: extra-model capabilities and infrastructure such as tools, memory, skills, interfaces, execution environments, or runtime governance;
\textbf{Loop}: explicit iterative control through action, observation, feedback, verification, recovery, or termination;
\textbf{Planning}: task decomposition, dependency structuring, scheduling, or allocation;
\textbf{Workflow}: construction, execution, or optimization of multi-step computational workflows;
\textbf{MAS}: multi-agent roles, allocation, team organization, communication, topology, or orchestration;
\textbf{State}: persistent runtime state, provenance, consistency, failure localization, or recovery;
\textbf{Self-Evolution}: experience-driven improvements that persist across executions;
\textbf{Ontology}: explicit ontology engineering or shared machine-interpretable semantics for system entities, relations, and constraints.
Symbols indicate substantive survey scope rather than paper quality; brief background mentions are not counted.
}

\end{table*}

\subsection{Distinction with Graph-based Approaches in Agents}

The coverage comparison above clarifies which parts of agent and system engineering are addressed by existing surveys, but topic coverage alone does not capture the main distinction of Graph Engineering. The key difference lies in the architectural role assigned to graph structures. Existing graph--agent approaches typically use graphs to enhance particular capabilities, such as reasoning, planning, memory, retrieval, tool organization, workflow execution, or multi-agent communication. In these settings, the graph is primarily a representation or computational mechanism supporting an agent capability.

Graph Engineering instead treats explicit graph structures as the organizational substrate of the intelligent system. Task Organization represents goals, subtasks, dependencies, and executable workflows; Agent Coordination represents capabilities, responsibilities, team structures, and communication relations; and Runtime State Management represents execution state, provenance, failures, and recovery dependencies. These structures are coupled: changes in task organization can alter capability requirements and agent allocation, changes in agent organization can affect communication and execution assumptions, and runtime evidence can trigger revisions to task and agent structures. System Evolution further turns execution experience into persistent structural improvements that can be validated, retained, reused, or rolled back across executions. Ontology Engineering complements these structures by providing shared, machine-interpretable definitions of system entities, relations, and constraints, thereby supplying a semantic foundation for their consistent interpretation and reuse.

The recent dynamic-graph view of self-evolving agents is particularly close to this perspective because it also treats agent components and relations as explicit graph objects that can change over time \cite{xu2026dynamicgraph}. The distinction is primarily one of organizing question and system scope. Dynamic graph transformation begins from persistent agent evolution and asks how memories, tools, skills, workflows, and inter-agent relations can be represented and rewritten as evolving graph structures. Graph Engineering begins from the organization of System Intelligence and asks how task structures, acting entities, and runtime states should be jointly represented, coordinated, governed, and improved. Evolution is therefore one dimension of Graph Engineering rather than its sole organizing axis.

This system-level view also clarifies the relationship between Graph Engineering and earlier engineering paradigms. \emph{Prompt Engineering} and \emph{Context Engineering} determine how model capabilities are elicited and what information is available at inference time. \emph{Harness Engineering} provides persistent and executable capabilities around the model, while \emph{Loop Engineering} organizes these capabilities into bounded, feedback-driven, and goal-directed execution. Graph Engineering addresses the next organizational scale: how multiple tasks, agents, resources, and evolving runtime states should be explicitly structured and coordinated as a coherent system. Ontology Engineering further provides a shared semantic model through which these system structures can be consistently interpreted and connected. In this sense, the progression from Model Intelligence to Individual Intelligence and ultimately System Intelligence is not defined by adding more components, but by expanding the engineering object from model behavior, to persistent agent execution, and finally to the explicit organization, evolution, and semantic grounding of system-level relationships.

\end{document}